\documentclass[useAMS,usenatbib]{mn2e}

\usepackage{graphicx,psfig,fancyhdr,subfig}
\usepackage{epsfig, psfig, epsf}
\usepackage{amsmath, cancel}
\usepackage{amssymb,float}
\usepackage{lscape}
\usepackage{deluxetable, lastpage, refcount}
\usepackage{wrapfig}
\usepackage{dblfloatfix}
\usepackage{wasysym}
\usepackage{threeparttable}
\usepackage{multirow}
\usepackage{mathtools}

\makeatletter

\title[PCA of CANDELS Morphology]{Beyond Spheroids and Discs: Classifications of CANDELS Galaxy Structure at 1.4 $<$ \MakeLowercase{z} $<$ 2 via Principal Component Analysis}

\author[M. Peth et al.]{Michael A. Peth$^{1}$\thanks{E-mail: mikepeth@jhu.edu},  Jennifer M. Lotz$^{2}$, Peter E. Freeman$^{3}$, Conor McPartland$^{4,2}$,
\newauthor S. Alireza Mortazavi$^{1}$, Gregory F. Snyder$^{2}$,  Guillermo Barro$^{5,6}$, Norman A. Grogin$^{2}$, 
\newauthor Yicheng Guo$^{5}$, Shoubaneh Hemmati$^{7}$, Jeyhan S. Kartaltepe$^{8,9}$, Dale D. Kocevski$^{10}$,  
\newauthor Anton M. Koekemoer$^{2}$, Daniel H. McIntosh$^{11}$, Hooshang Nayyeri$^{5,12}$, Casey Papovich$^{13}$, 
\newauthor Joel R. Primack$^{5}$ and Raymond C. Simons$^{1}$\\
$^{1}$Department of Physics and Astronomy, The Johns Hopkins University, 366 Bloomberg Center, Baltimore, MD 21218\\
$^{2}$Space Telescope Science Institute, 3700 San Martin Dr, Baltimore, MD 21218\\
$^{3}$Department of Statistics, Carnegie Mellon University, 5000 Forbes Avenue, Pittsburgh, PA 15213\\
$^{4}$Institute for Astronomy, University of HawaiÕi at Manoa, 2680 Woodlawn Drive, Honolulu, HI, 96822, USA\\
$^{5}$University of California Observatories/Lick Observatory, University of California, Santa Cruz, CA 95064\\
$^{6}$Department of Astronomy, University of California, 94720 Berkeley, USA\\
$^{7}$Department of Physics and Astronomy, University of California, Riverside, CA\\
$^{8}$National Optical Astronomy Observatory, 950 North Cherry Avenue, Tucson, AZ 85719\\
$^{9}$School of Physics and Astronomy, Rochester Institute of Technology, 84 Lomb Memorial Drive, Rochester, NY 14623\\
$^{10}$Department of Physics and Astronomy, Colby College, Waterville, ME\\
$^{11}$Department of Physics and Astronomy, University of Missouri-Kansas City, 5110 Rockhill Road, Kansas City, MO 64110\\
$^{12}$Department of Physics and Astronomy, University of California, Irvine, CA\\
$^{13}$George P. and Cynthia Woods Mitchell Institute for Fundamental Physics and Astronomy, Texas A$\&$M University, College Station, TX 77843}

\begin{document}
\date{Accepted ??. Received ??; in original form ??}
\pagerange{\pageref{firstpage}--\pageref{LastPage}} \pubyear{2015}
\maketitle
\label{firstpage}

\begin{abstract}
Important but rare and subtle processes driving galaxy morphology and star-formation may be missed by traditional spiral, elliptical, irregular or S\'ersic bulge/disk classifications.  To overcome this limitation, we use a principal component analysis of non-parametric morphological indicators (concentration, asymmetry, Gini coefficient, $M_{20}$, multi-mode, intensity and deviation) measured at rest-frame $B$-band (corresponding to HST/WFC3 F125W at 1.4 $< z <$ 2) to trace the natural distribution of massive ($>10^{10} M_{\odot}$) galaxy morphologies.  Principal component analysis (PCA) quantifies the correlations between these morphological indicators and determines the relative importance of each.  The first three principal components (PCs) capture $\sim$75 per cent of the variance inherent to our sample.  We interpret the first principal component (PC) as bulge strength, the second PC as dominated by concentration and the third PC as dominated by asymmetry.  Both PC1 and PC2 correlate with the visual appearance of a central bulge and predict galaxy quiescence.  PC1 is  a better predictor of quenching than stellar mass, as as good as other structural indicators (S\'ersic-n or compactness).  We divide the PCA results into groups using an agglomerative hierarchical clustering method.  Unlike S\'ersic, this classification scheme separates compact galaxies from larger, smooth proto-elliptical systems, and star-forming disk-dominated clumpy galaxies from star-forming bulge-dominated asymmetric galaxies.  Distinguishing between these galaxy structural types in a quantitative manner is an important step towards understanding the connections between morphology, galaxy assembly and star-formation.


\end{abstract}



\section{Introduction}  

Massive galaxies today form stars at a lower rate than in the past due to many factors. However, we do not have a complete accounting of the processes quenching the star-formation in galaxies.  An increase in the mass/number densities \citep{Tomczak14,vanderWel14a} of massive, red galaxies implies stars are not forming to the same extent they once were.  Each of these observations attempt to connect of observed color (or star-formation rate) and stellar masses to morphology.  The star-formation rate - stellar mass (SFR$-M_*$) relationship shows star-forming galaxies at $z\sim$ 0 follow a ``main sequence'' \citep{Brinchmann04,Wuyts11}.  Galaxies on the main sequence are bluer and have lower S\'ersic-indices than galaxies below the relation.  Massive galaxies with low SFRs are red and have high S\'ersic indices and bulge strengths.  The SFR$-M_*$ morphology relation has been shown to hold out to $z\sim$ 2.5 \citep{Wuyts11}.  However, bulge strength has been described as a ``necessary but not sufficient'' condition for quenching star-formation in $z \lesssim$ 2.2 galaxies \citep{Bell12}.

If the presence of a bulge is not sufficient to fully quench a galaxy, other factors such as size may be important for shutting down star-formation.  At redshifts $z \sim$ 1.5, galaxies of sufficiently high mass and small size are quenched \citep{Barro13}.  This suggests a relationship between so-called ``compactness'' ($\Sigma_{1.5} = M/r_e^{1.5}$) and the specific star-formation rate (sSFR) .  However, the number density of these compact galaxies has been decreasing with the age of the universe.  

The mechanisms for quenching star-formation and transforming the morphology of galaxies are not fully understood.  Proposed mechanisms include: major mergers (e.g. \citealp{Naab06,Hopkins10}); minor mergers (e.g. \citealp{HopkinsHernquist09,Villforth13a,Taniguchi99}); secular processes (for review see \citealp{Kormendy04,Cisternas11}); AGN feedback (e.g. \citealp{SilkRees98,Schawinski06}); and mass quenching \citep{Dekel06,Bell12}.  Comprehensive models of galaxy formation can yield a reasonable link between galaxy morphology and star formation (e.g. \citealp{Snyder15}) but we do not yet have a perfect accounting of how all these processes might contribute.  

As a result, two evolutionary tracks have been developed to explain the disappearance of compact, quenched galaxies: (1) major mergers at $z \sim$ 2-3 quickly cause a galaxy to quench, which later grow through minor mergers and gas accretion; (2) violent disk instabilities/secular processes/minor mergers at z$\sim$1.5 cause a slower decline in star-formation and simultaneous size growth before the quiescent phase.

To study the processes driving evolution, we need a method to effectively and efficiently characterize the structures and shapes of galaxies.  Visual classifications have been used since the discovery of galaxies, and have subsequently been adapted to fit modern surveys (e.g. Galaxy Zoo, \citealp{GalaxyZoo,Jeyhan14}).  Visual classifications can find subtle structural elements  possibly missed by an automated routine.  However, human classifications of galaxies can be very time consuming and subjective.  Additionally, galaxy structure at high redshift does not always correspond to the local Hubble sequence\citep{Bruce12,Bell12,Kriek09,Lee13}.  Disk-dominated galaxies can appear clumpy \citep{Forster09} and bulge-dominated galaxies can be compact, very red and massive, but possess no extended envelope (e.g. \citealp{vanDokkum08}).  


        
    GALFIT  \citep{Peng02,Peng10} is an automated technique often used to classify galaxies that models the light profile of galaxies with a S\'ersic profile ($r^{-1/n}$).  Disks have exponential light profiles ($n$=1), while ellipticals are best fit by a de Vaucouleurs profile ($n$=4). GALFIT is sensitive to small galaxies, can distinguish overlapping light profiles of nearby galaxies, incorporates the point spread function of a specific field/detector, and most importantly is easy to interpret.  However, GALFIT assumes a symmetric and smooth light profile, which at times can be problematic.  This assumption does not hold for irregular galaxies, merger remnants, and disk galaxies with bars or clumps.  GALFIT is typically used to calculate a single S\'ersic fit to the light profile of the galaxy. Two-component S\'ersic fits has also been used to combine disk and bulge components (e.g. \citealp{Simard11,Bruce14a,Bruce14b}); however, calculations can be quite CPU intensive, sometimes needing weeks to finish.
    
    Quantitative non-parametric morphological statistics characterize galaxy structure and do not assume an analytic light profile.  This fact allows us to apply automated characterization to irregular galaxies as well.  Examples of non-parametric morphological indicators include: concentration index ($C$, \citealp{Bershady00,Conselice03}), asymmetry ($A$, \citealp{Conselice00}), Gini coefficient ($G$, \citealp{Abraham03,Lotz04}), $M_{20}$  \citep{Lotz04}, and three new statistics from \citet{Freeman13}: Multimode ($M$), Intensity ($I$), and Deviation ($D$).  The $MID$ statistics have been found to be sensitive to mergers and clumpy star-formation, even at high redshift \citep{Freeman13}.
    
    
    However, for many galaxies these statistics can be strongly correlated.  Moreover, cosmological models of galaxy formation yield a picture in which these structures can evolve quickly along diverse paths, thereby motivating a broad deep classification system \citep{Snyder14}.  Therefore we require further analysis to understand the inherent relationships among these statistics, and between these statistics and galaxy assembly processes.
    
     Principal component analysis (PCA) is a simple way to reduce the dimensionality, break internal degeneracies and find the natural distributions of data in parameter space.  To eliminate degeneracies inherent in these morphological statistics we performed a PCA using 7 non-parametric morphology measurements on 1244 galaxies from 1.36 $< z <$ 1.97.  PCA has been shown to efficiently classify galaxies (e.g. \citealp{TP12}; the Zurich Estimator of Structural Types (ZEST), \citealp{Scarlata07}). A few studies immediately capitalized on the ZEST classifications to study the number density evolution of disk galaxies \citep{Sargent07}, the luminosity function evolution for elliptical galaxy progenitors \citep{Scarlata07b}, and the evolution of the galaxy merger rate to $z\sim$ 1 \citep{Kampczyk07}.

In this paper, we use PCA and hierarchical clustering to classify galaxies based on their structure. These classifications allow us to characterize galaxies by more subtle means than the traditional Hubble sequence scheme.  We can test the mechanisms which cause galaxies to reassemble and/or influence star-formation by tracking how morphologies change across time.  This places vital constraints on the physical mechanisms assembling galaxies and quenching star-formation.  

This paper is structured as follows: $\S$\ref{section:data} details the CANDELS data set and our sample selection; $\S$\ref{section:non-parametric-morphologies} defines the non-parametric morphologic measurements we perform, their associated error estimates and the principal component analysis as applied to our data set; $\S$\ref{section:groups} describes the results of our PCA, the clustering algorithm and convex hull method used for grouping galaxies, a test of the group, and PCA reliability and descriptions of the final galaxy groups; $\S$\ref{section:morphgroups} describes the general morphological characteristics of the galaxies in each group; $\S$\ref{section:discussion} details the relationship between stellar mass, quenching, and our group classifications.  Additionally, the disagreement between S\'ersic and visual classifications with PC-based classification (especially for compact/bulge-dominated galaxies) is discussed.

All magnitudes are quoted in the AB system.  A standard $\Lambda$CDM cosmology of $H_0$ = 70 km s$^{-1}$ Mpc$^{-1}$, $\Omega_M$ = 0.3, and $\Omega_{\Lambda}$ = 0.3 is used throughout this work.

\section{Data}\label{section:data}
	The Cosmic Assembly Near-IR Deep Extragalactic Legacy Survey (CANDELS, PIs: S. Faber and H. Ferguson; \citealt{Grogin11} and \citealt{Koekemoer11})  observed 5 heavily studied fields (of which we use UDS, GOODS-S and COSMOS) with the \textit{Hubble Space Telescope} (HST).  High resolution imaging by Wide Field Camera 3 (WFC3) in near-infrared bands F125W ($J$) and F160W ($H$), combined with observations from the Advanced Camera for Surveys (ACS) in visible bands F606W ($V$)  and F814W ($I_w$) constitute the new measurements in the CANDELS program.  For the purposes of our study, we initially focus only on the F125W WFC3 images.  Future work will study the evolution of galaxy morphology at a consistent rest-frame wavelengths.
	
	We use the CANDELS $H$-band (F160W) selected multi-wavelength catalogs (UDS, \citealp{Galametz13}; GOODS-S, \citealp{Guo13}; COSMOS, Nayyeri et al., in prep), photometric redshifts \citep{Dahlen13}, non-parametric morphologies (this work), S\'ersic parameters \citep{vanderWel12}, visual classifications \citep{Jeyhan14}, rest-frame photometry, and stellar masses (this work).  The limiting magnitude for HST/WFC3 F125W and F160W are 27.35 and 27.45 respectively with FWHM of $\sim$0.135'' and  $\sim$0.15'' respectively.  \citet{Galametz13} outlined the techniques used to create the photometric catalogs.   
	
	The photometric redshift catalogs of \citet{Dahlen13} are the combination of multiple different photometric redshift calculating codes and techniques which reduce the scatter of photometric redshifts (to $\sigma \sim$ 0.03, with an outlier fraction of 3 percent).  Throughout the rest of this paper, we use $z$ to denote the average photometric redshift in these CANDELS catalogs \citep{Mobasher15}.  
	
	Rest-frame $U-V-J$ colors were calculated by the sed-fitting code EAZY \citep{Brammer08}, using the empirical local galaxy templates of \citet{Brown14}.  Stellar masses were computed with FAST \citep{Kriek09},  assuming  \citet{BC03} delayed exponential star-formation histories, a \citet{Chabrier03} initial mass function,  \citet{Calzetti00} dust attenuation, and solar metallicities.
	
	
	
\subsection{Sample Selection Criteria}\label{section:sample_selection}
We select bright ($H <$ 24.5), massive ($M_* > 10^{10} M_{\astrosun}$) galaxies with 1.36 $< z <$ 1.97 galaxies measured in F125W ($J$).  This band approximately corresponds to rest-frame optical $B$-band at these redshifts. This redshift range provides a large sample of galaxies measured in a constant rest-frame waveband, and offers a high enough redshift to have a different morphological distribution from a local sample.  At this redshift and magnitude, the CANDELS surveys are mass-complete down to 10$^{10} M_{\astrosun}$ \citep{Wuyts11}.  In our sample of UDS, COSMOS and GOODS-S there are a total of 6269 galaxies with $H <$ 24.5 and $M_* > 10^{10} M_{\odot}$.  Of those galaxies 1539 are within our redshift range (1.36 $<$ z $<$ 1.97).  


The following affect our sample completeness: high signal-to-noise (per pixel) measurements (S/N $>$ 4), an internal morphology quality flag = 0, and a well measured concentration (i.e. C $\ne$ -99) requirement.  The quality flag requirement removes objects from the sample with discontiguous segmentation maps resulting from low surface brightness, and/or poor masking of bright neighbors.  In Appendix \ref{section:appendix} we include a brief discussion of galaxies with a quality flag = 1.  The concentration requirement removes the contamination from poorly measured galaxies on the overall PCA.  For some galaxies, $r_{20}$ (and thus $C$) can not be accurately measured because either the object is too small, or there is a bright point source disrupting the light profile (see $\S$\ref{section:concentration}).  The concentration requirement reduces the total of galaxies in the sample to 1482.  The FLAG requirement reduces the sample to 1250.  The signal-to-noise, FLAG and well measured concentration requirements together reduce our final sample to 1244 galaxies.

    \section{Morphological Measurements} \label{section:non-parametric-morphologies}
        

\subsection{Non-parametric Morphology}

We focus on non-parametric morphology statistics: concentration, asymmetry, Gini coefficient, $M_{20}$, along with three new statistics from \citealt{Freeman13}: multi-mode, intensity and deviation.  The code for calculating the morphological statistics (originally developed by \citealt{Lotz08a}) has been modified to include the new statistics and accommodate much larger input images.  The code is applied to the CANDELS F125W mosaics using the F160W detected catalogs and segmentation maps as the input.

    

    \subsubsection{Petrosian Radius}
    The Petrosian radius $r_p$ is the radius we set to where the surface brightness $\mu$ is 20 per cent of the the mean interior surface brightness (Eq. \ref{eq:petrosian_r}; \citealp{Petrosian76}).  The Petrosian radius is more robust to surface brightness dimming than isophotal sizes are.  We can measure the same physical portions for galaxies at a variety of redshifts (e.g. \citealp{Lotz04}).
    
    \begin{equation} \label{eq:petrosian_r}
    0.2 = \frac{\mu(r_p)}{\bar{\mu}(r<r_p)}
    \end{equation}
    

    \subsubsection{Concentration}\label{section:concentration}
The concentration index ($C$; \citealp{Bershady00,Conselice03}) is the ratio of the circular radius containing 80 per cent ($r_{80}$) of a galaxy's light (as measured within  1.5 Petrosian radii) to the radius containing 20 per cent ($r_{20}$) of the light (Eq. \ref{eq:concentration}).  A large concentration value indicates a majority of light is concentrated  at the center of the galaxy.  Elliptical galaxies and bulge-dominated spirals have high concentration values.  However, a spiral or irregular galaxy with diffuse light profile and weak/no bulge will have low concentration values.  

    \begin{equation}  \label{eq:concentration}
    	C = 5\log\left(\frac{r_{80}}{r_{20}}\right)
\end{equation}

For some galaxies $r_{20}$ (and thus $C$) can not be accurately measured because either the object is too small, or there is a bright point source disrupting the light profile.  These galaxies instead have unphysical concentration values ($C <$ 0) and are not included in the definition of our principal components (see $\S$\ref{section:sample_selection}).



    \subsubsection{Asymmetry}
    Asymmetry ($A$; \citealp{Conselice00}) measures the difference between the image of a galaxy ($I_{x,y}$) and the galaxy rotated by 180 degrees ($I_{180(x,y)}$; Eq. \ref{eq:asymmetry}).  This determines a ratio of the amount of light distributed symmetrically to all light from the galaxy.  $A$ is calculated from a sum of all pixels within 1.5 Petrosian radii from the center of the galaxy.   We then correct by $B_{180}$, which is the average asymmetry of the background.  An initial guess for the center of rotation is defined by the physical center, but is updated through an iterative process. This process continues until a global minimum value for $A$ is found \citep{Conselice00}.
    
      \begin{equation}  \label{eq:asymmetry}
    	A = \frac{\sum_{x,y} | I_{(x,y)} - I_{180(x,y)}|}{2\sum | I_{(x,y)}|} - B_{180}
\end{equation}
    
     Due to their uniform morphologies and lack of structure elliptical galaxies typically have small asymmetry values ($A$ $\sim$ 0.02). Meanwhile spiral galaxies usually have values between A $\sim$ 0.07 to 0.2 \citep{Conselice14}.  This statistic is most useful for identifying irregular galaxies because they appear lopsided or ragged.  Visually inspected merger remnants can have $A$ $\gtrsim$ 0.3 \citep{Conselice03}.  The asymmetry statistic is more sensitive to gas-rich mergers than to gas-poor or minor mergers  \citep{Lotz10a,Lotz10b}.

     If the local background is high and the galaxy is is sufficiently low surface brightness then negative $A$ values are measured. This is consistent with measurement errors (see $\S$\ref{section:error}). 


    \subsubsection{Gini Coefficient}
The Gini coefficient ($G$; \citealp{Lorenz1905,Abraham03,Lotz04}) is a statistic adapted from economics that measures the equality of light distribution in a galaxy.  The Gini coefficient is defined by the Lorenz curve of the galaxy's light distribution, and is not affected by spatial position.  This implies that only the amount of light distribution matters, which differentiates the Gini coefficient from the concentration statistic \citep{Conselice14}.

The pixels are ranked by increasing flux value, then $G$ is determined by Eq. \ref{eq:gini},  where $n$ is the number of pixels in the galaxy's segmentation map, $X_i$ is the pixel flux at the rank $i$ pixel and $\bar{X}$ is the mean pixel value.

\begin{equation}  \label{eq:gini}
    	G = \frac{1}{\bar{X}n(n-1)}\sum_i^n(2i-n-1)X_i
\end{equation}

A galaxy with equally distributed light will have a Gini coefficient approaching 0.  Conversely,  a galaxy with a large fraction of light concentrated on a few pixels will have a Gini coefficient closer to 1.  Elliptical galaxies and galaxies with bright nuclei have high Gini coefficients, while disks and galaxies with a uniform surface brightness will have low Gini coefficients.

\begin{figure*}
	\centering
	 \makebox[\textwidth]{\includegraphics[width=1.15\textwidth]
     {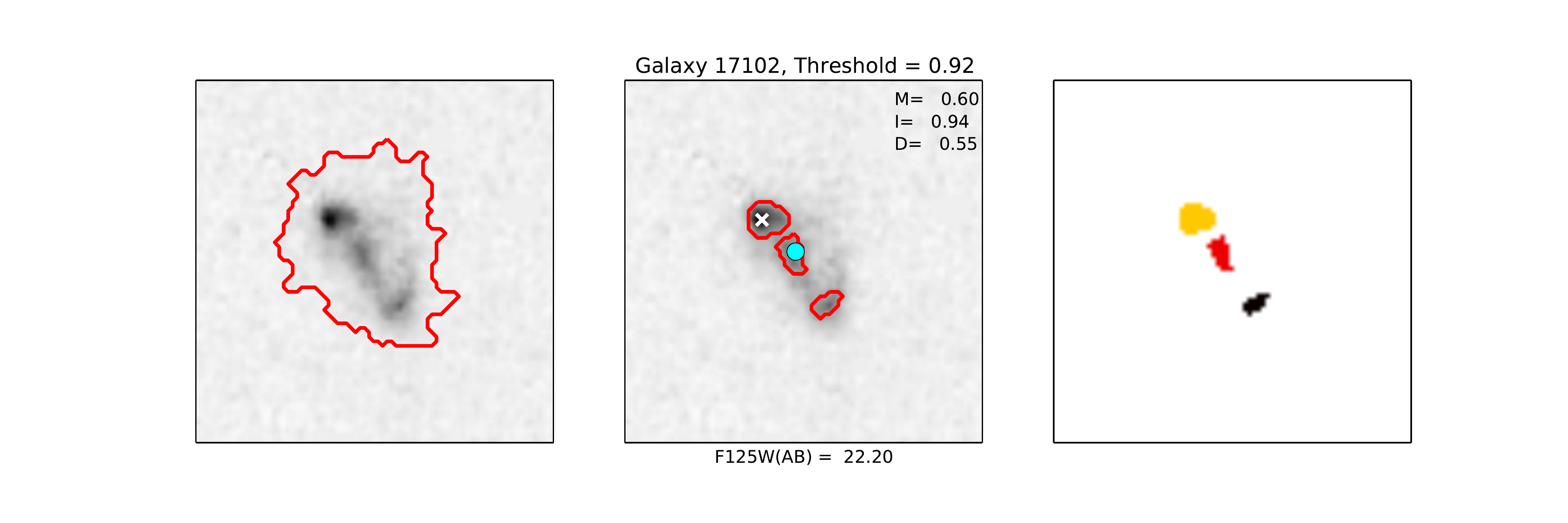}}
      \caption[Example of $MID$ statistics]
      {F125W (AB) = 22.2 CANDELS galaxy image is shown to demonstrate the $M$, $I$ and $D$ statistics.  The left panel shows the image of the galaxy outlined by the segmentation map created using our morphology code.  The middle panel shows red outlines describing the clumps found when calculating the $M$ statistic.  The white $X$ displays the location of the brightness distribution peak, and the cyan circle represents the location of the intensity centroid used to calculate the $D$ statistic ($\S$\ref{section:deviation}).  The right panel color codes the clumps for easy identification.  This galaxy is highly disturbed and is broken into 3 bright regions, with the brightness peak well separated from the intensity centroid.  The threshold value ($q_l$) in this case is 0.92, which represents the threshold where the $M$ statistic was maximized.}
      \label{fig:mid_example1}
    \end{figure*}


    \subsubsection{M$_{20}$}

    The second order moment of the brightest regions of a galaxy ($M_{20}$; \citealp{Lotz04}) traces the spatial distribution of any bright clumps.  When used in tandem with the Gini coefficient, $M_{20}$ can be an effective tool for differentiating galaxies with bright off-center clumps (such as irregular galaxies) from those with one bright central region (such as the bulge of a spiral galaxy).  We define the regions representing the brightest 20 per cent of the galaxy (Eq. \ref{eq:flux20}), and then calculate the spatial distribution of those pixels as an offset from the central pixel.  The center is defined as the position minimizing $M_{tot}$.
        \begin{equation}  \label{eq:flux20}
    	\sum_i f_i < 0.2 f_{tot}
\end{equation}

\begin{equation}  \label{eq:mtot}
	M_{tot} = \sum_i^n M_i = \sum_i^n f_i \left [(x_i - x_c)^2 + (y_i - y_c)^2\right]
\end{equation}

Finally we calculate the second order moment (Eq. \ref{eq:m20}).

\begin{equation}  \label{eq:m20}
	M_{20} = \log\left(\frac{\sum_i M_i}{M_{tot}}\right)
\end{equation}

Values for the $M_{20}$ statistic are generally between -0.5 and -2.5.  Elliptical galaxies have $M_{20}$ closer to -2.5 signifying a lack of bright-off center clumps. Meanwhile disk galaxies can have $M_{20} >$ -1.6 when, for example, bright star-forming knots are present.  Similar to concentration, $M_{20}$ is biased low for galaxies where the brightest 20 per cent light is unresolved.

     \subsubsection{Multi-mode}
    The multi-mode ($M$) statistic is the ratio, in pixels, of the two brightest regions of a galaxy (adapted from \citealp{Freeman13}).  Bright regions are determined via a threshold method where  $q_l$ represents the normalized flux value, and $l$ per cent of pixel fluxes are less than $q_l$.  This creates a new binary image $g_{i,j}$ where 1 represents fluxes larger than $q_l$ and 0 represents fluxes less than $q_l$ (Eq. \ref{eq:g_multimode}).
    
    \begin{equation}  \label{eq:g_multimode}
    	g_{i,j} =
  \begin{cases}
   1 & f_{i,j} \geq q_l  \\
   0       & \text{otherwise}
  \end{cases}
\end{equation}

The number of pixels in contiguous groups of pixels with value 1 are then sorted in descending order by area.  The 2 largest groups ($A_{l,(2)}$ and $A_{l,(1)}$)  define an area ratio $R_l$: 

\begin{equation}  \label{eq:r_multimode}
	R_l = \frac{A_{l,(2)}}{A_{l,(1)}} 
\end{equation}

The previous two steps are recomputed for various normalized flux levels $l$, and the $M$ statistic is the maximum $R_l$ value (Eq. \ref{eq:m_multimode}).  Values approaching 1 represent multiple nuclei, while values near 0 are single nuclei systems.  

\begin{equation}  \label{eq:m_multimode}
	M = \max R_l
\end{equation}

This formulation is slightly revised from \citet{Freeman13} to limit the $M$ statistic to values between 0 and 1.   \citet{Freeman13} multiplies Eq. \ref{eq:r_multimode} by an additional factor of $A_{l,(2)}$ to limit the effect of hot pixels.  However, this adds a size dependent factor to the calculation.  Because we wish to measure $M$ values for galaxies at a variety of angular distance scales, it is important to have a size independent measure.  For illustrative purposes, Fig. \ref{fig:mid_example1} shows an example of how the $MID$ statistics are calculated.  In small galaxies that are poorly resolved $A_{l,(1)}$ is very small (approaching zero) and we set M=-99.  We have tested the result of setting M=-99 values to M=0 but find the PC weights and group assignments are very similar to the original values.


    \subsubsection{Intensity}  
    Intensity ($I$) is the ratio, in flux, of the two brightest regions \citep{Freeman13}.  The galaxy image is first smoothed by a symmetric bivariate Gaussian kernel.  Regions are defined using maximum gradient paths, where the surrounding eight pixels of every pixel are inspected and the path of maximal intensity increase is followed until a local maximum is reached.  Regions consist of pixels linked to a unifying local maximum.  The fluxes within these groups are summed and sorted into descending order (by total flux) leading to our intensity ratio:
    
    \begin{equation}\label{eq:intensity}
    	I = \frac{I_{(2)}}{I_{(1)}}
\end{equation}

Similar to the $M$ statistic, elliptical galaxies with a bright bulge have $I \sim$ 0, while disk galaxies with bright clusters of star-formation are more likely to have $I$ values approaching 1.

    
    \subsubsection{Deviation} \label{section:deviation}
    Deviation ($D$) measures the distance between the intensity centroid of a galaxy and the center of the brightest region (\citealp{Freeman13}, Eq. \ref{eq:centroid} and Eq.\ref{eq:deviation}).  Disk and spheroidal galaxies have deviation values near 0 because their central bulges typical possess the brightest pixels.  On the other hand, a high deviation value indicates a galaxy has bright star forming knots significantly separated from the intensity centroid (e.g. Fig. \ref{fig:mid_example1}).
    
    \begin{equation} \label{eq:centroid}
    (x_{cen},y_{cen}) = \left ( \frac{1}{n_{seg}}\sum_i\sum_j i f_{i,j} ,  \frac{1}{n_{seg}}\sum_i\sum_j j f_{i,j} \right )
    \end{equation}
    
    The deviation statistic $D$ is the Euclidean distance (in pixels) between the intensity centroid and brightest pixel scaled by a crude estimate of a galaxy's radius based upon the number of pixels comprising the galaxy.

    \begin{equation} \label{eq:deviation}
    	D = \sqrt{\frac{\pi}{n_{seg}}}\sqrt{(x_{cen} - x_{l_{(1)}})^2 + (y_{cen} - y_{l_{(1)}})^2}
\end{equation}

 \subsection{Morphological Principal Components}
    Principal component analysis (PCA) is a linear transformation of multivariate data.  This defines a set of uncorrelated axes, called principal components (PCs), which are ranked by the variance they capture \citep{Pearson1901, astroMLText}.  A linear combination of the original data and eigenvector solutions (also called weights) project the original data onto the PCs.     Principal component analysis is a simple way to reduce the dimensionality and find the natural distributions of data in parameter space.  PCA is able to determine the correlations between the input data and can find relationships missed by other means.
         
    We begin by ``whitening'' the data, i.e. we subtract the mean of each morphological measurement and divide by the standard deviation of each feature.  By dividing our data by feature variance we remove the effects of mixed units.  We calculate the singular value decomposition ({\bf $x_{ij} = V\Sigma V^T$}, SVD) of the ``whitened'' data matrix ({\bf x$_{ij}$}).  An SVD decomposes the original data into a diagonal matrix (${\bf \Sigma}$) containing eigenvalues ($e$)  and a non-diagonal matrix {\bf V} containing the expansion coefficients (aka weights).  The eigenvalues determine how important each principal component is to explaining the original data set.  The eigenvectors are rank ordered by their associated eigenvalue.  We then project our ``whitened'' data onto our new eigenbasis to calculate the principal component scores, which inform us how similar are data points to each other ($PC_i$, Eq. \ref{eq:pca}).


 \begin{equation} \label{eq:pca}
	PC_i = \sum_{j=1}^N V_{ji}x_{j}(i=1,...,N)
\end{equation}

Table \ref{table:pc_weights} shows the correlations and importance of different statistics across the eigenvector solutions of the principal component analysis.  The scree value ($e^2$/$\sum e^2$) represents the amount of variance captured by a single principal component.  The scree values demonstrate that the first 3 PCs account for $>$75 per cent of the variance in the data.  The fact that PC1 only accounts for 40 per cent of the variance shows that more than a single parameter is needed to define a galaxy.  The error estimates are the result of the scattering method described in $\S$\ref{section:error}.
  
\begin{figure}
  \centering
    \includegraphics[width=0.5\textwidth]
          {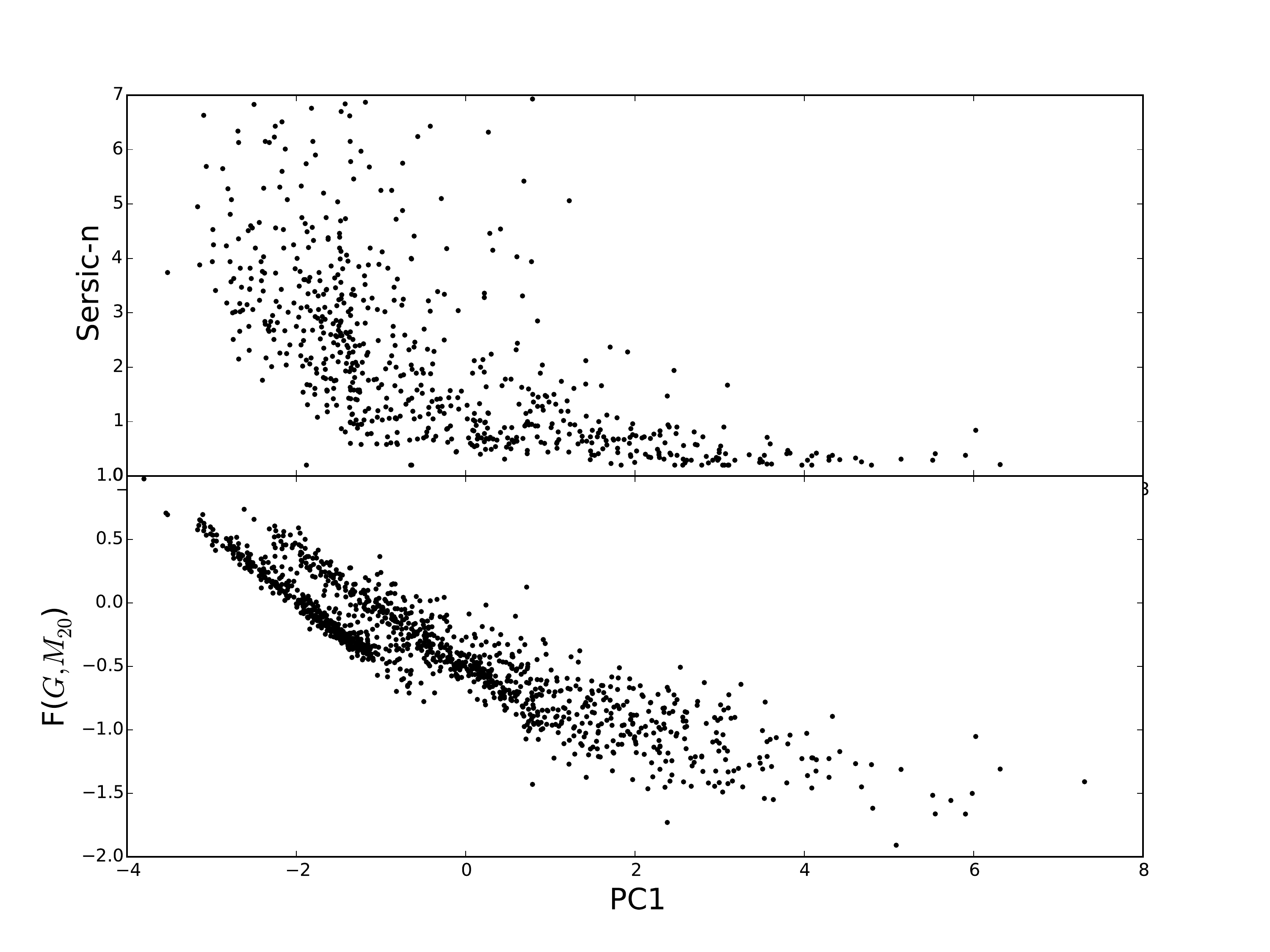}
            \caption[PC1 vs Sersic Index and Gini-$M_{20}$ Bulge Strength]
 {PC1 v. Sersic Index and PC1 v. the Gini-$M_{20}$ bulge strength metric ($F$, \citealp{Snyder15}).  PC1 is anti-correlated with S\'ersic index and the Gini-$M_{20}$ Bulge factor, $F$, and thus low PC1 values are indicative of a strong central bulge.  Small galaxies can have M = -99 which shifts PC1 and leads to the two parallel stripes.  See $\S$\ref{section:morphgroups} for more on how group 6 galaxies are different from the remainder of the sample. }
      \label{fig:pc1_fgm20_n}
    \end{figure}

PC1 is highly dependent upon $M$, $I$, $D$, $M_{20}$ and the Gini coefficient.  We interpret PC1 as a ``bulge strength'' indicator given the correlation with $G-M_{20}$ and the  importance of the $MID$ statistics.  Fig. \ref{fig:pc1_fgm20_n} shows the relationship between PC1, S\'ersic index and the Gini-$M_{20}$ ``bulge strength'' (Eq. \ref{eq:f} and \ref{eq:f_gm20}) the vector of correlations between Gini and $M_{20}$; \citealp{Snyder15}).  Galaxies with low PC1 values have high Sersic indices and high F indicative of strong bulges, while galaxies with higher PC1 values have progressively smaller bulges and more prevalent disc properties (see $\S$\ref{section:morphgroups} for more on the physical and visual properties of specific groups).  We observe two correlations between $F$ and PC1 which corresponds to different groups of galaxies.  Additionally, the two parallel stripes of data seen in Fig. \ref{fig:pc1_fgm20_n} are the result of M=-99 outlier values shifting PC1.  We have tested the result of setting M=-99 values to M=0 and find that the PC eigenweights and the group classifications are very similar to our original values.

\begin{equation}
\label{eq:f}
F = -0.693M_{20}+4.95G-3.85
\end{equation}

\begin{equation}
\label{eq:f_gm20}
F(G,M_{20}) = \begin{dcases*}
        |F|  & G $\geq$ 0.14$M_{20}$ + 0.778 \\
        -|F| & G $<$ 0.14$M_{20}$ + 0.778
        \end{dcases*}
\end{equation}

PC2 is highly dependent upon concentration, and is larger for galaxies with bright centers and extended envelopes.  PC3 is dominated by asymmetry and is larger for disturbed galaxies.  The other principal components are harder to interpret, but are also less important as evidenced by their lower scree values.  It is interesting to note PC1 defines a bulge strength but is not dependent on concentration (Eq. \ref{eq:concentration}). Concentration for very small ($r_e <$ 2 kpc), high Sersic ($n >$ 2.5) galaxies is strongly biased down (see Appendix \ref{section:c_n}).  This bias is potentially important for $\sim$14 per cent of our sample. 
  


    We performed tests on how PCA results are affected by whitening the data set using the interquartile range (IQR) statistic instead of a standard deviation. The eigenvectors calculated using either whitening method are mainly consistent.  However, we chose to use the standard deviation to whiten our data because the PC weights are more volatile when calculated with an IQR whitened data set.  In particular, the weight in PC3 describing concentration has a variance nearly nine times larger when calculated for an IQR-scaled data set compared to a standard deviation-scaled data set.
    




\begin{table*}
\centering

\caption{PC Weights with error estimates based on a bootstrap scattering method}
\begin{tabular}{lrrrrrrr}
\hline
Parameter & PC1 & PC2 & PC3 & PC4 & PC5 & PC6 & PC7 \\
\hline
Scree value & 0.41 &  0.19 &  0.15 &  0.08 & 0.06 & 0.06 & 0.05 \\
Concentration & -0.06 $\pm$ $~$ 0.02 & 0.74 $\pm$ 0.01 & -0.35 $\pm$ 0.03 & 0.19 $\pm$ 0.04 & -0.31 $\pm$ 0.11 & 0.03 $\pm$ 0.12 & -0.43 $\pm$ 0.07 \\ 
M$_{20}$ & 0.48 $\pm$ $<$0.01 & -0.03 $\pm$ 0.02 & -0.12 $\pm$ 0.02 & 0.16 $\pm$ 0.07 & -0.67 $\pm$ 0.19 & 0.07 $\pm$ 0.19 & 0.52 $\pm$ 0.09 \\ 
Gini & -0.45 $\pm$ $~$  0.01 & 0.27 $\pm$ 0.02 & 0.12 $\pm$ 0.02 & 0.45 $\pm$ 0.05 & 0.11 $\pm$ 0.16 & -0.46 $\pm$ 0.13 & 0.53 $\pm$ 0.07 \\ 
Asymmetry & 0.00 $\pm$ $<$0.01 & 0.41 $\pm$ 0.03 & 0.82 $\pm$ 0.02 & -0.31 $\pm$ 0.03 & -0.18 $\pm$ 0.05 & 0.18 $\pm$ 0.05 & 0.06 $\pm$ 0.03 \\ 
Multi-mode & 0.38 $\pm$ $<$0.01 & 0.45 $\pm$ 0.02 & -0.27 $\pm$ 0.02 & -0.30 $\pm$ 0.07 & 0.56 $\pm$ 0.11 & 0.14 $\pm$ 0.15 & 0.40 $\pm$ 0.07 \\ 
Intensity & 0.49 $\pm$ $<$0.01 & 0.04 $\pm$ 0.01 & 0.13 $\pm$ 0.01 & -0.13 $\pm$ 0.03 & 0.02 $\pm$ 0.10 & -0.82 $\pm$ 0.15 & -0.24 $\pm$ 0.07 \\ 
Deviation & 0.43 $\pm$ $<$0.01 & 0.00 $\pm$ 0.01 & 0.30 $\pm$ 0.01 & 0.73 $\pm$ 0.04 & 0.31 $\pm$ 0.15 & 0.25 $\pm$ 0.10 & -0.18 $\pm$ 0.06 \\ 
\hline
  \label{table:pc_weights}
  \end{tabular} 

\end{table*}

\begin{figure*}
\centering
     \makebox[\textwidth]{\includegraphics[width=1.15\textwidth]
     {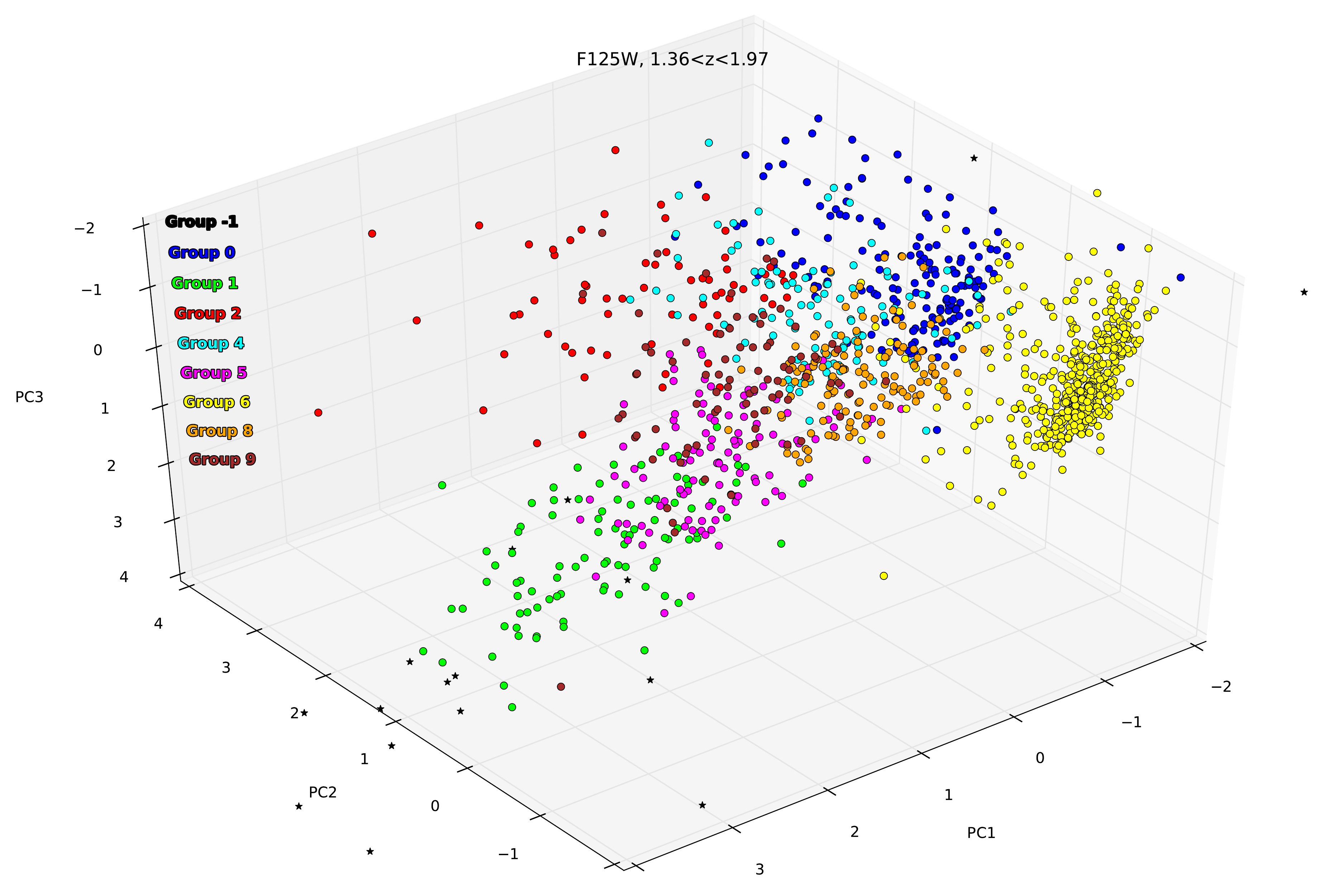}}
          
            \caption[3D PC1-PC2-PC3 color-coded by group ]
      {PC1 v. PC2 v. PC3  for our sample of $M_*$ $>10^{10}$ $M_{\odot}$, 1.36 $< z <$ 1.97 galaxies, color-coded by their hierarchical cluster definitions.  PC1 anti-correlates with bulge strength,  PC2 is dominated by concentration, and PC3 is dominated by asymmetry (see Table \ref{table:pc_weights}).  Group -1 galaxies ($black$ $stars$) are outliers from remaining groups, initially they comprised groups 3 and 7. }
      \label{fig:f125w_cluster}
    \end{figure*}




    \begin{figure}
       \centering
     \includegraphics[width=0.45\textwidth]
	{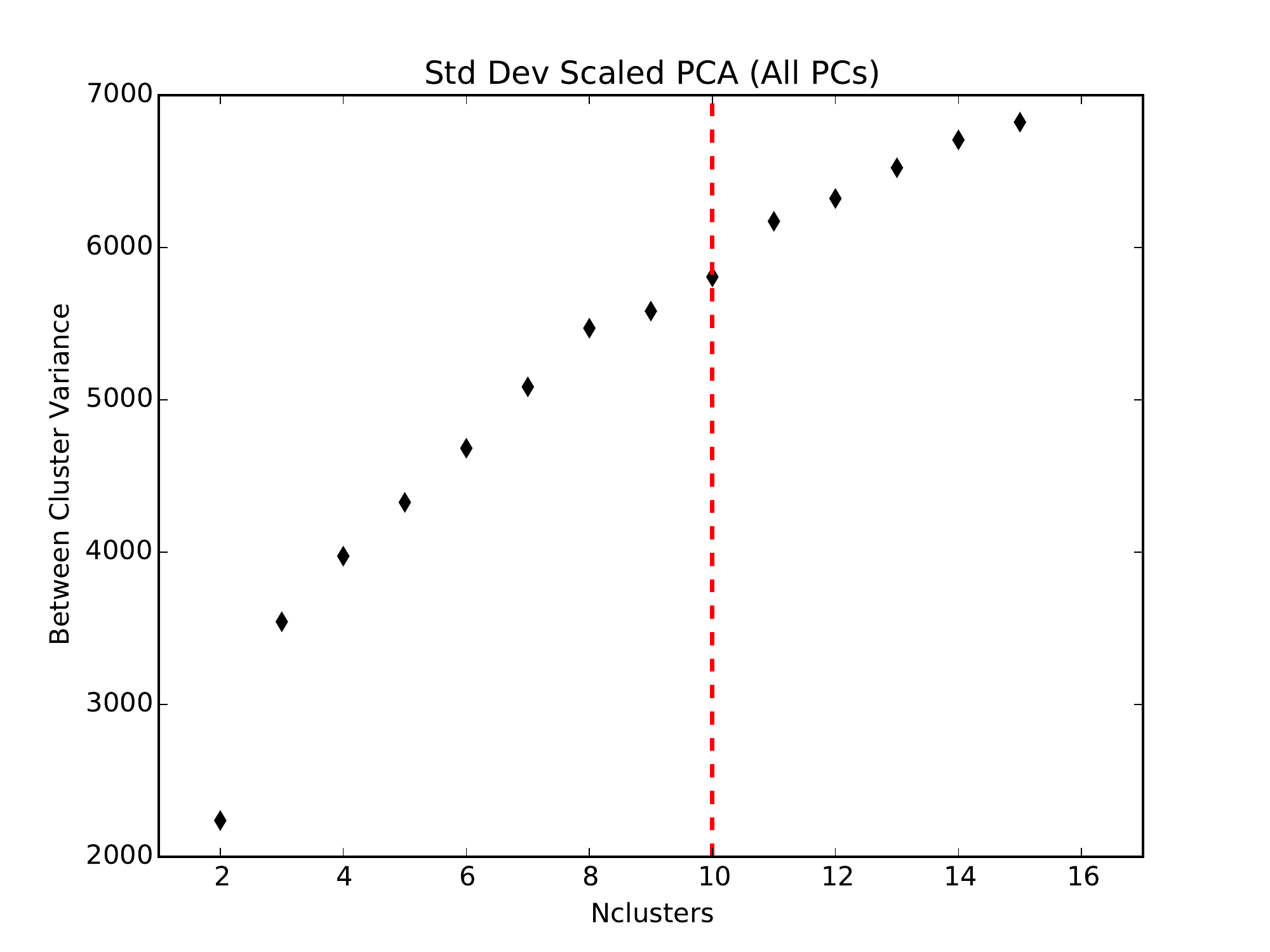}
      \caption[Between cluster variance vs. N clusters]
      {The amount of between-cluster variance as a function of the number of clusters grouped by the Ward hierarchical agglomerative clustering routine.  The between-cluster variance is the sum of the distances from the centroid of each cluster to the centroid of all the data.  Eventually this value grows to the total variance in the data when the number of clusters equals the number of data points.}
     \label{fig:cluster-variance}
    \end{figure}

   
\begin{figure*}
	\centering
		\subfloat[Magnitude vs $\Delta$(GOODS - UDF)]{
			\includegraphics[width=\textwidth]{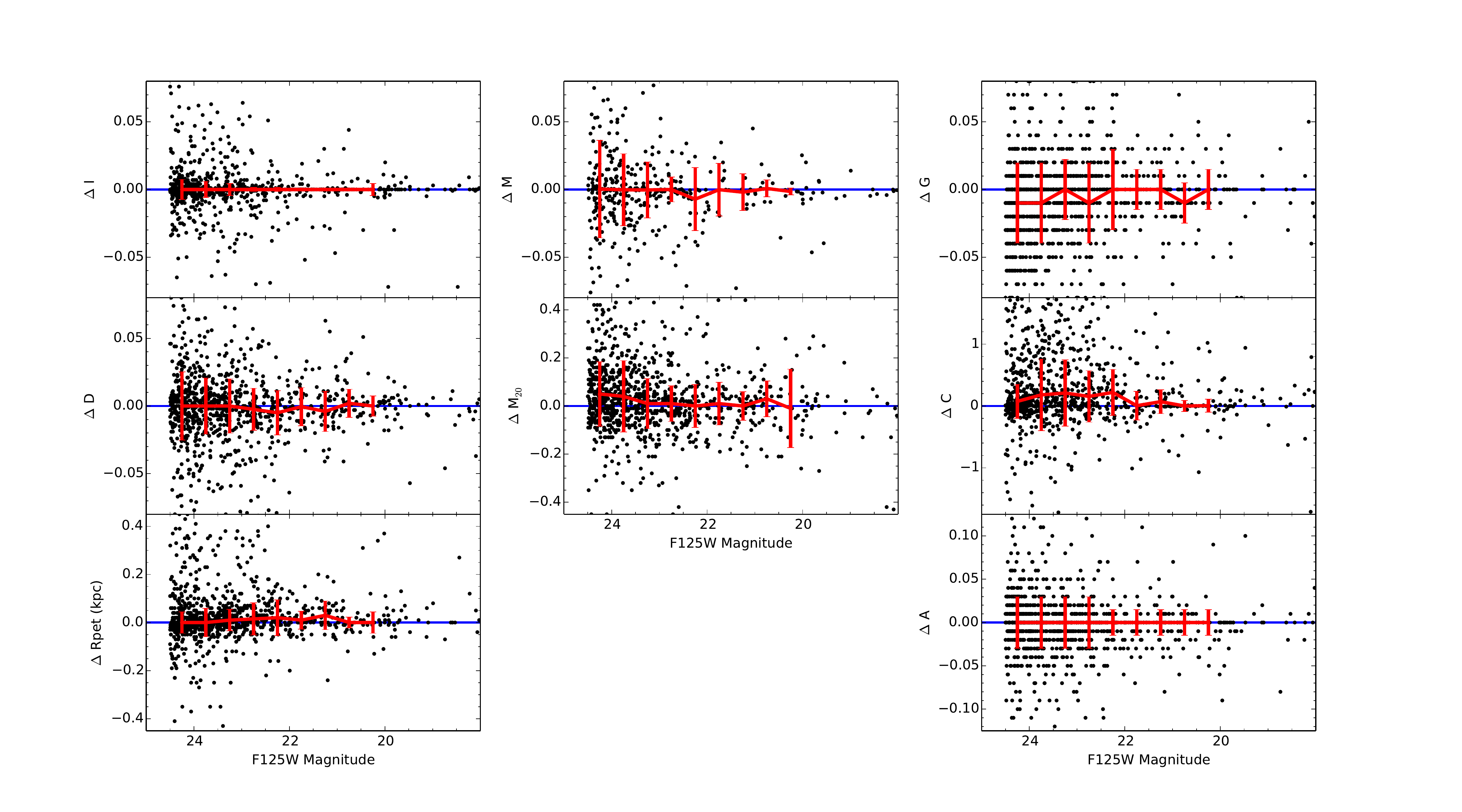}}
			
		\subfloat[Surface Brightness vs $\Delta$(GOODS - UDF)]{
			\includegraphics[width=\textwidth]{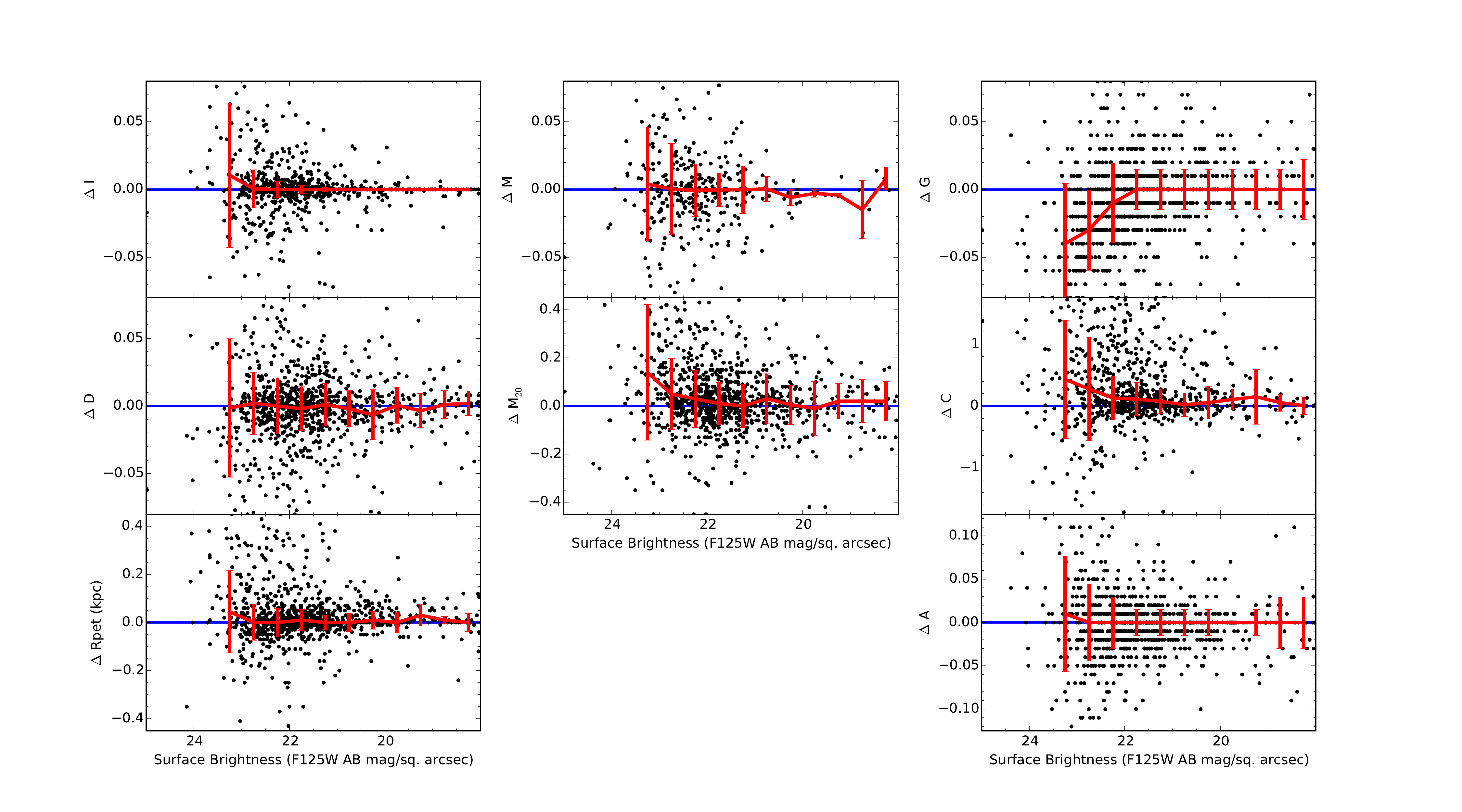}}
		   
	\caption[Magnitude vs. $\Delta$(GOODS - UDF)  morphological statistics]
	{Magnitude and Surface Brightness vs. $\Delta$[GOODS - UDF]  morphological statistics as measured in wide-field imaging of GOODS-S compared to the deep imaging of UDF.  Red error bars represent the median $\Delta$ morphology value binned in magnitude (or surface brightness) bins of 0.5.  Error bars represent the median absolute deviation of each bin corresponding to a 1$\sigma$ deviation.}
	\label{fig:errors}
\end{figure*}

\section{PCA-Morphology Group Properties} \label{section:groups}
\subsection{Defining PCA morphology groups}\label{section:ward}

Studies using PCA usually only select the top eigenvectors to represent the data.  However, this is not a requirement of the analysis. In our case, the number of variables is not very large and thus retaining the entire parameter space is not a computationally expensive procedure.  We aim to reconstruct the full set of galaxy morphological correlations at other redshift ranges by using all 7 PC dimensions to represent the data set.  The correlations from the higher PC eigenvectors may be important at different redshifts.  When the goal of PCA is to cluster data then reducing the number of features based on the amount of variance captured is not the only option \citep{JoliffeBook,Benhur}.  In these cases more principal components can better recreate the original data set.


The morphologies of galaxies are not inherently discrete, but rather lie on a continuum.  However, it is often useful to bin galaxies into discrete morphological groups.  Fig. \ref{fig:f125w_cluster} shows the distribution of galaxies when projected onto the first three principal axes.  Except for a large distinct cluster of data points, most of our sample are not well separated, requiring the need for an objective data dependent grouping method.  

To classify galaxies in distinct groups, we employ the Ward hierarchical agglomerative clustering routine of {\tt scikit-learn} \citep{scikit-learn}.  Hierarchical clustering (specifically agglomerative clustering) treats each galaxy as its own cluster, which are then merged with nearby clusters while minimally increasing the in-cluster variance.  Mergers of adjacent clusters continue until the desired number of groupings are attained.   We define 10 groups, 2 of which are very sparsely populated, with only a combined 12 galaxies.  The sparsely populated clusters consist of extreme outliers from the other 8 clusters.  For this reason, we group all outliers into a single cluster.  



Fig. \ref{fig:cluster-variance} shows the amount of between-cluster variance calculated for various numbers of clusters.  Typically, the optimal number of clusters chosen corresponds to the turnover in this distribution (the point where the increase in between-cluster variance begins to slow; \citealp{Everitt+Hothorn}) which occurs at $\sim$10 clusters.  Increasing the number of clusters any further does not provide any more discriminatory power and only complicates interpretations of the final results.  We must note that there is no definitive criterion to help define how many clusters are to be defined in the data set.


The hierarchical clustering algorithm defines the groups based on the distribution of the data.  In order to reproduce the same group definitions for new objects with potentially different distributions (e.g. different redshifts), we use a convex hull method to define the original group boundaries in principal component space.  A convex hull defines the smallest area containing a set of points.  We define convex hulls using the 10 clusters determined by Ward's method for our $z\sim 1.5$ galaxy sample.  In practice we disregard the 2 sparsely populated clusters and instead group all of those galaxies into the outlier class.  

We use all 7 PCs to define the convex hull.  Calculating convex hulls in 7-dimensional space is computationally intensive and currently impossible for large data sets,  thus we outline a simple workaround. We define a convex hull based on 2 PC dimensions at a time and test whether a galaxy falls within the boundaries of a  group using all combinations of 2 PC dimensions.  The group a galaxy falls in the most times is determined to be its group.  If more than one group is equally likely, the smallest distance from the galaxy's position in PC space to the center of the possible groups is used to determine group membership.  Galaxies that are misclassified following the convex hull method generally exist on the boundaries of a convex hull.  We present the python code determining the group membership based on convex hull groupings\footnote{https://github.com/mikepeth/PyML}.




%
%

\begin{figure*}
\centering
	\makebox[\textwidth]{\includegraphics[width=1.1\textwidth]
	{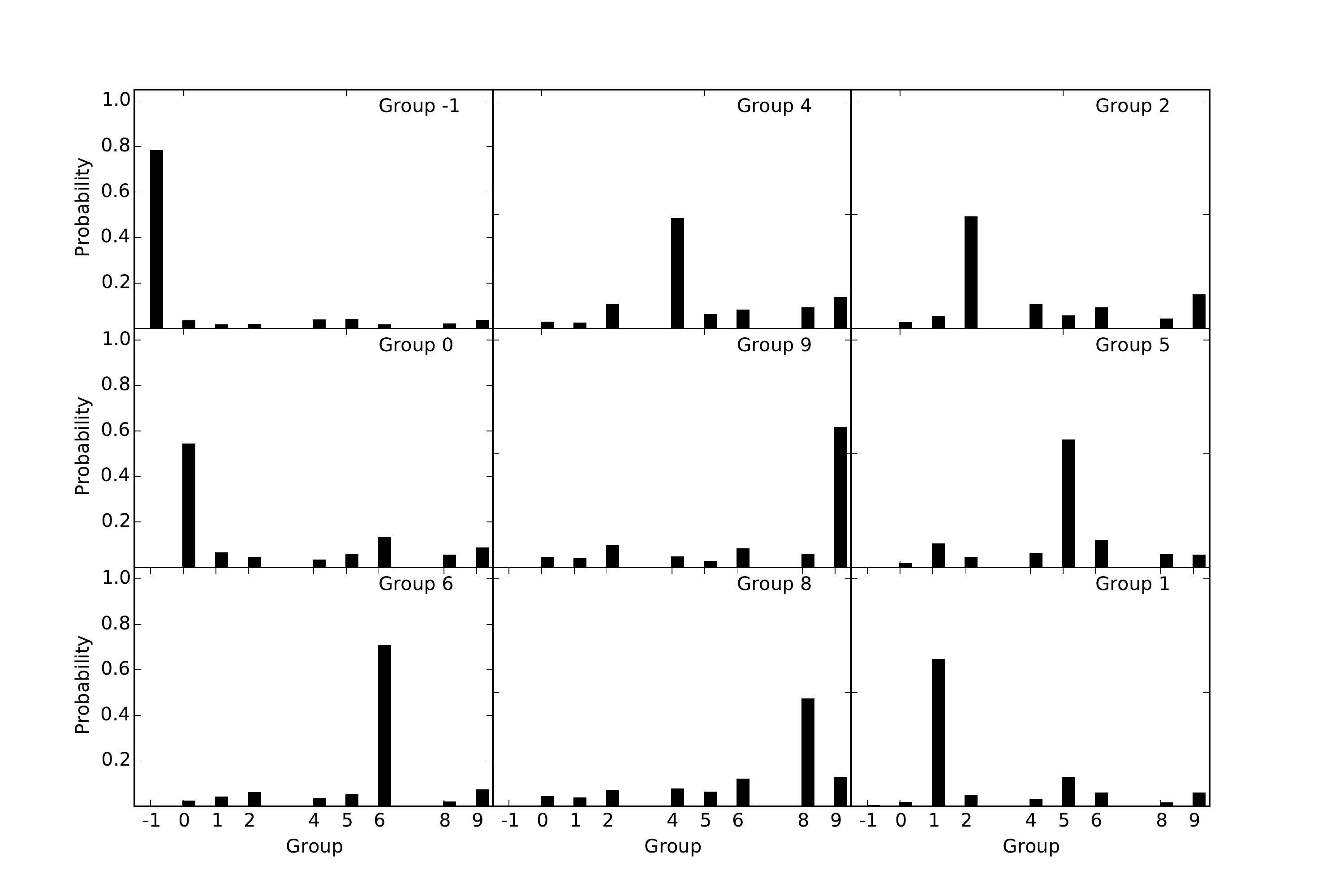}}
      
      \caption[Group classification uncertainty]
      {Group classification uncertainty, based on bootstrapped morphology measurement errors. Each galaxy's non-parametric morphologies are randomly scattered based on gaussians with widths based on errors found in Fig. \ref{fig:errors}.  The principal components and group membership are redetermined 250 times.  The resulting MC group distributions for each originally defined group are shown. Groups 1, 6, and -1 are the most robust to measurement errors,  whereas half of Groups 2, 4, 5 and 8 galaxies are scattered into other groups.   The panels are roughly arranged by PC1 (increasing left to right) and PC2 (increasing bottom to top, except for group -1).}
      \label{fig:prob_dist}
    \end{figure*}

 \begin{figure*}
        \centering
      \makebox[\textwidth]{\includegraphics[width=1.15\textwidth]
      {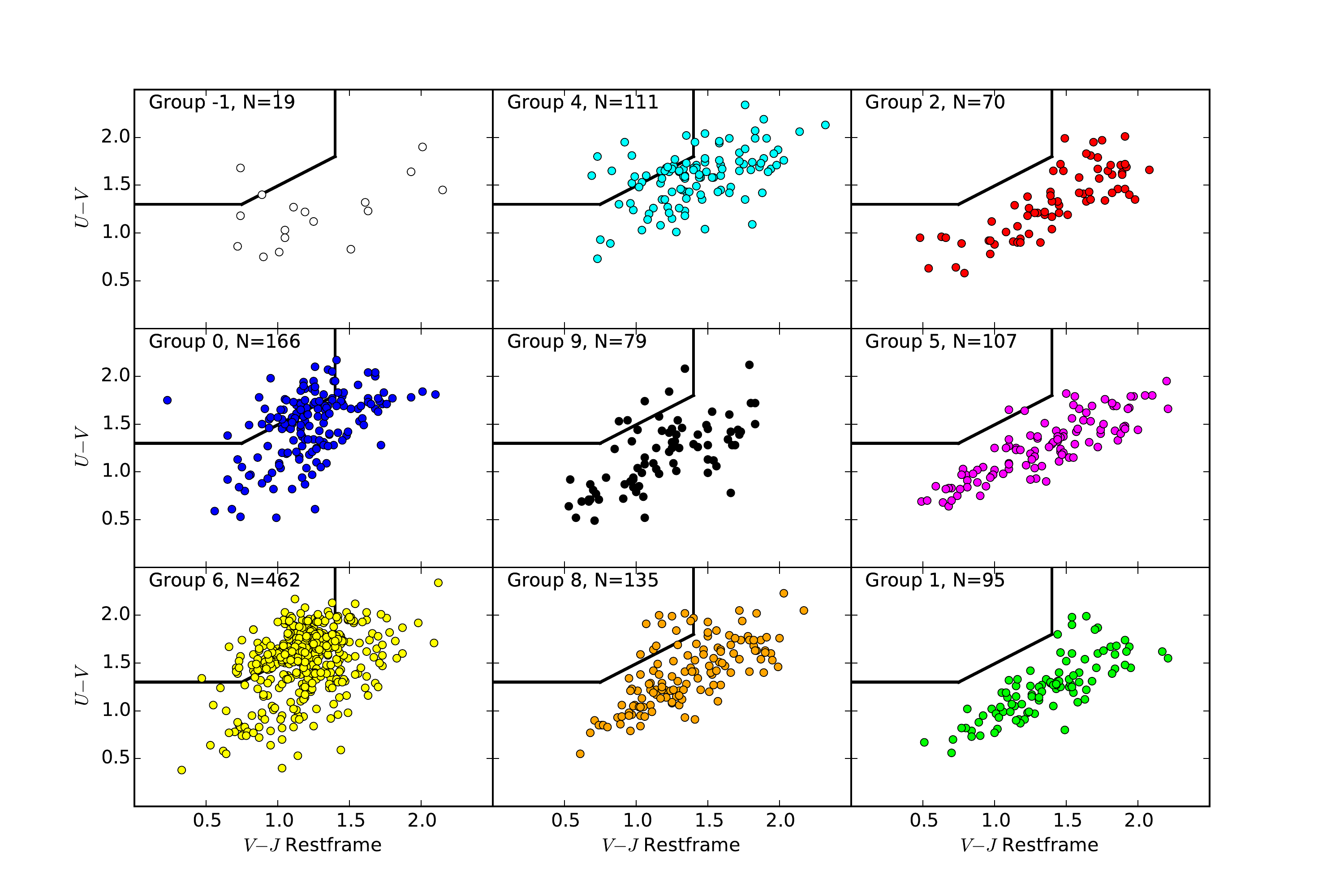}}
      
      \caption[Rest-frame $UVJ$]
      {Rest-frame $UVJ$ diagram for $M_* >10^{10}$ $M_{\odot}$, 1.36 $< z <$ 1.97 galaxies for each  group.  A $UVJ$ diagram is used to separate quenched galaxies from star-forming galaxies \citep{Williams09}.  Quenched galaxies reside in the upper left trapezoid.  Star-forming galaxies follow a sequence of increasing dust from the bottom left to the upper right.  The panels are roughly arranged by PC1 (increasing left to right) and PC2 (increasing bottom to top, except for group -1).  The majority of quenched galaxies are in group 6, with some quenched galaxies found in groups 0 and 8.  As PC1 increases we observe a decrease in the fraction of quenched galaxies.}
      \label{fig:uvj_f125w}
    \end{figure*}
    

\subsection{Morphological Error Estimation} \label{section:error}
The Hubble Ultra Deep Field (UDF) consists of deep imaging on a portion of the shallower GOODS-S field \citep{Koekemoer11}.  We measure the same galaxies using different depth images to the determine reliability of morphological measurements as a function of signal-to-noise and magnitude.  The non-parametric morphologies of galaxies are measured both in the deep UDF region and the GOODS-S observations.  We calculate the differences of GOODS-S morphologies from UDF morphologies.  We then bin galaxies in magnitude (or surface brightness) to find the average difference and median absolute deviation, which we define as the error for that morphological measurement.  

Fig. \ref{fig:errors} shows that larger and brighter galaxies are (unsurprisingly) well measured morphologically.  The median absolute deviations (red error bars) show a majority of galaxies have statistics that do not vary widely between shallow and deep images.  In general, the morphological offsets seen in Fig. \ref{fig:errors} are very small.  (For similar study see Fig. 6 in \citealp{Grogin11}.)






Now that we have calculated the principal morphological components and resulting morphology groups,  we can test their robustness to measurement errors.  We use Monte Carlo resampling test to randomly scatter our initial morphological measurements by Gaussians with sigma equal to the median absolute deviation for each morphological measurement (Fig. \ref{fig:errors}).  We then perform a principal component analysis for this new data set and repeat this process 250 times.  We project the scattered data on the original PC weights and then classify the galaxies based on the originally defined convex hulls ($\S$\ref{section:ward}) each time.  The group with a plurality of the reclassifications is defined to be the ``Monte Carlo'' (MC) group.  Fig. \ref{fig:prob_dist} can thus be seen as the probability distribution function for a galaxy of a certain group to be classified into a group via the convex hull method.  Group 6 is the most robustly classified group.  Only group 4 galaxies are re-classified as such following the Monte Carlo scattering to less than a majority of times (however still a large plurality of times).  The plots are separated by group and ordered roughly by PC1 horizontally and PC2 vertically.  The largest PC1 values and smallest PC2 values are in group 6 galaxies.   
  
  Table \ref{table:pc_weights} shows that the most important principal components (PC1-3) have typical resampled deviations $\leq$ 10 - 15 per cent of their weights. Higher principal component dimension display greater variability, but are also less important to our group classifications. 
  
Every galaxy has a MC reclassification with a probability associated with it and the group with a plurality of the reclassifications is defined as the MC resampled group.  Regardless of the probability, the reclassification is either the same or different from the original group designation.  This similarity or difference in classification determines the completeness and purity of the classification scheme.  The MC resampled classifications are 90.8 per cent complete and 90.4 per cent pure relative to original group classifications.  The completeness and purity are highest when all 7 PCs are used to define the groups instead of only 3 PCs.  Representing the data set with 3 PCs slightly drops the completeness and purity scores to 88.3 and 89.4 per cent.  In contrast, the completeness and purity values significantly drop to 25.9 and 20.3 percent when PCs are calculated from IQR-scaled data.  The volatility of the eigenvectors calculated from an IQR-scaled data set is the cause of these poor reclassification results.  This evidence leads to our conclusion that using all 7 standard-deviation scaled PC eigenvectors will result in more definitive groups.

Note that this does not include the systematic biases e.g. those due to the PSF.  This bias likely important for groups 6 and 0.




    
     \begin{figure*}
        \centering
      \makebox[\textwidth]{\includegraphics[width=1.1\textwidth]
      {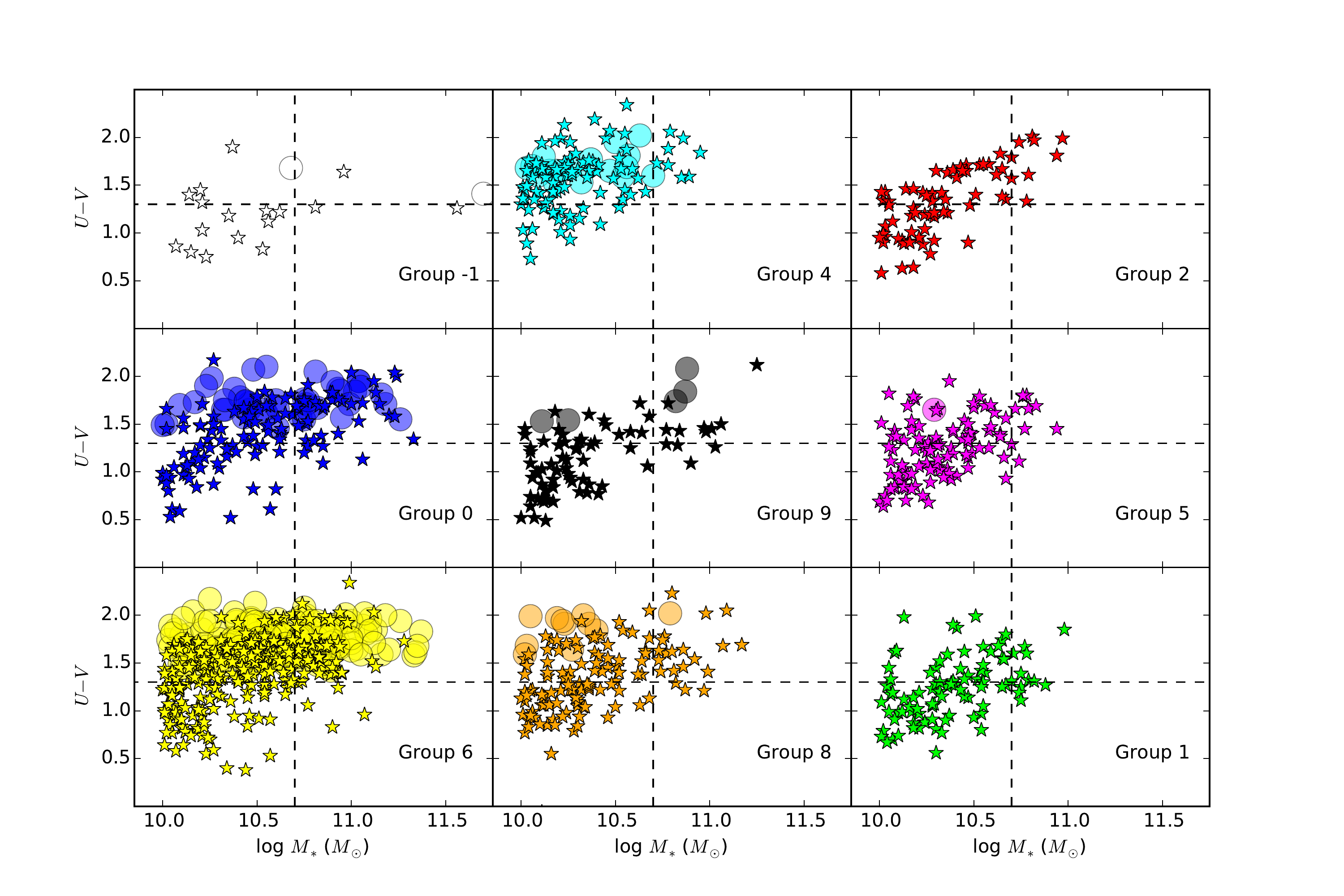}}
      
      \caption[Rest-frame $U-V$ vs. Stellar Mass]
      {Rest-frame $U-V$ vs. Stellar Mass diagram for $M_* >10^{10}$ $M_{\odot}$, 1.36 $< z <$ 1.97 galaxies for each cluster group.  Galaxies classified by $UVJ$ as star-forming ($stars$) and quenched ($circles$) are shown for each group.  The dashed line in $U-V$ represents the approximate dividing line between quenched and star-forming galaxies.  Groups 6, 0, and 9 have the greatest fractions of galaxies with large masses (dashed line, $M_* > 5 \times 10^{10} M_{\astrosun}$).  The panels are roughly arranged by PC1 (increasing left to right) and PC2 (increasing bottom to top, except for group -1).}
      \label{fig:uv-Mass}
    \end{figure*}
    		
\begin{figure*}
    \centering
      \makebox[\textwidth]{\includegraphics[width=0.95\textwidth]
      {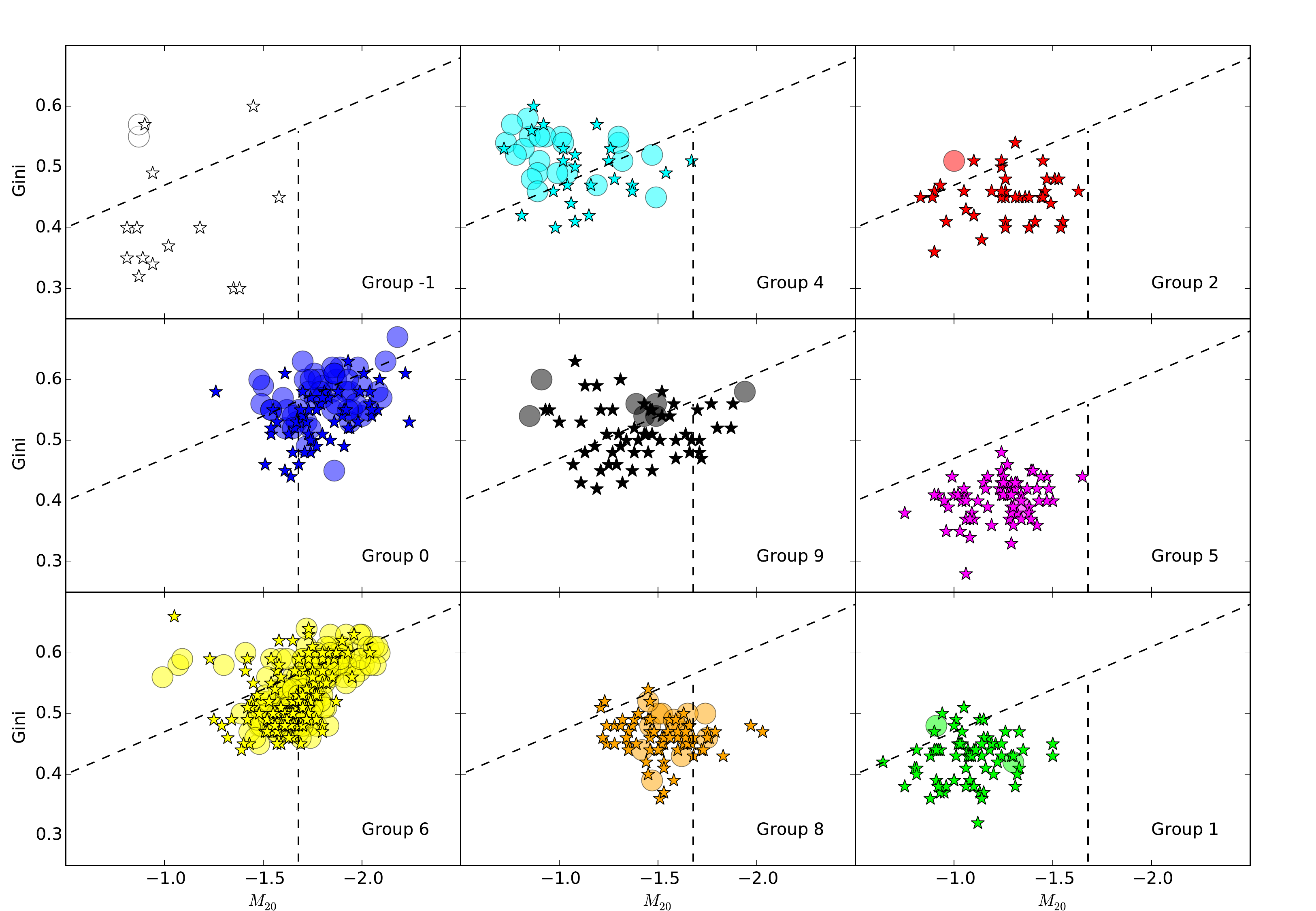}}
      \caption[Gini$--M_{20}$]
      {$G-M_{20}$ for each group.  Overplotted are the dividing lines between: mergers ({\it top left corner}), bulge-dominated ({\it right-most region}), and disk-dominated ({\it bottom left region}) modified from \citet{Lotz04}.  Group 0 fully occupies the bulge-dominated region of the plot. Symbols same as Fig. \ref{fig:uv-Mass}.  The panels are roughly arranged by PC1 (increasing left to right) and PC2 (increasing bottom to top, except for group -1).}
      \label{fig:gini_m20}
    \end{figure*}
    
   \begin{figure*}
   \centering
 \makebox[\textwidth]{\includegraphics[width=1.05\textwidth]
 {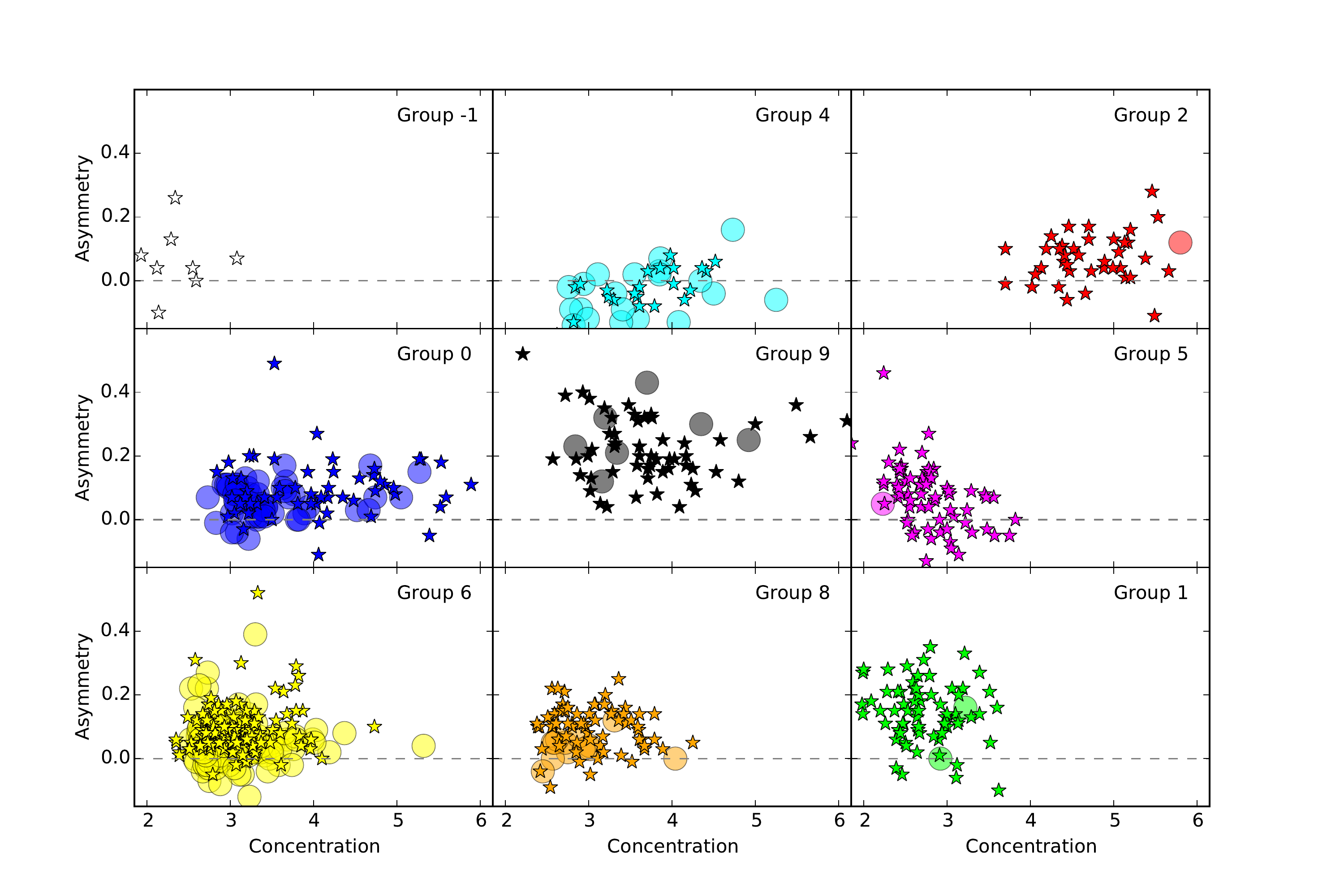}}
      \caption[Concentration - Asymmetry]
      {Concentration - Asymmetry for each  group.  Plotting symbols same as Fig. \ref{fig:uv-Mass}.  Groups 9 and 1 have the highest asymmetry, while group 0 has the highest concentration.  The panels are roughly arranged by PC1 (increasing left to right) and PC2 (increasing bottom to top, except for group -1).}
      \label{fig:c_a}
    \end{figure*}
%
%
%

 \begin{figure*} 
    \centering
 \makebox[\textwidth]{\includegraphics[width=0.95\textwidth]
 {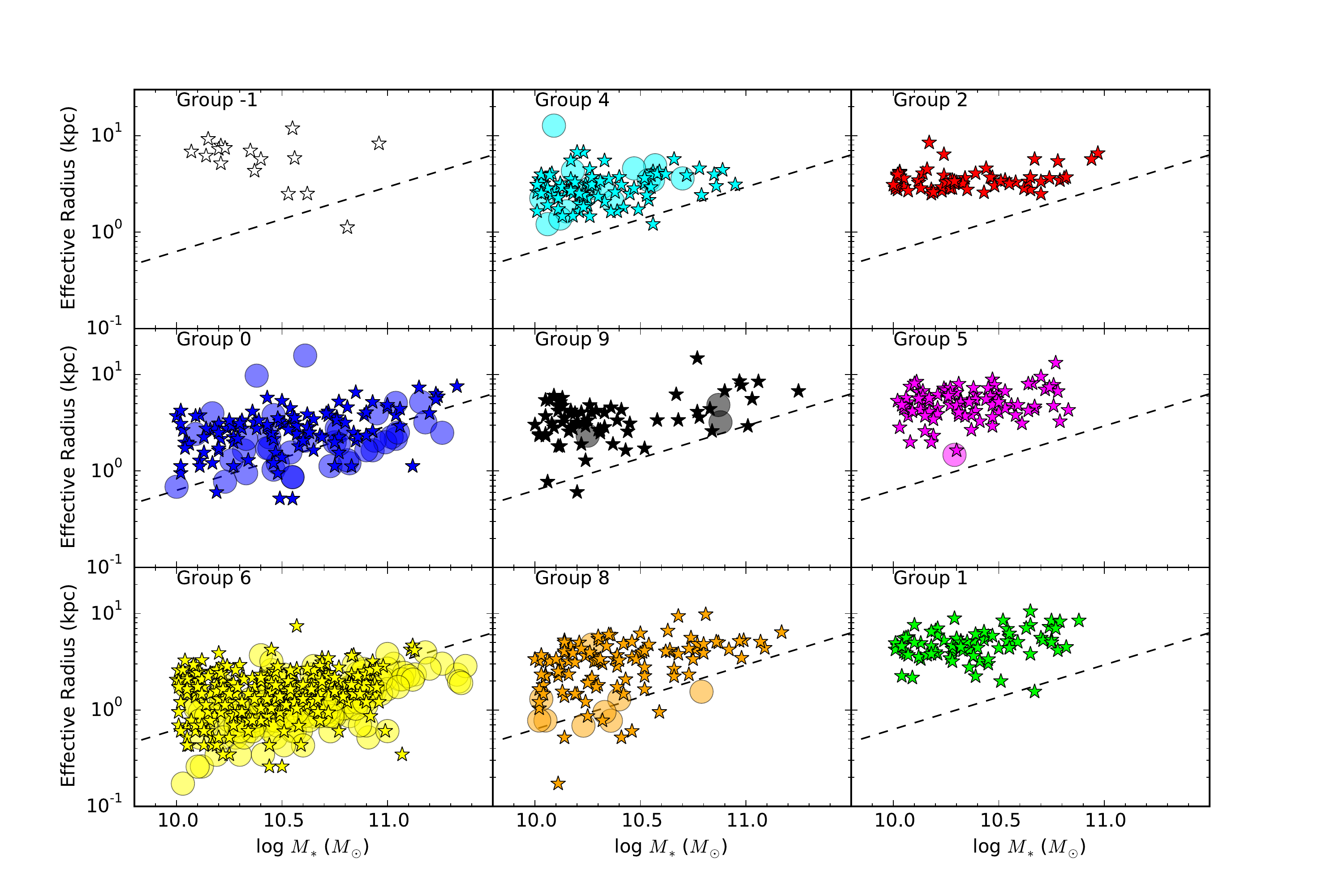}}
      \caption[Effective radii (kpc) - $M_*$]
      {Effective radii (kpc, as measured in WFC3 F160W by \citealp{vanderWel12}) vs. stellar mass for each group.  Dotted lines represent the ``compact'' criteria ($M/r_e^{1.5} <  10.3 M_{\odot}$ kpc$^{-1.5}$) of  \citet{Barro13}.  Almost all group 6 galaxies are very compact, with most galaxies smaller than 2 kpc.  Groups 0 and 8 have a number of borderline compact galaxies.  The remaining groups have only a few compact galaxies at most.  The panels are roughly arranged by PC1 (increasing left to right) and PC2 (increasing bottom to top, except for group -1).} 
      \label{fig:size_mass}
    \end{figure*}

    \section{PCA Morphology Groups at $z \sim 1.5$}\label{section:morphgroups}
    
%
        The connection between morphology and star-formation has been well studied \citep{Wuyts11,Kriek09,Brinchmann04}.  Late-type galaxies are typically still actively forming stars, whereas early-type galaxies have had their star-formation quenched.  However, there are examples of red, quenched disks and blue, star-forming ellipticals which are important rare ``transitional'' classes.  
        
        We use a $UVJ$ color-color diagram (Fig. \ref{fig:uvj_f125w}) to classify galaxies as ``star-forming'' and ``quenched'' using the bimodality of these two types of galaxies seen in $U-V$ and $V-J$ rest-frame colors  \citep{Labbe05,Wuyts07,Williams09}.  Star-forming galaxies follow a sequence determined by dust extinction.  The panels are arranged in Fig.  \ref{fig:uvj_f125w}  so that PC1 increases along the x-axis and PC2 increases along the y-axis.  Most groups are primarily comprised of star-forming galaxies.  Groups with lower PC1 values are more compact and quenched.  Similarly, a $UV-$Mass diagram separate star-forming from quenched galaxies (Fig. \ref{fig:uv-Mass}).  Again galaxies with lower PC1 values are more massive and more quenched.
%
%
%

         Previous studies (e.g. \citealp{Lotz04,Conselice00,Lee13}) utilize $G-M_{20}$ (Fig. \ref{fig:gini_m20}) or Concentration-Asymmetry (Fig. \ref{fig:c_a}) diagrams to classify galaxies into early and late-type categories.  In our study we use these tools to reinforce how effective our PCA groups are at separating different classes of galaxies.  In Fig. \ref{fig:gini_m20} the dotted lines signify classification regions adapted from \citet{Lotz04} for z$\sim$1-2 galaxies observed by $HST$.  Mergers are in the upper left region, late-type galaxies are in the lower region and early-type galaxies are in the wedge-shaped region on the rightmost portion of the $G-M_{20}$ diagram. $C-A$ diagrams (for review see \citealp{Conselice14}) have been used  to differentiate giant ellipticals (which live in regions of large $C$ and small $A$) from spirals (with progressively smaller $C$ and larger $A$) and from mergers (which are the most asymmetric but the least centrally concentrated).
         
        
        For our group descriptions in the following sections we will refer heavily to Fig. \ref{fig:uvj_f125w} - \ref{fig:size_mass}, the example galaxies of Fig. \ref{fig:f125w_stamps_group6} - \ref{fig:f125w_stamps_group-1} and Tables \ref{table:mass_class} - \ref{table:quenched_classes}.  For these figures the locations of each subplot represents the approximate position of that group in PCA space.  From left to right, PC1 increases which is indicative of an increase in bulge strength.  From bottom to top, PC2 increases thus concentration increases.
        
        
        Tables \ref{table:mass_class}-\ref{table:quenched_classes} describe the group demographics in terms of stellar mass, visual classification \citep{Jeyhan14}, S\'ersic indices \citep{vanderWel12} and quenched fraction.  These demographics are both listed in terms of the original group (as determined by the hierarchical clustering method, left columns) and in terms of the MC group (determined using the scattering method, right columns). The agreement between the galaxy demographics in the original groups and scattered MC groups shows the group characteristics are quite robust to noise.  Table \ref{table:mass_class} shows that high PC1 (disk-dominated) groups have very few high mass galaxies.  Meanwhile, low PC1 (compact/bulge-dominated) groups have a larger fraction of high mass galaxies.  
        
        
        We use the CANDELS visual classifications \citep{Jeyhan14} to determine the demography of the PCA groups in disk, spheroidal and irregular galaxy classes.  For a galaxy to be counted as a ``disk'', ``spheroid'' or ``irregular'' it must have been classified by at least two-thirds of the classifiers as such, and less than one-third as the other classes.  A ``disk+spheroid'' is classified as {\it both} a disk and a spheroid by at least two-thirds of the classifiers.  The ``other'' class represents everything that does not belong to the other 4 categories.  The fractions of galaxies in each morphological type are shown in Table \ref{table:visual_classes}.  
        
          S\'ersic fits have been used extensively to classify galaxies into early- and late-type categories \citep{vanderWel12,vanDokkum10,Patel11,Peng02}.  Typically, $n$=2.5 is used to divide late-type ($n <$ 2.5) and early-type ($n >$ 2.5) galaxies.  Table \ref{table:sersic_classes} shows the percentage of galaxies representing a certain classification for each group as a percentage of the group population \citep{vanderWel12}.  Similar to visual classification, the percentage of galaxies with disk-dominated morphologies decreases with decreasing PC1 values.
          
                         
       Table \ref{table:quenched_classes} and  Fig. \ref{fig:uvj_f125w} show that in this redshift range (1.36 $<$ z $<$ 1.97) and mass range ($\gtrsim 10^{10}M_{\odot}$) only 23 per cent of galaxies are quenched.  Table \ref{table:quenched_classes} shows that the quenched fraction for a group is anti-correlated to PC1 and PC2.  

       Fig. \ref{fig:size_mass} shows the effective radii (kpc) - stellar mass relation for each group.  In this figure, PC1 and PC2 are strongly correlated a galaxy's compactness. Group 6 galaxies are by far the most compact, with the largest fraction of quenched galaxies.  As PC1 and PC2 increase the number of quenched galaxies in each group decreases.
      
\begin{table*}
\centering

			\caption{Group percentages by mass range for both original group and ``MC Group''}
\begin{threeparttable}
		\begin{tabular}{cccccccccc}
	
	\hline	
	
	%
& Group & \multicolumn{2}{c}{10.0 $< \log M_* <$ 10.5} & \multicolumn{2}{c}{10.5 $<  \log M_* <$ 11.0} & \multicolumn{2}{c}{ $\log M_* > $ 11.0} & \multicolumn{2}{c}{Group Percentage} \\
\hline
\multirow{3}{*}{Low PC1} & 6 &49.1$_{-3.1}^{+3.3}$ & 51.0$_{-3.5}^{+3.7}$ & 45.2$_{-3.0}^{+3.2}$ & 43.8$_{-3.2}^{+3.5}$ & 5.6$_{-1.1}^{+1.3}$ & 5.1$_{-1.1}^{+1.4}$ & \bf{37.3 (462)} & \bf{31.1 (385.2)} \\ 
& 0 &46.1$_{-5.0}^{+5.6}$ & 42.1$_{-6.7}^{+7.8}$ & 41.8$_{-4.8}^{+5.4}$ & 45.4$_{-6.9}^{+8.0}$ & 12.1$_{-2.6}^{+3.2}$ & 12.4$_{-3.7}^{+4.8}$ & \bf{13.3 (165)} & \bf{7.0 (87.2)} \\ 
& 9 &71.8$_{-9.2}^{+10.4}$ & 65.7$_{-6.0}^{+6.6}$ & 23.1$_{-5.2}^{+6.5}$ & 29.4$_{-4.0}^{+4.6}$ & 5.1$_{-2.5}^{+3.9}$ & 4.9$_{-1.7}^{+2.3}$ & \bf{6.3 (78)} & \bf{13.4 (165.4)} \\ 
\hline 
& & 50.9$_{-0.1}^{+0.3}$ & 53.6$_{-1.1}^{+1.3}$  & 42.0$_{-0.1}^{+0.3}$ & 40.3$_{-1.0}^{+1.1}$  & 7.1$_{-0.1}^{+0.3}$ & 6.1$_{-0.4}^{+0.5}$ \\ 
\hline
\multirow{2}{*}{Mid PC1} & 4 &76.6$_{-7.9}^{+8.8}$ & 70.1$_{-7.8}^{+8.8}$ & 23.4$_{-4.4}^{+5.3}$ & 28.2$_{-5.0}^{+6.0}$ & 0.0$_{-0.4}^{+1.6}$ & 1.7$_{-1.3}^{+2.4}$ & \bf{9.0 (111)} & \bf{8.4 (103.7)} \\ 
& 8 &70.9$_{-6.9}^{+7.7}$ & 70.6$_{-7.5}^{+8.4}$ & 26.9$_{-4.3}^{+5.0}$ & 27.5$_{-4.7}^{+5.6}$ & 2.2$_{-1.3}^{+2.1}$ & 1.9$_{-1.3}^{+2.3}$ & \bf{10.8 (134)} & \bf{9.2 (114.0)} \\ 
\hline 
& & 73.5$_{-0.4}^{+0.9}$ & 70.4$_{-3.7}^{+4.1}$  & 25.3$_{-0.3}^{+0.8}$ & 27.8$_{-2.3}^{+2.8}$  & 1.2$_{-0.2}^{+0.7}$ & 1.8$_{-0.6}^{+1.1}$ \\ 
\hline
\multirow{3}{*}{High PC1} & 1 &66.3$_{-8.0}^{+9.0}$ & 66.1$_{-6.5}^{+7.2}$ & 33.7$_{-5.7}^{+6.7}$ & 31.9$_{-4.6}^{+5.3}$ & 0.0$_{-0.5}^{+1.9}$ & 2.0$_{-1.2}^{+2.0}$ & \bf{7.7 (95)} & \bf{11.4 (140.9)} \\ 
& 2 &74.3$_{-9.8}^{+11.3}$ & 67.3$_{-7.3}^{+8.1}$ & 25.7$_{-5.8}^{+7.3}$ & 30.7$_{-4.9}^{+5.8}$ & 0.0$_{-0.7}^{+2.6}$ & 2.0$_{-1.3}^{+2.3}$ & \bf{5.6 (70)} & \bf{9.3 (115.4)} \\ 
& 5 &76.6$_{-8.1}^{+9.0}$ & 64.2$_{-7.2}^{+8.0}$ & 23.4$_{-4.5}^{+5.4}$ & 32.5$_{-5.1}^{+6.0}$ & 0.0$_{-0.5}^{+1.7}$ & 3.2$_{-1.7}^{+2.6}$ & \bf{8.6 (107)} & \bf{9.2 (114.1)} \\ 
\hline 
& & 72.4$_{-0.4}^{+0.8}$ & 65.9$_{-2.1}^{+2.4}$  & 27.6$_{-0.3}^{+0.7}$ & 31.7$_{-1.5}^{+1.7}$  & 0.0$_{-0.2}^{+0.7}$ & 2.4$_{-0.4}^{+0.7}$ \\ 
\hline
Outliers &  -1 &58.8$_{-18.0}^{+24.1}$ & 50.2$_{-19.3}^{+27.3}$ & 41.2$_{-15.2}^{+21.3}$ & 37.4$_{-16.8}^{+24.9}$ & 0.0$_{-2.9}^{+10.6}$ & 12.5$_{-10.2}^{+18.9}$ & \bf{1.4 (17)} & \bf{1.0 (12.9)} \\

\hline 
\bf{N Galaxies} & & \multicolumn{2}{c}{\bf{746}} & \multicolumn{2}{c}{\bf{440}} & \multicolumn{2}{c}{ \bf{58}} & \multicolumn{2}{c}{\bf{1244}} \\
\hline 
%

\end{tabular}
\begin{tablenotes}
            \item Note: The left hand columns for each mass range represent the demographics based upon the original group based on hierarchical clustering.  The right hand columns are based on the total group probabilities based on the scattering technique classifications.
        \end{tablenotes}
     \end{threeparttable}
\label{table:mass_class}
\end{table*}

\begin{table*}
\centering
		 \makebox[\linewidth]{

\begin{minipage}{200mm}
\caption{Demographics of Visual Classifications of Groups}\label{table:visual_classes}
\begin{threeparttable}
\begin{tabular}{ccrrrrrrrrrrrr}
\hline

& Group & \multicolumn{2}{c}{Disks} & \multicolumn{2}{c}{Spheroids} & \multicolumn{2}{c}{Irregulars} & \multicolumn{2}{c}{D+Sph} & \multicolumn{2}{c}{Other}\\
\hline
\multirow{3}{*}{Low PC1} & 6 & 12.5$_{-1.9}^{+2.3}$ & 19.0$_{-0.3}^{+0.8}$  & 52.5$_{-4.0}^{+4.3}$ & 47.4$_{-0.5}^{+0.9}$  & 1.0$_{-0.6}^{+0.9}$ & 1.6$_{-0.2}^{+0.7}$  & 25.6$_{-2.8}^{+3.1}$ & 23.7$_{-0.4}^{+0.8}$  & 8.5$_{-1.6}^{+1.9}$ & 8.3$_{-0.3}^{+0.7}$ \\ 
& 0 & 31.0$_{-5.0}^{+5.9}$ & 33.8$_{-1.1}^{+3.2}$  & 34.5$_{-5.3}^{+6.2}$ & 31.8$_{-1.1}^{+3.2}$  & 0.9$_{-1.0}^{+2.0}$ & 1.8$_{-0.8}^{+3.0}$  & 32.7$_{-5.2}^{+6.0}$ & 31.5$_{-1.1}^{+3.2}$  & 0.9$_{-1.0}^{+2.0}$ & 1.1$_{-0.8}^{+3.0}$ \\ 
& 9 & 41.2$_{-8.6}^{+10.6}$ & 42.9$_{-0.8}^{+1.9}$  & 13.7$_{-5.1}^{+7.1}$ & 23.8$_{-0.7}^{+1.8}$  & 23.5$_{-6.6}^{+8.6}$ & 10.7$_{-0.6}^{+1.8}$  & 11.8$_{-4.7}^{+6.8}$ & 15.9$_{-0.6}^{+1.8}$  & 9.8$_{-4.3}^{+6.4}$ & 6.8$_{-0.5}^{+1.8}$ \\ 
\hline 
& & 20.0$_{-0.1}^{+0.4}$ & 27.1$_{-1.2}^{+1.4}$  & 43.9$_{-0.2}^{+0.4}$ & 39.3$_{-1.4}^{+1.7}$  & 3.4$_{-0.1}^{+0.4}$ & 3.9$_{-0.5}^{+0.7}$  & 25.8$_{-0.1}^{+0.4}$ & 22.9$_{-1.1}^{+1.3}$  & 6.8$_{-0.1}^{+0.4}$ & 6.9$_{-0.6}^{+0.9}$ \\ 
\hline 
\multirow{2}{*}{Mid PC1} & 4 & 51.4$_{-8.2}^{+9.6}$ & 49.7$_{-1.1}^{+3.0}$  & 17.1$_{-4.8}^{+6.2}$ & 20.8$_{-0.9}^{+2.9}$  & 0.0$_{-0.7}^{+2.6}$ & 2.4$_{-0.8}^{+2.8}$  & 24.3$_{-5.7}^{+7.1}$ & 21.0$_{-0.9}^{+2.9}$  & 7.1$_{-3.1}^{+4.6}$ & 6.1$_{-0.8}^{+2.8}$ \\ 
& 8 & 73.8$_{-9.0}^{+10.1}$ & 58.5$_{-1.1}^{+2.8}$  & 9.5$_{-3.3}^{+4.5}$ & 19.2$_{-0.9}^{+2.6}$  & 3.6$_{-2.1}^{+3.4}$ & 2.9$_{-0.7}^{+2.5}$  & 6.0$_{-2.6}^{+3.9}$ & 12.9$_{-0.8}^{+2.6}$  & 7.1$_{-2.9}^{+4.1}$ & 6.6$_{-0.8}^{+2.6}$ \\ 
\hline 
& & 63.6$_{-0.6}^{+1.4}$ & 54.3$_{-5.2}^{+5.9}$  & 13.0$_{-0.4}^{+1.2}$ & 19.9$_{-3.1}^{+3.9}$  & 1.9$_{-0.3}^{+1.2}$ & 2.7$_{-1.2}^{+2.0}$  & 14.3$_{-0.4}^{+1.2}$ & 16.7$_{-2.9}^{+3.6}$  & 7.1$_{-0.4}^{+1.2}$ & 6.3$_{-1.8}^{+2.6}$ \\ 
\hline 
\multirow{3}{*}{High PC1} &1 & 72.1$_{-9.8}^{+11.3}$ & 64.2$_{-0.9}^{+2.2}$  & 1.5$_{-1.6}^{+3.3}$ & 11.6$_{-0.6}^{+1.9}$  & 16.2$_{-4.7}^{+6.2}$ & 8.9$_{-0.6}^{+1.9}$  & 1.5$_{-1.6}^{+3.3}$ & 9.0$_{-0.6}^{+1.9}$  & 8.8$_{-3.5}^{+5.1}$ & 6.4$_{-0.6}^{+1.9}$ \\ 
& 2 & 76.3$_{-13.6}^{+16.2}$ & 56.0$_{-1.1}^{+2.8}$  & 7.9$_{-4.6}^{+7.4}$ & 21.8$_{-0.9}^{+2.6}$  & 5.3$_{-3.8}^{+6.7}$ & 3.7$_{-0.7}^{+2.5}$  & 7.9$_{-4.6}^{+7.4}$ & 12.5$_{-0.8}^{+2.5}$  & 2.6$_{-2.9}^{+5.9}$ & 6.1$_{-0.7}^{+2.5}$ \\ 
& 5 & 95.5$_{-11.5}^{+13.0}$ & 61.5$_{-1.1}^{+2.8}$  & 0.0$_{-0.8}^{+2.7}$ & 17.6$_{-0.8}^{+2.6}$  & 3.0$_{-2.2}^{+3.9}$ & 4.3$_{-0.7}^{+2.5}$  & 0.0$_{-0.8}^{+2.7}$ & 12.5$_{-0.8}^{+2.5}$  & 1.5$_{-1.6}^{+3.4}$ & 4.1$_{-0.7}^{+2.5}$ \\ 
\hline 
& & 82.0$_{-0.6}^{+1.3}$ & 60.9$_{-3.1}^{+3.5}$  & 2.3$_{-0.3}^{+1.1}$ & 16.5$_{-1.6}^{+2.0}$  & 8.7$_{-0.3}^{+1.1}$ & 5.9$_{-1.0}^{+1.4}$  & 2.3$_{-0.3}^{+1.1}$ & 11.1$_{-1.3}^{+1.8}$  & 4.7$_{-0.3}^{+1.1}$ & 5.6$_{-1.0}^{+1.4}$ \\ 
\hline 
Outliers & -1 & 40.0$_{-19.8}^{+30.5}$ & 32.1$_{-6.3}^{+21.9}$  & 30.0$_{-17.4}^{+28.2}$ & 19.4$_{-6.2}^{+21.8}$  & 20.0$_{-14.5}^{+25.5}$ & 18.8$_{-6.2}^{+21.8}$  & 0.0$_{-5.0}^{+18.0}$ & 19.4$_{-6.2}^{+21.8}$  & 10.0$_{-10.9}^{+22.3}$ & 10.2$_{-6.1}^{+21.8}$ \\ 
\hline
\bf{Total Fraction} & & \multicolumn{2}{c}{\bf{42$\%$ (337)}} & \multicolumn{2}{c}{\bf{29$\%$ (233)}} & \multicolumn{2}{c}{ \bf{4$\%$ (36)}} & \multicolumn{2}{c}{\bf{18$\%$ (147)}} & \multicolumn{2}{c}{\bf{6$\%$ (52)}} \\
\hline

\end{tabular}

\begin{tablenotes}
            \item Note: Visual classification from \citet{Jeyhan14} for UDS and GOODS-S (no classifications for COSMOS galaxies).  For a galaxy to be visually classified 2/3 observers need to agree.  `Other' classification refers to galaxies failing the 2/3 agreement requirement. The left hand columns for each visual classification represent the demographics based upon the original group based on hierarchical clustering.  The right hand columns are based on the total group probabilities based on the scattering technique classifications.    
        \end{tablenotes}
        
     \end{threeparttable}
\end{minipage}

}

\end{table*}
              
\begin{table*}
\centering


\caption{Demographics of S\'ersic Classifications of Groups}
\begin{threeparttable}
\begin{tabular}{cccccccccc}
\hline
& Group & \multicolumn{2}{c}{0 $< n < $ 1} & \multicolumn{2}{c}{1 $< n < $ 2.5} & \multicolumn{2}{c}{2.5 $< n < $ 4} & \multicolumn{2}{c}{$n > $ 4}\\
\hline
\multirow{3}{*}{Low PC1} & 6 & 36.6$_{-3.3}^{+3.6}$ & 13.3$_{-0.3}^{+0.8}$  & 19.5$_{-2.4}^{+2.7}$ & 31.2$_{-0.4}^{+0.9}$  & 22.1$_{-2.6}^{+2.9}$ & 34.6$_{-0.4}^{+0.9}$  & 13.4$_{-2.0}^{+2.3}$ & 20.9$_{-0.4}^{+0.9}$ \\ 
& 0 & 30.7$_{-5.0}^{+5.9}$ & 6.1$_{-1.0}^{+3.3}$  & 18.1$_{-3.8}^{+4.7}$ & 25.0$_{-1.1}^{+3.4}$  & 18.1$_{-3.8}^{+4.7}$ & 28.4$_{-1.2}^{+3.5}$  & 25.3$_{-4.5}^{+5.4}$ & 40.5$_{-1.2}^{+3.5}$ \\ 
& 9 & 53.2$_{-9.8}^{+11.7}$ & 30.0$_{-0.8}^{+2.1}$  & 26.6$_{-7.0}^{+9.0}$ & 38.0$_{-0.8}^{+2.2}$  & 2.5$_{-2.4}^{+4.6}$ & 12.7$_{-0.7}^{+2.0}$  & 7.6$_{-3.8}^{+5.9}$ & 19.3$_{-0.7}^{+2.1}$ \\ 
\hline 
& & 37.0$_{-0.2}^{+0.4}$ & 16.3$_{-1.0}^{+1.3}$  & 19.9$_{-0.1}^{+0.4}$ & 32.0$_{-1.4}^{+1.7}$  & 19.0$_{-0.1}^{+0.4}$ & 28.4$_{-1.4}^{+1.6}$  & 15.6$_{-0.1}^{+0.4}$ & 23.3$_{-1.2}^{+1.5}$ \\ 
\hline 
\multirow{2}{*}{Mid PC1} & 4 & 47.7$_{-7.9}^{+9.3}$ & 23.2$_{-1.1}^{+3.2}$  & 25.2$_{-5.8}^{+7.2}$ & 43.7$_{-1.2}^{+3.3}$  & 8.1$_{-3.3}^{+4.8}$ & 16.0$_{-1.0}^{+3.2}$  & 10.8$_{-3.8}^{+5.3}$ & 17.1$_{-1.0}^{+3.2}$ \\ 
& 8 & 60.0$_{-8.1}^{+9.2}$ & 38.6$_{-1.1}^{+3.0}$  & 20.0$_{-4.7}^{+5.9}$ & 37.4$_{-1.1}^{+3.0}$  & 4.4$_{-2.3}^{+3.6}$ & 13.2$_{-0.9}^{+2.9}$  & 3.0$_{-1.9}^{+3.2}$ & 10.8$_{-0.9}^{+2.9}$ \\ 
\hline 
& & 54.4$_{-0.6}^{+1.3}$ & 31.2$_{-4.4}^{+5.2}$  & 22.4$_{-0.4}^{+1.2}$ & 40.5$_{-5.0}^{+5.8}$  & 6.1$_{-0.4}^{+1.2}$ & 14.6$_{-3.0}^{+3.9}$  & 6.5$_{-0.4}^{+1.2}$ & 13.8$_{-3.0}^{+3.8}$ \\ 
\hline 
\multirow{3}{*}{High PC1} & 1 & 83.2$_{-10.6}^{+12.0}$ & 68.7$_{-1.1}^{+2.5}$  & 3.2$_{-2.2}^{+3.8}$ & 12.6$_{-0.7}^{+2.3}$  & 0.0$_{-0.7}^{+2.6}$ & 8.8$_{-0.7}^{+2.2}$  & 1.1$_{-1.4}^{+3.1}$ & 9.9$_{-0.7}^{+2.3}$ \\ 
& 2 & 71.4$_{-13.1}^{+15.8}$ & 39.2$_{-1.1}^{+3.1}$  & 12.9$_{-5.7}^{+8.5}$ & 33.9$_{-1.1}^{+3.1}$  & 1.4$_{-2.3}^{+5.4}$ & 14.3$_{-0.9}^{+3.0}$  & 1.4$_{-2.3}^{+5.4}$ & 12.6$_{-0.9}^{+3.0}$ \\ 
& 5 & 84.1$_{-10.8}^{+12.3}$ & 55.2$_{-1.2}^{+3.1}$  & 4.7$_{-2.7}^{+4.3}$ & 16.5$_{-0.9}^{+2.9}$  & 0.0$_{-0.8}^{+2.7}$ & 14.5$_{-0.9}^{+2.9}$  & 0.0$_{-0.8}^{+2.7}$ & 13.9$_{-0.9}^{+2.9}$ \\ 
\hline 
& & 80.9$_{-0.6}^{+1.3}$ & 55.7$_{-3.4}^{+3.9}$  & 5.9$_{-0.3}^{+1.1}$ & 20.2$_{-2.1}^{+2.6}$  & 0.3$_{-0.3}^{+1.0}$ & 12.2$_{-1.6}^{+2.1}$  & 0.7$_{-0.3}^{+1.0}$ & 12.0$_{-1.6}^{+2.1}$ \\ 
\hline 
Outliers & -1 & 78.9$_{-27.4}^{+37.7}$ & 60.8$_{-6.6}^{+22.2}$  & 0.0$_{-5.0}^{+18.0}$ & 19.6$_{-6.2}^{+22.0}$  & 0.0$_{-5.0}^{+18.0}$ & 0.0$_{-6.1}^{+21.8}$  & 15.8$_{-13.1}^{+24.3}$ & 19.5$_{-6.2}^{+22.0}$ \\ 
\hline
\bf{Total Fraction} & & \multicolumn{2}{c}{\bf{56$\%$ (630)}} & \multicolumn{2}{c}{\bf{19$\%$ (213)}} & \multicolumn{2}{c}{ \bf{13$\%$ (150)}} & \multicolumn{2}{c}{\bf{12$\%$ (131)}} \\
\hline

\end{tabular}
\begin{tablenotes}
            \item Note: The left hand columns for each S\'ersic-index range represent the demographics based upon the original group based on hierarchical clustering.  The right hand columns are based on the total group probabilities based on the scattering technique classifications.  Due to the small sizes of certain galaxies, not every galaxy has a measured S\'ersic fit.
        \end{tablenotes}
     \end{threeparttable}
  \label{table:sersic_classes}
\end{table*}

\begin{table*}
	\caption{$UVJ$ Quenched Fractions of Groups}
	\begin{threeparttable}
	\begin{tabular}{ccrrrr}

\hline
& Group & \multicolumn{2}{c}{Quenched} & \multicolumn{2}{c}{Star-Forming} \\
\hline
\multirow{3}{*}{Low PC1}  & 6 & 43.5$_{-2.9}^{+3.1}$ & 39.3$_{-3.0}^{+3.3}$  & 56.5$_{-3.3}^{+3.5}$ & 60.5$_{-3.8}^{+4.0}$ \\ 
& 0 & 25.9$_{-3.8}^{+4.4}$ & 27.0$_{-5.3}^{+6.5}$  & 74.1$_{-6.4}^{+7.0}$ & 73.0$_{-8.7}^{+9.8}$ \\ 
& 9 & 7.6$_{-3.0}^{+4.4}$ & 15.1$_{-2.9}^{+3.5}$  & 92.4$_{-10.3}^{+11.6}$ & 84.9$_{-6.8}^{+7.4}$ \\ 
\hline 
& & 35.4$_{-0.1}^{+0.3}$ & 31.3$_{-0.8}^{+1.0}$  & 64.6$_{-0.1}^{+0.3}$ & 68.5$_{-1.2}^{+1.4}$ \\ 
\hline 
\multirow{2}{*}{Mid PC1} & 4 & 15.3$_{-3.6}^{+4.5}$ & 17.2$_{-3.9}^{+4.9}$  & 84.7$_{-8.3}^{+9.2}$ & 82.8$_{-8.5}^{+9.5}$ \\ 
& 8 & 8.1$_{-2.4}^{+3.1}$ & 13.9$_{-3.4}^{+4.2}$  & 91.9$_{-7.9}^{+8.6}$ & 86.1$_{-8.3}^{+9.1}$ \\ 
\hline 
& & 11.4$_{-0.2}^{+0.8}$ & 15.5$_{-1.7}^{+2.2}$  & 88.6$_{-0.4}^{+0.9}$ & 84.5$_{-4.0}^{+4.5}$ \\ 
\hline 
\multirow{3}{*}{High PC1}  & 1 & 0.0$_{-0.5}^{+1.9}$ & 8.3$_{-2.3}^{+3.1}$  & 100.0$_{-9.8}^{+10.8}$ & 91.7$_{-7.7}^{+8.4}$ \\ 
& 2 & 0.0$_{-0.7}^{+2.6}$ & 15.0$_{-3.5}^{+4.3}$  & 100.0$_{-11.4}^{+12.8}$ & 84.9$_{-8.2}^{+9.0}$ \\ 
& 5 & 0.9$_{-1.0}^{+2.1}$ & 13.9$_{-3.4}^{+4.2}$  & 99.1$_{-9.2}^{+10.1}$ & 86.1$_{-8.3}^{+9.1}$ \\ 
\hline 
& & 0.4$_{-0.2}^{+0.7}$ & 12.1$_{-0.9}^{+1.2}$  & 99.6$_{-0.4}^{+0.8}$ & 87.8$_{-2.4}^{+2.7}$ \\ 
\hline 
& -1 & 10.5$_{-7.6}^{+13.4}$ & 14.2$_{-9.3}^{+15.7}$  & 89.5$_{-20.9}^{+26.2}$ & 85.8$_{-21.7}^{+27.7}$ \\ 
\hline
\bf{Total Fraction} & & \multicolumn{2}{c}{\bf{23$\%$ (281)}} & \multicolumn{2}{c}{\bf{77$\%$ (962)}}\\
\hline

\end{tabular}
\begin{tablenotes}
            \item Note: Quenched/star-forming classifications based on Fig. \ref{fig:uvj_f125w}. The left hand columns for quenched/star-forming classifications represent the demographics based upon the original group based on hierarchical clustering.  The right hand columns are based on the total group probabilities based on the scattering technique classifications.
        \end{tablenotes}
     \end{threeparttable}
      \label{table:quenched_classes}
\end{table*}

\subsection*{Group 6} 

Constituting 37 per cent of the entire sample, group 6 is by far the most populated group at $z \sim 1.5$ (example postage stamps in Fig. \ref{fig:f125w_stamps_group6}) .  Group 6 galaxies are characterized by their compact sizes ($r_e \sim$ 1.57 $\pm$ 0.81 kpc) and smooth features.  Many of these galaxies are barely resolved by {\it HST WFC3} which leads to their structureless appearance.  Therefore, the structural properties of this group should be interpreted with caution, since it is possible that unresolved features in these galaxies would cause them to be classified as a different group if we had access to higher resolution observations.  43 per cent of the group is quenched, which represents 72 per cent of all quenched galaxies at this redshift.  Groups 0 and 4 are the only other group with a $>$10 per cent fraction of quenched galaxies.

Group 6 galaxies also dominate the high mass galaxies at this epoch, constituting 48 per cent of galaxies with $5 \times 10^{10} M_{\odot} < M_{*} < 10^{11} M_{\odot}$ and 49 per cent of galaxies with $M_{*} > 10^{11} M_{\odot}$. 

Group 6 galaxies have low concentrations ($C \sim$ 3.03 $\pm$ 0.40), moderate Gini coefficients ($G \sim$ 0.53 $\pm$ 0.05), low $M_{20}$ ($\sim$ -1.67 $\pm$ 0.17),  extremely low MID values ($I \sim$ 0.00 $\pm$ 0.02, $D \sim$ 0.06 $\pm$ 0.04),  and low asymmetry values ($A \sim$ 0.05 $\pm$ 0.06). The $G-M_{20}$ diagram classifies the majority of these galaxies as borderline disk/spheroidal (with occasional irregular classification).  However, $M_{20}$ values are potentially biased because the 20 per cent light is not resolved.  These galaxies have large average S\'{e}rsic indices ($\bar{n} \sim$ 3.11).  

Group 6 is comprised of the highest percentage of visually identified spheroids (52 percent) and disk+spheroids (26 percent), and also has the lowest percentage of disks (13 percent) of any group.

Upwards of 26 per cent of group 6 galaxies are small ($r_e <$ 2 kpc) with high Sersic ($n >$ 2.5) which could result in an underestimation of concentration and PC2 values.  These galaxies would instead be classified into group 0.  See Appendix B for more discussion.


\subsection*{Group 0} 

Group 0 galaxies are characterized by a strong bulge component which is surrounded by a faint smooth extended component (example postage stamps in Fig. \ref{fig:f125w_stamps_group0}). A significant fraction of group 0 galaxies are quenched galaxies (26 percent; Table \ref{table:quenched_classes}).  Although group 0 galaxies make up only 13 per cent of the galaxies in the sample,  they constitute 38 per cent of the galaxies more massive than $10^{11} M_{\odot}$ (Table \ref{table:mass_class}).

These galaxies have high concentration values ($C \sim$ 3.80 $\pm$ 0.78), low $M_{20}$ ($\sim$ -1.80 $\pm$ 0.17), high Gini coefficients ($G \sim$ 0.55 $\pm$ 0.04), low deviations ($D \sim$ 0.06 $\pm$ 0.04), low multi-modes ($M \sim$ 0.03 $\pm$ 0.04), low intensities ($I \sim$ 0.03 $\pm$ 0.04) and low asymmetries ($A \sim$ 0.06 $\pm$ 0.07).  This group of galaxies is the only class to fall almost entirely into the spheroidal region of the $G-M_{20}$ diagram. 

Visually, these galaxies have a large disk+spheroid fraction (33 percent), a large spheroid fraction (35 percent), a moderate disk fraction (31 percent) and a very low irregular fraction (1 percent).  Parametric fits find that group 0 galaxies have moderately sized effective radii ($r_e \sim$  3.13 $\pm$ 1.97 kpc) and large average S\'{e}rsic indices ($\bar{n} \sim$ 3.87). The visual classifications and distribution of S\'{e}rsic indices agree with $G-M_{20}$ measurements and thus describe the prototypical group 0 galaxy as bulge-dominated with a faint disk component or extended envelope.


\subsection*{Group 9} 
Group 9 is characterized by their asymmetric, irregular morphologies and strong bulge component (example postage stamps in Fig. \ref{fig:f125w_stamps_group9}).  
These galaxies make up a significant portion of the $M_* > 10^{11} M_{\odot}$ galaxies (15 percent).  However, most of these galaxies are lower mass ($M_* <  3 \times 10^{10} M_{\odot}$).  Only 8 per cent of group 9 galaxies are quenched.

These galaxies have moderate concentrations ($C \sim$ 3.70 $\pm$ 0.70), moderate Gini coefficient (G $\sim$ 0.52 $\pm$ 0.05), moderate $M_{20}$ ($\sim$ -1.40 $\pm$ 0.27), moderate MID values ($M \sim$ 0.14 $\pm$ 0.14, $I \sim$ 0.21 $\pm$ 0.18, D $\sim$ 0.19 $\pm$ 0.09) and high asymmetry ($A \sim$ 0.21 $\pm$ 0.10).   These galaxies lie along the $G-M_{20}$ merger/disk galaxy dividing line and also overlap with the spheroidal region.

Group 9 galaxies have large radii ($r_{e} \sim$ 3.67 $\pm$ 1.64 kpc) and moderately low average S\'{e}rsic indices ($\bar{n} \sim$ 2.11).  

This group is the most visually irregular group (24 percent), and has a relatively low disk fraction (41 percent), spheroid fraction (13 percent) and disk+spheroid fraction (11 percent).  These statistics and visual classifications imply many galaxies have bright off-center clusters, in addition to bright central bulges. 


\subsection*{Group 4} 
Group 4 galaxies are low-mass, smooth, extended galaxies with moderate central concentrations (example postage stamps in Fig. \ref{fig:f125w_stamps_group4}). Although mostly star-forming, group 4 contains some quenched galaxies ($\sim$11 percent).  Some galaxies are extended and also quenched; meaning they are rare ``red disk'' population.    None of the group 4 galaxies are more massive than $M_* > 10^{11} M_{\odot}$.   Primarily these galaxies are lower mass ($M_* < 3 \times 10^{10} M_{\odot}$).


Group 4 has moderate concentrations ($C \sim$ 3.53 $\pm$ 0.66), moderate Gini coefficients ($G \sim$ 0.49 $\pm$ 0.04 ),  high $M_{20}$ ($\sim$ -1.11 $\pm$ 0.24), low intensities ($I \sim$ 0.05 $\pm$ 0.05), small multi-mode values ($ M \sim 0.07 \pm$ 0.07) , low deviations ($D \sim$ 0.10 $\pm$ 0.07), and low asymmetry ($A \lesssim$ 0). 

Group 4 galaxies have moderate effective radii ($r_e \sim$ 3.13 $\pm$ 1.63 kpc) and medium average S\'{e}rsic indices ($\bar{n} \sim$ 2.68).  

Group 4 members are primarily visually classified as disks (51 percent) or disk+spheroids (24 percent) and are less classified as spheroids (17 percent) or irregulars (0 percent). 

\subsection*{Group 8} 
Group 8 galaxies are an interesting class of bulge+disk systems with dominant and smooth disks (example postage stamps in Fig. \ref{fig:f125w_stamps_group8}).  This
class is dominated by low-mass star-forming galaxies,  but also includes low-mass ($< 3 \times 10^{10} M_{\odot}$) quenched galaxies ($\sim$ 8 percent).  Very few galaxies have stellar masses $> 5 \times 10^{10} M_{\odot}$. 



Group 8 galaxies have small concentrations ($C \sim$ 3.05 $\pm$ 0.43), moderate Gini coefficients ($G \sim$ 0.46 $\pm$ 0.03), moderate $M_{20}$ ($\sim$ -1.56 $\pm$ 0.17),  low but non-zero MID values ($M \sim$ 0.06 $\pm$ 0.06, $I \sim$ 0.10 $\pm$ 0.11, $D \sim$ 0.09 $\pm$ 0.05), and low asymmetry values ($A \sim$ 0.08 $\pm$ 0.07). On the $G-M_{20}$ diagram these galaxies fall within the disk-dominated region but are close to the spheroidal/disk dividing line. 

S\'ersic fits to this class find moderate sizes ($r_e \sim$ 3.48 $\pm$ 1.89 kpc) and low average S\'{e}rsic indices ($\bar{n} \sim$ 1.46).  

Group 8 is dominated by visually-classified disks (74 percent) with only a modest fraction of spheroids (10 percent).  A small number of galaxies are quenched and compact which overlaps with groups 0 and 6.

\subsection*{Group 1} 
Group 1 galaxies are primarily large disks and irregulars with bright off-center star-forming knots (example postage stamps in Fig. \ref{fig:f125w_stamps_group1}).   None of these galaxies are quenched based on their $UVJ$ colors.   The distribution of masses is heavily weighted towards lower mass galaxies with very few objects more massive than 3$\times 10^{10}$ $M_{\odot}$.


Group 1 galaxies have low concentration values ($C \sim$ 2.76 $\pm$ 0.40), low Gini coefficients ($G \sim$ 0.43 $\pm$ 0.04), high $M_{20}$ ($\sim$ -1.07 $\pm$ 0.17),  moderately high asymmetry values ($A \sim$ 0.13 $\pm$ 0.11), large multi-mode values ($M \sim$ 0.40 $\pm$ 0.27), high deviations ($D \sim$ 0.37 $\pm$ 0.13) and large intensities ($I \sim$ 0.61 $\pm$ 0.24).  The high $A$ and $MID$ statistics indicate many of these galaxies have bright off-center clusters and are potentially irregular.

The visual classifications and S\'{e}rsic indices primarily classify this group as disk galaxies and/or irregulars.  Group 1 is dominated by visually-classified disks (72 percent) and has a relatively large fraction of irregulars (16 percent). This group has very few spheroids or bulge-dominated disk galaxies. Their effective radii are large for this redshift and mass ($r_e \sim$ 5.35 $\pm$ 1.43 kpc).  This group has low average S\'{e}rsic indices ($\bar{n} \sim$ 0.63) imply a large disk and irregular population.  


\subsection*{Group 2} 
Group 2 galaxies are primarily low-mass , star-forming, smooth disk galaxies with high central concentrations and few visually detected star-forming knots (example postage stamps in Fig. \ref{fig:f125w_stamps_group2}). None of these galaxies are quenched.  The mass distribution for this group is a steeply declining function where there are only a few galaxies with masses $> 3 \times 10^{10} M_{\odot}$. 

Group 2 galaxies have large concentrations ($C \sim$ 4.81 $\pm$ 0.62), low Gini coefficients ($G \sim$ 0.45 $\pm$ 0.04 ), moderate $M_{20}$ ($\sim$ -1.20 $\pm$ 0.24), low asymmetry ($A \sim$ 0.06 $\pm$ 0.08), low deviations ($D \sim$ 0.16 $\pm$ 0.09), moderate multi-modes ($M \sim$ 0.16 $\pm$ 0.21),  and a wide spread of intensity values ($I \sim$ 0.29 $\pm$ 0.29).  On the $G-M_{20}$ diagram these galaxies fall within the  disk-dominated and irregular portion of the diagram. However, their high $C$ values suggest a bright nuclear component. 

The visual classifications show this group is dominated by disks (76 percent) and only small fractions of irregular galaxies ($\sim$5 percent) and disk+spheroid galaxies ($\sim 8$ percent).  They have mid-sized effective radii ($r_e \sim$ 3.52 $\pm$ 0.83 kpc) and mid-to-low average S\'{e}rsic indices ($\bar{n} \sim$ 1.10). 


\subsection*{Group 5} 
Group 5 galaxies are primarily low-mass ($M_* < 3 \times 10^{10} M_{\odot}$), star forming, extended disk galaxies with a weak bulge component (example postage stamps in Fig. \ref{fig:f125w_stamps_group5}).  This group has a negligible fraction of quenched galaxies ($\sim$ 1 percent).

Group 5 is mostly comprised of low concentration values ($C \sim$ 2.87 $\pm$ 0.42), low Gini coefficients ($G \sim$ 0.40 $\pm$ 0.03), low/moderate $M_{20}$ ($\sim$ -1.20 $\pm$ 0.19), a wide spread in multi-mode ($M \sim$ 0.26 $\pm$ 0.24),  large intensity values ($I \sim$ 0.52 $\pm$ 0.28), low deviation values ($D \sim$ 0.12 $\pm$ 0.06),  and low asymmetry values ($A \sim$ 0.03 $\pm$ 0.12).  On the $G-M_{20}$ diagram these galaxies fall solidly within the disk-dominated region.

The defining feature of this group is its large typical size ($r_e \sim$ 5.47 $\pm$ 1.81 kpc).  Group 5 galaxies have very low average S\'{e}rsic indices ($\bar{n} \sim$ 0.65); implying a disk-dominated/irregular population.  

Visual classification indicate group 5 is comprised almost entirely of disks (95 percent), and a few irregulars (3 percent).  This group has no visually identified bulge-dominated or spheroidal galaxies and are not clumpy.


\subsection*{Group -1}  \label{outliers}

The original groups 3 and 7 were comprised of only a few galaxies each (19 in total, example postage stamps in Fig. \ref{fig:f125w_stamps_group-1}).  They were outliers from the remaining groups and are combined into a single outlier group.  These galaxies are most likely outliers because they have at least one poorly measured (or missing) morphological parameter (especially the multi-mode statistic).  These galaxies have a low surface brightness, very large radii ($r_{e} \sim$ 6.73 $\pm$ 2.30 kpc), low concentration ($C \sim$ 2.21 $\pm$ 0.74), high intensity ($I \sim$ 0.44 $\pm$ 0.39 ), high $M_{20}$ ($\sim$ -0.99 $\pm$ 0.26), low Gini coefficient ($G \sim$ 0.41 $\pm$ 0.10), extremely high deviations ($D \sim$ 0.69 $\pm$ 0.49) and high multi-modes ($M \sim$ 0.53 $\pm$ 0.39).   The deviation values can separate group -1 galaxies from all the other groups.


  
    \section{Discussion} \label{section:discussion}

      The spatial distribution of light for galaxies is a snapshot of the orbital paths of the constituent stars, gas, and dust.  The morphology of a galaxy informs us of the merger and gas-accretion history in ways integrated colors, spectral-energy distributions and stellar mass cannot directly probe. 
      
	Using a S\'ersic index, bulge-dominated galaxies are traditionally defined to have $n >$ 2.5 (e.g. \citealp{Bruce14a}).  For the purposes of of our PC classifications we define galaxies with low PC1 values as bulge-dominated (the constituents of groups 0, 6 and 9).  These two definitions lead to differences in the characteristics of what are defined as `bulge-dominated' and we will explore these differences in the following sections.
    
        The connection between morphology and star-formation has been well studied \citep{Wuyts11,Kriek09,Brinchmann04}.  Late-type galaxies are typically still actively forming stars, whereas early-type galaxies have had their star-formation quenched.  However, there are examples of red, quenched disks and blue, star-forming ellipticals which are important rare ``transitional'' classes.  In our study we delve deeper into the correlations between morphological type and star-formation and how the connection between them is not always clear-cut.
    
    



\subsection{Stellar Mass - Quenching Connection for groups}

 \begin{figure}
 \centering
    \includegraphics[width=0.5\textwidth]
    {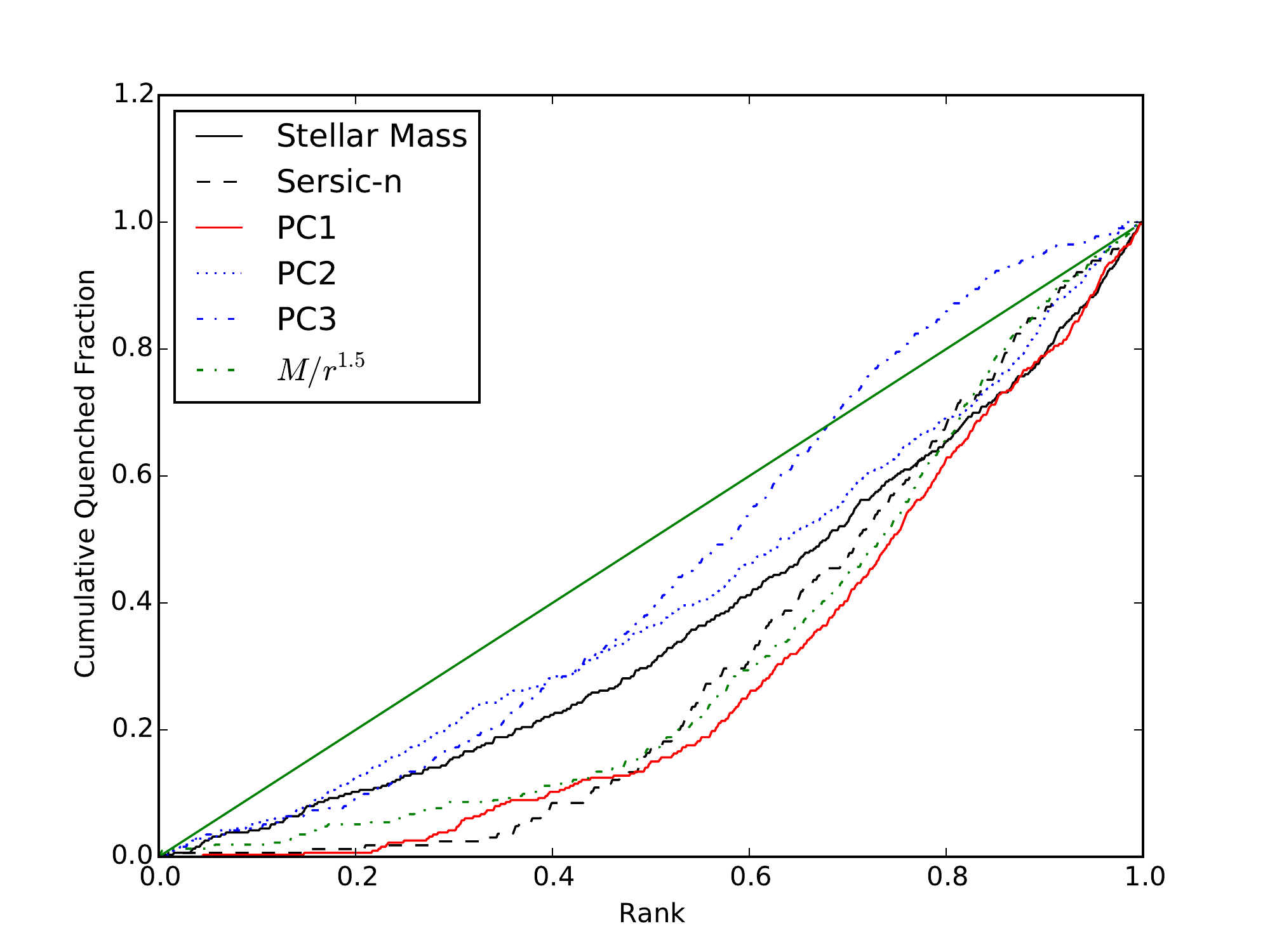}
      \caption[Cumulative quenched fraction rank]
      {Cumulative quenched fraction rank ordered by various metrics: PC1, PC2, PC3, stellar mass, S\'ersic-n and ``compactness''.  The green solid line represents no correlation between quenched fraction and rank.  S\'ersic-n and PC1 have a similar CQF shape, where n is less contaminated by quenched galaxies at low values but PC1 is less contaminated at high values.}
      \label{fig:cumPC}
    \end{figure}


Fig. \ref{fig:cumPC} shows the cumulative distribution of the quenched fraction rank-ordered by ``compactness'' ($M/r_e^{1.5} <  10.3 M_{\odot}$ kpc$^{-1.5}$;  \citealp{Barro13}), stellar mass, S\'ersic-$n$, PC1, PC2 and PC3. For every galaxy we assign a 0 to star-forming galaxies and 1/$n_{quenched}$ for quenched galaxies (as determined by the $UVJ$ diagram, Fig. \ref{fig:uvj_f125w}) and then these values are cumulatively summed.  We observe a flat trend in stellar mass and PC2, and a much steeper trend in PC1 and S\'ersic-n.  However, S\'ersic-n has previously been shown to correlate well with quenching (e.g. \citealp{Wuyts11,Bell12}).  The similarities in steepness between the PC1 and S\'ersic-n curves show PC1 is an equivalently useful predictor of quenching.




We also investigate the relationship between quenching and PC1 through the color-mass relation.  In Fig. \ref{fig:uv-Mass} we observe a correlation between the increase in the fraction of massive galaxies ($> 5 \times 10^{10} M_{\odot}$) for a specific group and the magnitude of PC1 (bulge strength).  The amount of quenched galaxies correlates more strongly with PC1 (bulge strength) than PC2 (concentration).  Similar results have been found for $z\sim$ 1-2 galaxies \citep{Bell12,Barro13,Lang14}.  Unsurprisingly, the most massive galaxies are also the most likely to be quenched.  For instance, group 6 has the largest amount of red galaxies and many massive galaxies ($> 5 \times 10^{10} M_{\odot}$). The only other groups with a substantial number of quiescent galaxies are groups 0 (26 percent) and 4 (15 percent).   Group 0 galaxies are primarily bulge-dominated with a faint disk.  However, group 6 galaxies are more massive ($> 5 \times 10^{10} M_{\odot}$) than group 0 galaxies.   Furthermore, a much larger percentage of these massive galaxies in group 6 are quenched.  Group 9 galaxies are slightly less massive ($\log M_* \sim$ 10.4), but still generally have a strong bulge component (as determined by PC1).  


Group 4 and 8 galaxies fall between the extremes of the bulge-dominated groups (0, 6 and 9) and the disk-dominated groups (1, 2 and 5) in stellar mass, bulge strength and quenched fraction (see Tables \ref{table:mass_class} - \ref{table:quenched_classes}).  The galaxies of groups 4 and 8 are more bulge-dominated than the disk-dominated galaxies which would explain the larger quenched fraction. Groups 4 and 8 galaxies are not as massive as those in the bulge-dominated groups 0, 6 and 9 (Table \ref{table:mass_class}) and are not quenched to the same extent either (Table \ref{table:quenched_classes}).

\subsection{The relationship between PCA Classes and Visual/S\'ersic Classifications}

PCA, in conjunction with our group finding algorithm, provides a distinct picture of galaxy structure from S\'ersic index and visually based classifications.  This classification scheme also separates quenched compact galaxies (group 6)  from larger, smooth proto-elliptical systems (group 0), and star-forming disk-dominated clumpy galaxies (group 1) from star-forming bulge-dominated asymmetric galaxies (group 9).  Separating clumpy star-formers and bulge dominated star forming galaxies has great importance for understanding the mechanisms that formed these galaxies and the potential avenues for evolution available to them.

   Based upon the visual and S\'ersic classifications, our groups belong to 3 distinct types: the ``disk-dominated'' galaxies of groups 1, 2, and 5; the ``compact/bulge-dominated'' galaxies of groups 0, 6, and 9; and the ``intermediate'' galaxies of groups 4 and 8.  For the purposes of our discussion we refer the reader to Figs. \ref{fig:uvj_f125w} - \ref{fig:size_mass}, the example galaxies of Figs. \ref{fig:f125w_stamps_group6} - \ref{fig:f125w_stamps_group-1} and Tables \ref{table:mass_class} - \ref{table:quenched_classes}.
   



\subsubsection{The Compact and Bulge-Dominated Galaxies: Groups 0, 6 and 9}


Galaxies in groups 0, 6, and 9 display a variety of visual classifications, but have a single unifying characteristic: many of these galaxies are bulge-dominated.  Group 6 galaxies are very small and compact ($r_e \sim$ 1.57 $\pm$ 0.81 kpc) with no discernible stellar envelope.  Group 0 galaxies are slightly larger ($r_e \sim$  3.13 $\pm$ 1.97 kpc) than group 6, and display evidence for an extended stellar envelope.   Groups 0 and 6 display some distinguishing characteristics as well.  Group 6 galaxies lower measured concentrations ($C \sim$ 3.04 $\pm$ 0.40) than those in group 0 ($C \sim$ 3.80 $\pm$ 0.78). The small sizes and lower concentrations for group 6 galaxies are due to the fact that $r_{20}$ measurements are near or below the resolution limit of the survey.   Additionally, $r_{80}$ measurements are very small for group 6 compared to galaxies in all other groups (see Appendix \ref{section:c_n}).

The size-mass (Fig. \ref{fig:size_mass}) relation for these two groups is different as well.  Group 6 galaxies are smaller but overlap in masses with group 0 galaxies.  Thus many more group 6 galaxies are compact using the \citet{Barro13} definition.  Compact galaxies in group 6 are quenched, whereas the quenched galaxies of group 0 are more extended.



Group 6 galaxies are visually classified as bulge-dominated (either pure spheroid or disk+spheroid morphology) $>$78  per cent of the time.  However, a S\'ersic cut of $n > 2.5$ yields only 35 per cent.   Similarly for group 0 galaxies, 66 per cent of galaxies are visually classified as bulge-dominated, but a S\'ersic classification only indicates 43 per cent are bulge-dominated galaxies.  Meanwhile, group 9 galaxies are the most visually disturbed group (26 per cent irregular) and have bright central bulges determined by PC1.  

     Classifications based on PCs provide a slightly different picture from those based on S\'ersic-$n$  or visual inspection.  A PC classification determines $\sim$57 per cent of galaxies are compact/bulge-dominated (groups 0, 6 and 9) while visual classifications determine $\sim$47 per cent of galaxies are bulge-dominated (either pure spheroids or disk+spheroids) and S\'ersic indices classify $\sim$25 per cent of galaxies as bulge-dominated ($n >$ 2.5).   The differences between the classification schemes are subtle but important because they mean each is probing a slightly different subset of galaxies.

     The compact/bulge-dominated nature and high masses of these 3 groups could imply an evolutionary connection.  In this scenario, galaxies begin as group 6 galaxies, a naked core with no extended envelope or structure.  Following a gas-rich merger disturbed tidal features become visible and the galaxy becomes classified as group 9.  After a sufficient time for the gas to settle in a disk or spheroidal envelope ($\gtrsim$ 1.5 Gyr) the galaxy would appear as a group 0 galaxy.  In this scenario, the quenched galaxies of group 6 have star-formation reignited following the merger, only to once again fade during the disk settlement period.  Mergers would thus be a major mechanism for triggering disk growth.
     

 


 \subsubsection{The Disk-dominated Galaxies: Groups 1, 2 and 5}
 
 Groups 1, 2 and 5 all have an overwhelmingly large percentage of visually classified disk galaxies (72 per cent,  76 per cent, and 96 per cent respectively).  S\'ersic classifications largely agree with the visual classifications for these groups.  The only difference is that S\'ersic classifications yield more disk-dominated galaxies (1 $< n <$ 2.5) than visual classifications would indicate.  Non-parametric morphologies determine these disk galaxies have varying degrees of clumpiness and disturbances.  

   
       Group 1 galaxies are the most disturbed of the ``disk-dominated'' groups.  They have the largest asymmetries ($A \sim$ 0.13 $\pm$ 0.11), multi-modes ($M \sim$ 0.40 $\pm$ 0.27), intensities ($I \sim$ 0.61 $\pm$ 0.24) and deviations ($D \sim$ 0.37 $\pm$ 0.13).  They are more often visually classified as irregular (16 percent), but have a weaker bulge component (indicated by their larger $M_{20}$ values, $\sim$ -1.07 $\pm$ 0.17) than groups 2 and 5.
		
     Of the remaining disk-dominated groups, group 5 galaxies have much higher $M$ and $I$ statistics ($M \sim$ 0.26 $\pm$ 0.24 and $I \sim$ 0.52 $\pm$ 0.28) than those in group 2 ($M \sim$ 0.16 $\pm$ 0.21 and $I \sim$ 0.29 $\pm$ 0.29).  However, these two groups have similar asymmetry values ($A \sim$ 0.05), $M_{20}$ values ($\sim$ -1.2), and deviations ($D \sim$ 0.1).  
     
     The disk-dominated galaxies of groups 1, 2 and 5 are on average less massive, bluer in $U-V-J$ and larger than the compact/bulge-dominated galaxies of groups 0, 6 and 9.

\subsubsection{The Intermediate Galaxies: Groups 4 and 8}

Groups 4 and 8 represent an intermediate PC class between the compact/bulge-dominated morphologies of groups 0, 6 $\&$ 9 and the disk-dominated groups 1,2 $\&$ 5.  Group 4 and 8 both have a population of quenched galaxies.  However, the quenched galaxies of group 8 are smaller than those of group 4.

Both groups 4 and 8 have a large fraction of galaxies with $n <$ 2.5 (72 per cent and 80 per cent, respectively).  However, group 8 galaxies are more likely to be visually classified as disks  than group 4 galaxies (74 per cent compared to 51 percent).  Meanwhile, group 4 galaxies are more likely be visually classified as bulge-dominated (41 per cent compared to 15 percent).  However, the differences between groups should taken with caution as the small numbers of galaxies in these groups reduces the significance of the percentages.

For groups 4 and 8 the classifications based upon non-parametric morphologies do not always agree with classifications based on S\'ersic indices or visual inspection.  Group 8 has a much smaller average $M_{20}$ value ($M_{20} \sim$ -1.56 $\pm$ 0.17) than group 4 ($M_{20} \sim$ -1.11 $\pm$ 0.24).  This indicates the bulges of group 8 galaxies are large and possibly dominate the morphology.  However, S\'ersic indices and visual classifications would suggest there is no sizable bulge component for most of these galaxies.  Group 4 galaxies have high concentrations, low S\'ersic indices and are the least well defined group by bootstrap measures (see Fig. \ref{fig:prob_dist}).  Meanwhile, the $G-M_{20}$ diagram suggests a population of irregular galaxies while visual classifications find no irregular galaxies.  The bright nuclear components may be the result of an AGN or starburst activity.




\subsubsection{Comparing the Irregular Galaxies of Groups 1 and 9}

The galaxies of groups 1 and 9 are the most likely to be classified visually as irregular.  While group 1 is defined by star-forming disk-dominated clumpy galaxies, group 9 is defined by star-forming bulge-dominated asymmetric galaxies with tidal features.   These subtle morphological differences are missed by S\'ersic index, $C-A$ and $Gini-M_{20}$ based classifications and potentially offer clues as to the formation and evolutionary tracks of these galaxies.

Group 9 galaxies display tidal features and irregular disks but their strong central bulge is missed by S\'ersic fits.  Group 9, itself, shows the power of our PCA classifications to find interesting subtypes of galaxy morphology.  Group 9 galaxies are visually classified as disks (41 percent), irregulars (23 percent) and bulge-dominated disks (12 percent).  However, small PC1 values would indicate group 9 galaxies posses a strong central bulge.  Meanwhile, group 1 galaxies much more likely to be visually classified as a pure disk galaxy (72 percent), slightly less likely to be irregular (16 percent) and are not bulge-dominated (0 percent).    Group 1 galaxies also have higher PC1 values, indicating a weaker bulge component.  Using S\'ersic index classifications, both groups 1 and 9 have a very large fraction of these galaxies are disk-dominated (85 percent) as opposed to bulge-dominated (15 percent). Groups 1 and 9 would be considered very similar in a S\'ersic classification and the differences between these groups are more subtle.
 
 We observe subtle differences between these two groups in many statistics; group 9 galaxies are more asymmetric (0.22 $\pm$ 0.10 vs. 0.13 $\pm$ 0.11) and have lower $M_{20}$ values (-1.40 $\pm$ 0.27 vs. -1.07 $\pm$ 0.17) than galaxies found in group 1.  Group 9 galaxies are also more concentrated (3.70 $\pm$ 0.70 vs. 2.76 $\pm$ 0.40).  Meanwhile $M$, $I$ and $D$ statistics all display an increased enhancement in group 1 galaxies because these statistics probe the existence of off-center clumps. 
 
 Based on these differences it is possible these two types of galaxies have experienced different formation scenarios or exist at different stages along their evolution.  Group 9 galaxies have a large central bulge which could be the result of either a merger or the accretion of many star-forming clumps in the disk.  Meanwhile, group 1 galaxies are still clumpy and have small central bulges.  Different levels of the amount of violent disk instabilities (VDI; \citealp{Dekel09,Guo15}) is a possible explanation for the segregation of groups 1 and 9.  Group 9 galaxies have a larger bulge, possibly grown by the migration of clumps to the central galaxy regions following repeated VDIs.  Meanwhile, group 1 galaxies, which still have bright clumps in the disk (as evidenced by enhanced $MID$ statistics) have yet to experience as many VDIs and thus the central bulge remains smaller.

\section{Summary}


We use a principal component analysis of non-parametric morphology measurements ($G$, $M_{20}$, $C$, $A$, $M$, $I$ and $D$) and agglomerative hierarchical clustering to group galaxies into a more descriptive schema than the traditional spiral, elliptical, and irregular categories.  The PCA weights we calculate (Table \ref{table:pc_weights}) show that non-parametric morphological correlations vary in importance: PC1 is based upon $M$,$I$,$D$,$M_{20}$ and Gini thus it is interpreted as a bulge strength indicator; PC2 is dominated by concentration; and PC3 is dominated by asymmetry; the remaining PCs are less important and difficult to interpret. 

The size-mass relation is dependent on PC1 and PC2.  Galaxies with high PC1 values (stronger bulges) are generally more compact and quiescent than galaxies with high PC2 values.  We determine PC1 is a valid predictor of whether a galaxy is quenched.

\begin{figure*}
      \includegraphics[scale=0.125]
      {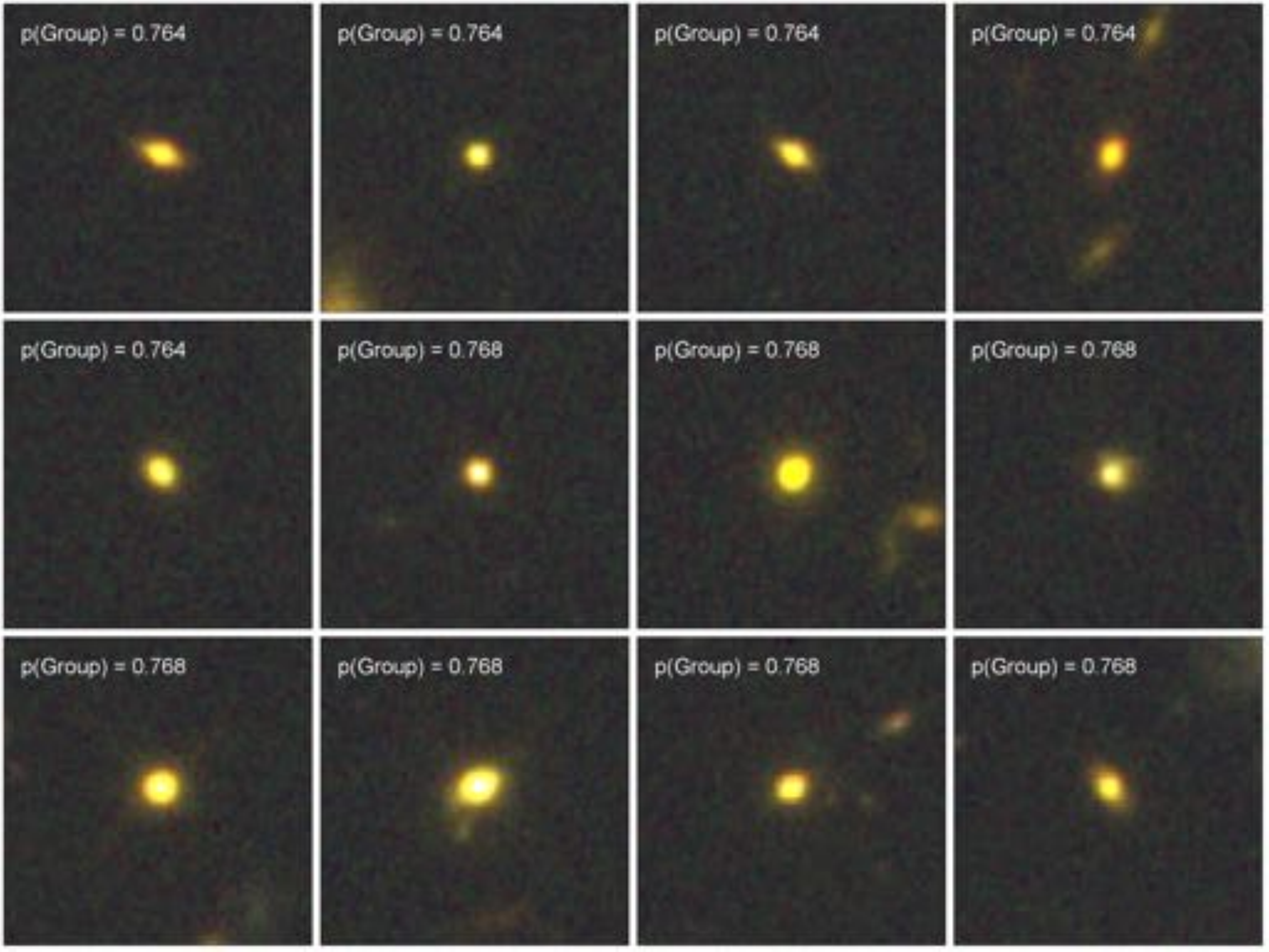}
      \centering
      \caption[Group 6 F125W 1.36 $<$ z $<$ 1.97 galaxies]
      {Group 6 F125W 1.36 $<$ z $<$ 1.97 galaxies, shown in F160W/F125W/F814W  RGB 6''x6'' postage stamps.    p(Group) represents the percentage of times a galaxy is classified into group 6 after the scattering test.  Very compact and small spheroidal galaxies. This group contains the largest spheroidal and quenched fraction.  Many of these galaxies are barely resolved which leads to their structureless appearance.}
      \label{fig:f125w_stamps_group6}
    \end{figure*}

\begin{figure*}
      \includegraphics[scale=0.125]
      {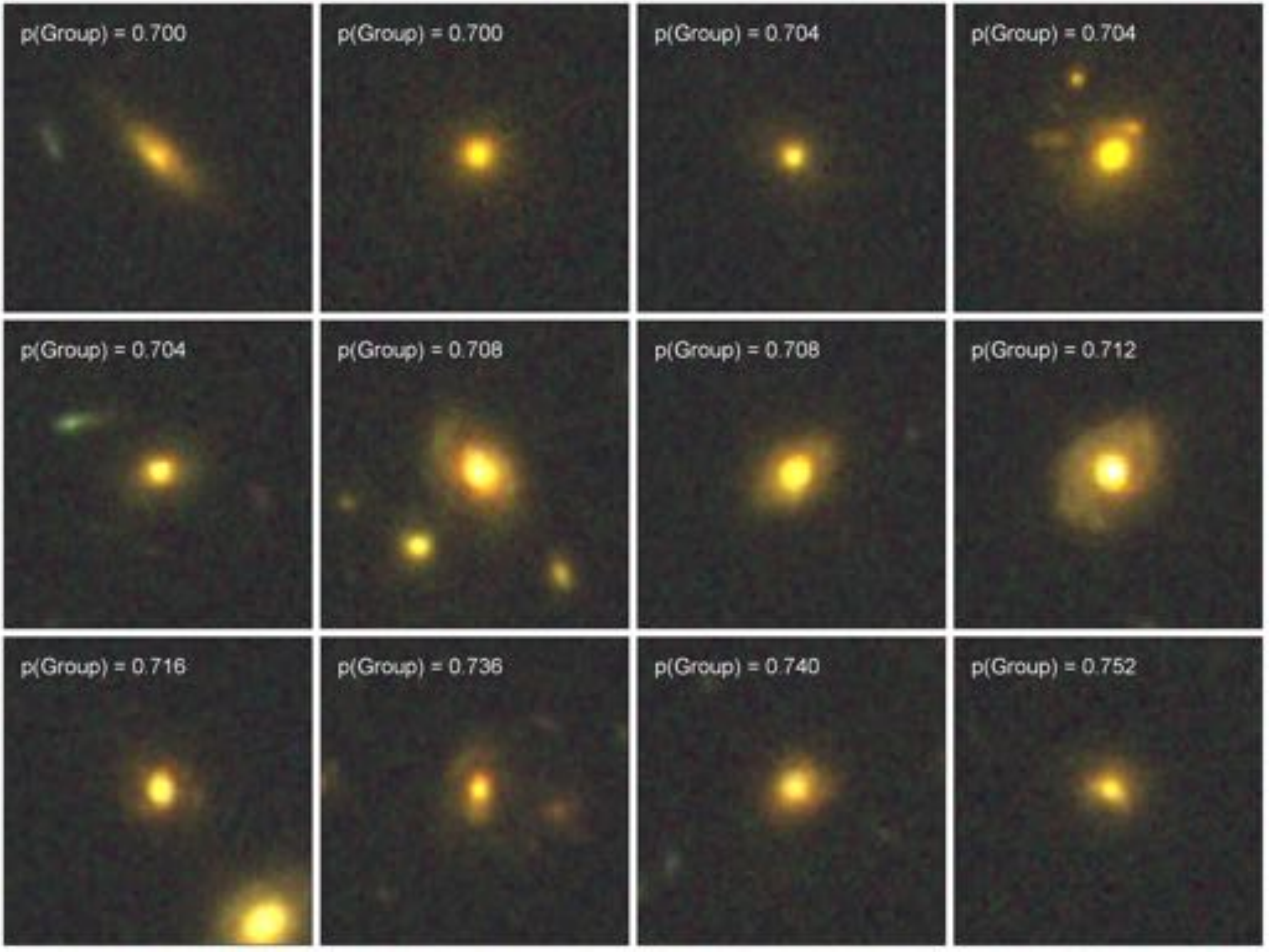}
      \centering
      \caption[Group 0 F125W 1.36 $<$ z $<$ 1.97 galaxies]
      {Group 0: These galaxies are characterized by a strong bulge component surrounded by a fainter smooth disk.}
      \label{fig:f125w_stamps_group0}
    \end{figure*}

\begin{figure*}
      \includegraphics[scale=0.125]
      {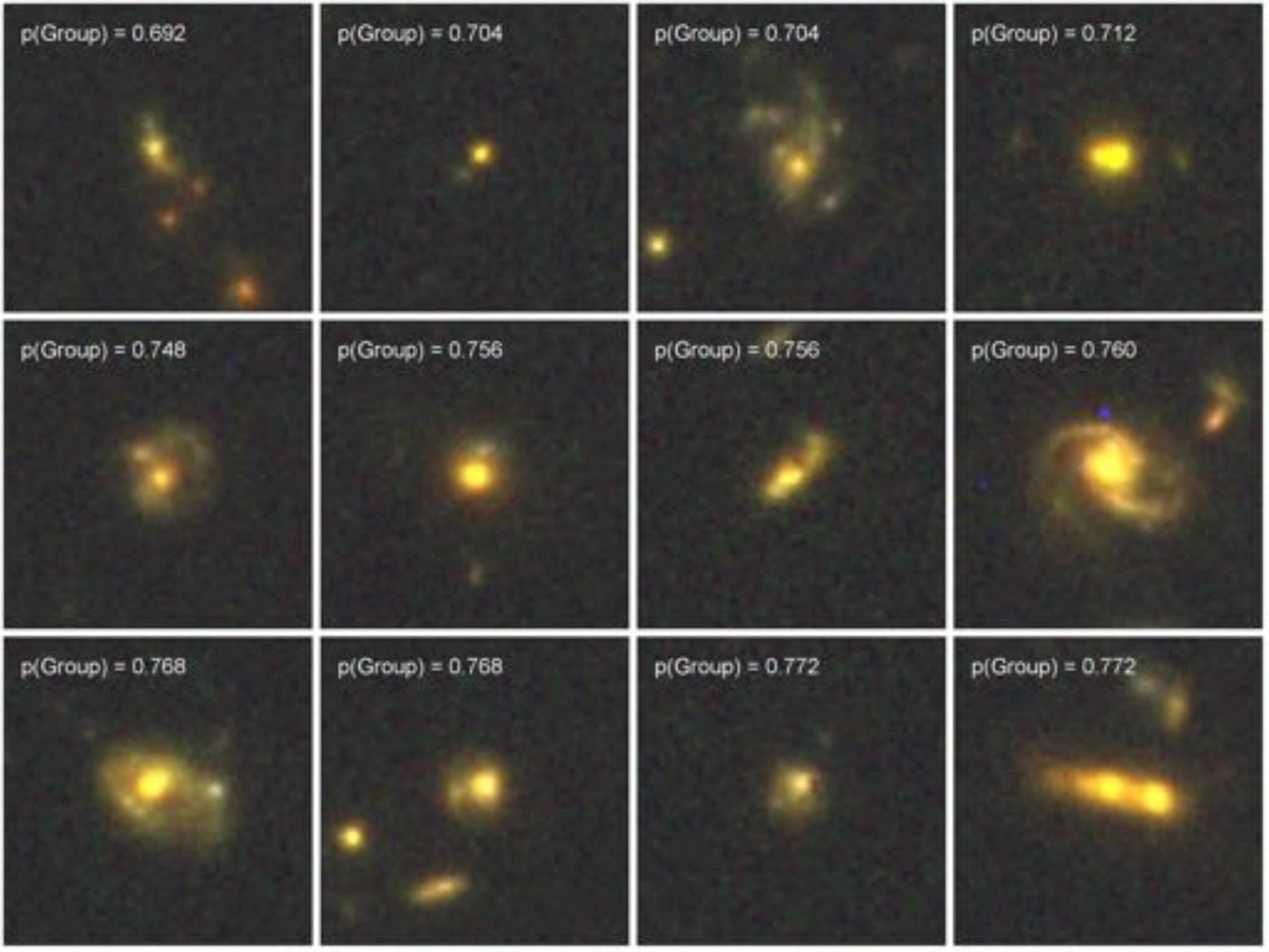}
      \centering
      \caption[Group 9 F125W 1.36 $<$ z $<$ 1.97 galaxies]
      {Group 9: These galaxies are characterized by their asymmetric, irregular morphologies and strong bulge component.}
      \label{fig:f125w_stamps_group9}
    \end{figure*}

\begin{figure*}
      \includegraphics[scale=0.125]
      {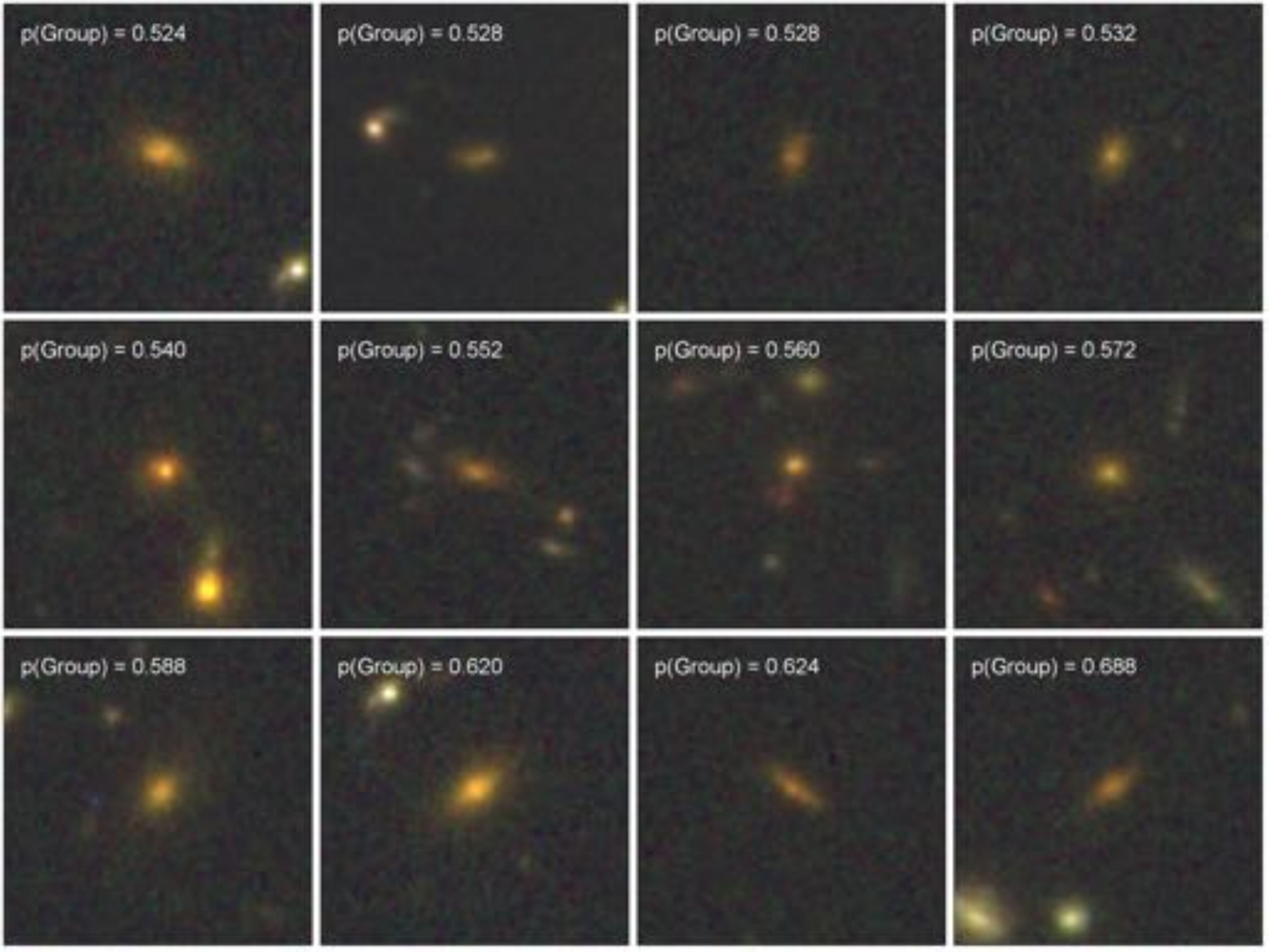}
      \centering
      \caption[Group 4 F125W 1.36 $<$ z $<$ 1.97 galaxies]
      {Group 4: These galaxies consist of low-mass smooth galaxies with moderate central concentrations.}
      \label{fig:f125w_stamps_group4}
    \end{figure*}

\begin{figure*}
      \includegraphics[scale=0.125]
      {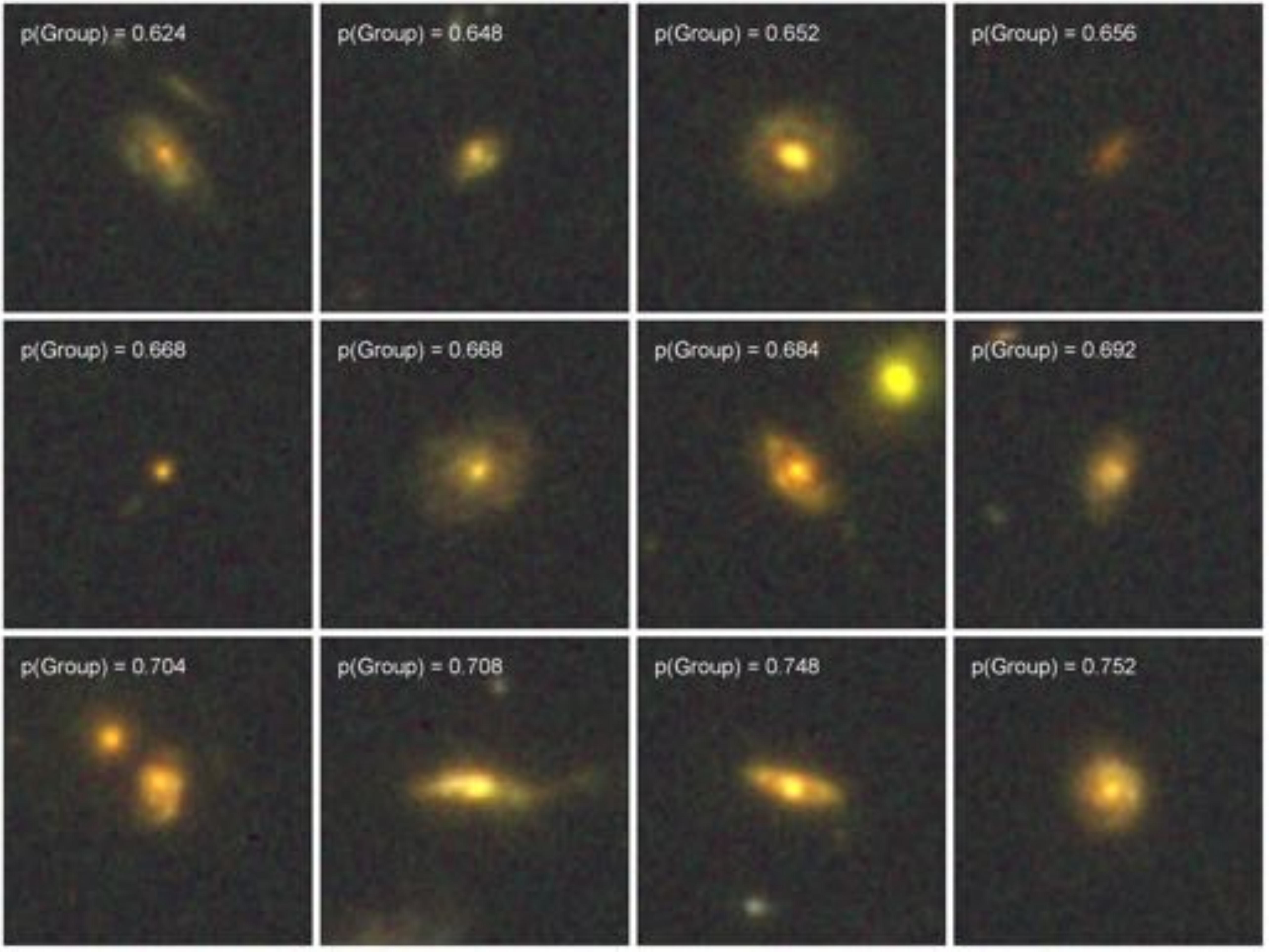}
      \centering
      \caption[Group 8 F125W 1.36 $<$ z $<$ 1.97 galaxies]
      {Group 8: These galaxies represent class of bulge+disk systems with dominant smooth disks.}
      \label{fig:f125w_stamps_group8}
    \end{figure*}

\begin{figure*}
      \includegraphics[scale=0.125]
      {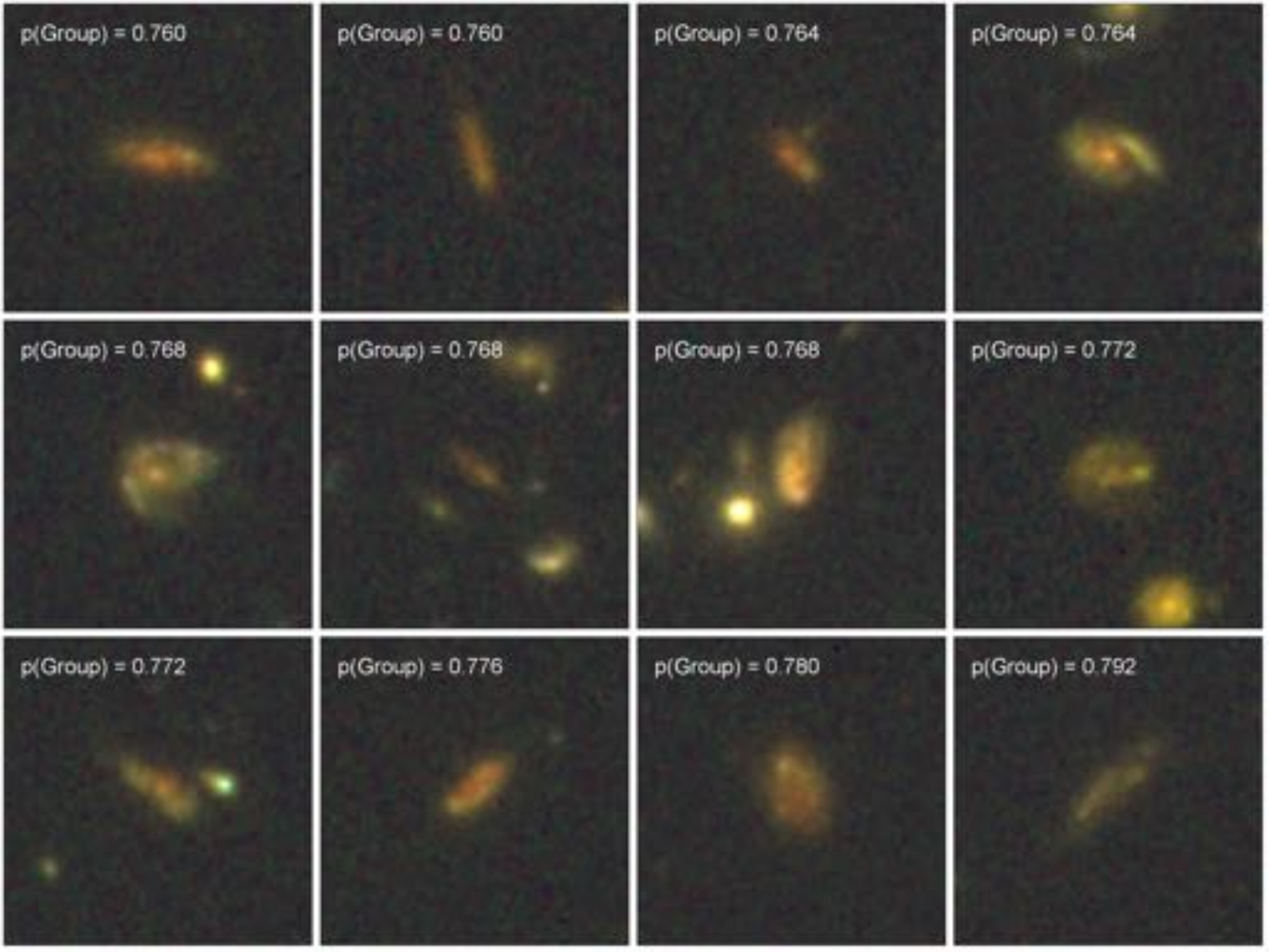}
      \centering
      \caption[Group 1 F125W 1.36 $<$ z $<$ 1.97 galaxies]
      {Group 1: These galaxies are primarily large disks and irregulars with bright off-center star-forming knots.}
      \label{fig:f125w_stamps_group1}
    \end{figure*}

\begin{figure*}
      \includegraphics[scale=0.125]
      {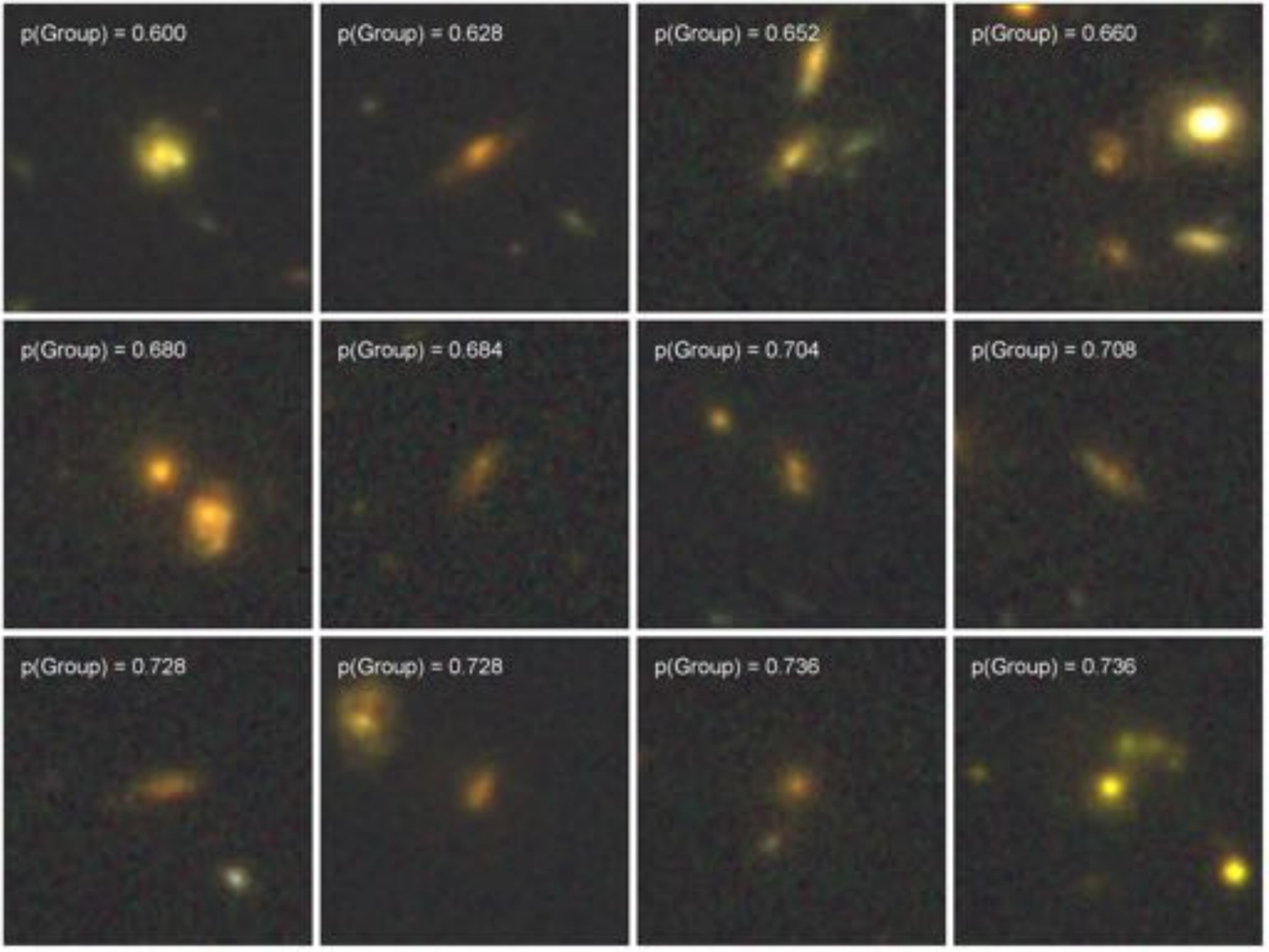}
      \centering
      \caption[Group 2 F125W 1.36 $<$ z $<$ 1.97 galaxies]
      {Group 2: These galaxies appear to be primarily low-mass star-forming disk galaxies with higher central concentrations and few detected star-forming knots. }
      \label{fig:f125w_stamps_group2}
    \end{figure*}

\begin{figure*}
      \includegraphics[scale=0.125]
      {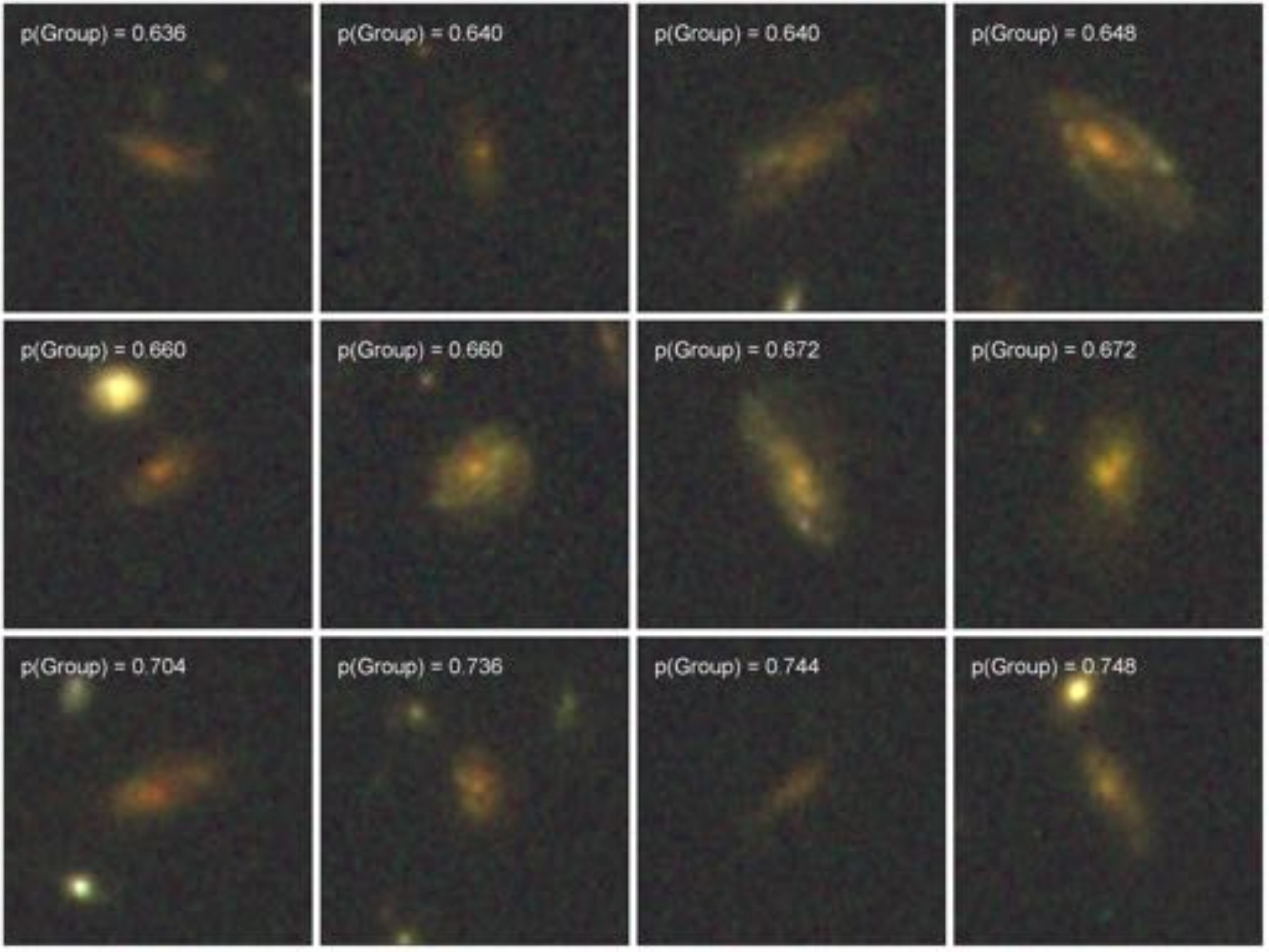}
      \centering
      \caption[Group 5 F125W 1.36 $<$ z $<$ 1.97 galaxies]
      {Group 5: Many of these galaxies are low-mass extended star forming disk galaxies with weak (if any) bulge components.}
      \label{fig:f125w_stamps_group5}
    \end{figure*}

\begin{figure*}
      \includegraphics[scale=0.125]
      {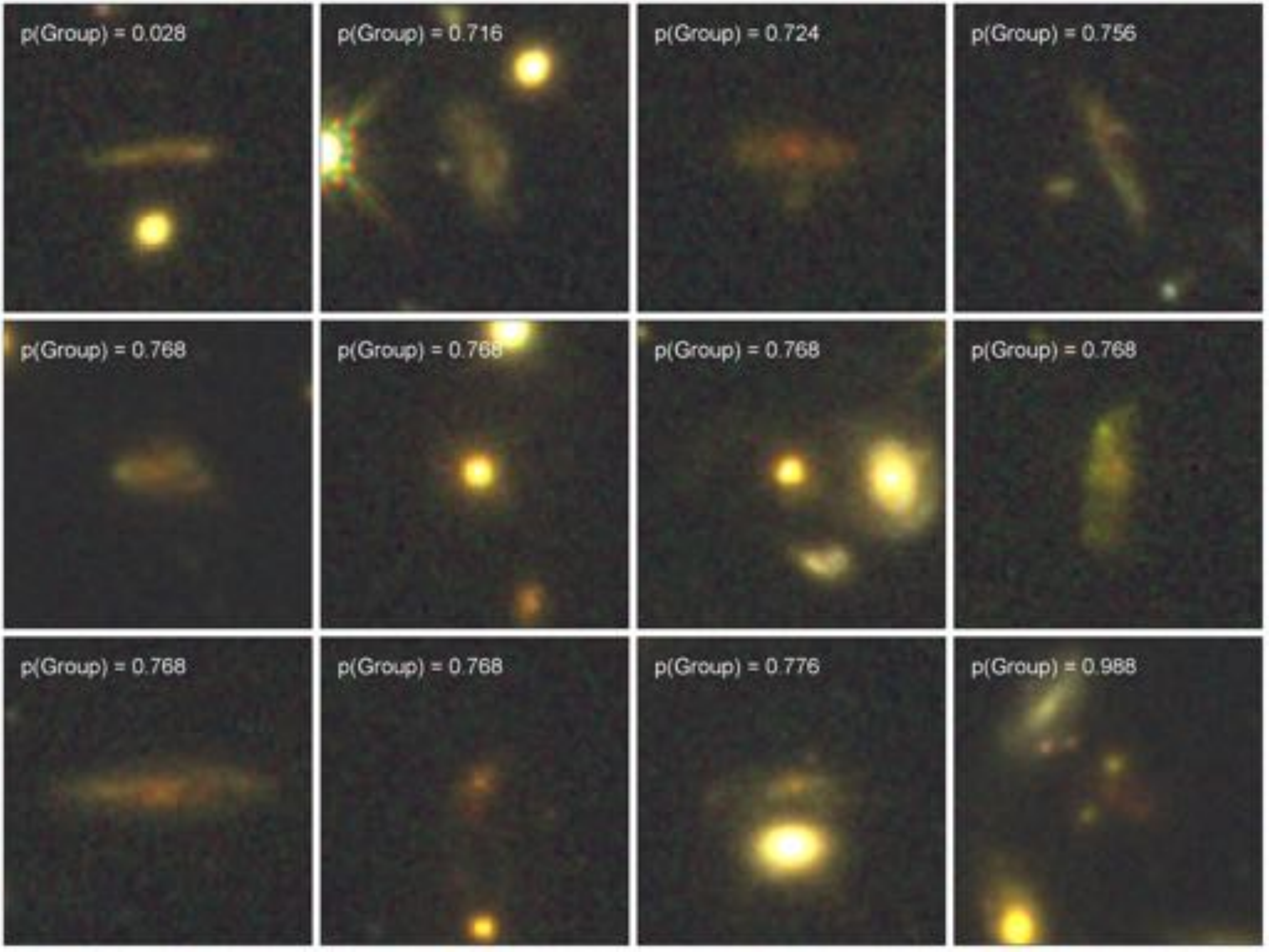}
      \centering
      \caption[Group -1 F125W 1.36 $<$ z $<$ 1.97 galaxies]
      {Group -1: Low surface brightness galaxies originally in groups 3 and 7, and only have a combined 19 galaxies which are outliers from all other groups.}
      \label{fig:f125w_stamps_group-1}
    \end{figure*}

 We observe segregations of galaxy morphology by group and describe those results as follows:

    	\begin{itemize}
	
	\item {\bf Compact or Bulge-dominated/low PC1, $\sim$ 57 per cent} 
	\begin{itemize}
	\item {\bf Group 6}: Most populated group ($\sim$ 37 per cent of sample, examples seen in Fig. \ref{fig:f125w_stamps_group6}).  Very compact and most massive galaxies; and contains the largest spheroidal (based on S\'ersic and visual classifications) and quenched fraction. 
	\item {\bf Group 0}: Large bulge+disk population, has prominent bulge with faint disk component. ($\sim$ 13 per cent,  Fig. \ref{fig:f125w_stamps_group0}).  Contains a sizable fraction of massive and quenched galaxies, not to the same extent as group 6 however.
	\item {\bf Group 9}: Large and massive galaxies with a substantial irregular population.  Visually, these galaxies posses tidal tails, bright star-forming knots and a large bulge. ($\sim$ 6 per cent,  Fig. \ref{fig:f125w_stamps_group9})
	\end{itemize}
	
	\item {\bf Bulge+Disk/intermediate PC1, $\sim$20 per cent}
	\begin{itemize}
	\item {\bf Group 4}: Smaller and less massive bulge-dominated disk galaxies with high Gini, S\'ersic index and concentration values. ($\sim$ 9 per cent,  Fig. \ref{fig:f125w_stamps_group4})
	\item {\bf Group 8}: Slightly larger bulge+disk systems. ($\sim$ 11 per cent,  Fig. \ref{fig:f125w_stamps_group8})
	\end{itemize}
	
	\item {\bf Disk-dominated/high PC1, $\sim$ 22 per cent}
	\begin{itemize}
	\item {\bf Group 1}: Large galaxies with prominent (albeit) irregular disks. ($\sim$ 8 per cent,  Fig. \ref{fig:f125w_stamps_group1})
	\item {\bf Group 2}: Compact and small disks galaxies. ($\sim$ 6 per cent, Fig. \ref{fig:f125w_stamps_group2})
	\item {\bf Group 5}: Large and low mass disk galaxies with evidence of disturbances and interactions. ($\sim$ 9 per cent, Fig. \ref{fig:f125w_stamps_group5})
	\end{itemize}
	
	\item {\bf Group -1}: Low surface brightness galaxies ($\sim$ 1 per cent, Fig. \ref{fig:f125w_stamps_group-1}) with outlier PC values.
	\end{itemize}

	The PC classification scheme separates quenched compact galaxies from larger, smooth proto-elliptical systems, and star-forming disk-dominated clumpy galaxies from star-forming bulge-dominated asymmetric galaxies.  Additionally, classifications based on PCs provide a different picture from those based on S\'ersic-n  or visual inspection.  A PC classification determines $\sim$51 per cent of galaxies are compact or bulge-dominated (groups 0, 6 and 9) while visual classifications determine $\sim$39 per cent of galaxies are bulge-dominated (either pure spheroids or disk+spheroids) and S\'ersic indices classify $\sim$20 of galaxies as bulge-dominated ($n>$ 2.5).  
	
	
	In the future we will extend our PCA classifications to different redshifts.  We will use the classifications defined here to study the evolution of star-formation for a variety of morphological types.  Star-formation can be quenched in many ways and with a reliable morphology classification for different epochs we can begin to answer the question:  whether star-formation quenching is occurring at the same time as the bulge is forming? A temporal connection between these two could have important consequences on how galaxies have been quenching star-formation.  

\section*{Acknowledgements}
This work is based on observations taken by the CANDELS Multi-Cycle Treasury Program with the NASA/ESA Hubble Space Telescope.  Support for HST Programs GO-12060 and GO-12099 was provided by NASA through grants from the Space Telescope Science Institute, which is operated by the Association of Universities for Research in Astronomy, Inc., under NASA contract NAS5-26555.  Support also came from the HST Archival program through HST-AR-12856.  We also acknowledge support from the STScI DDRF.  We would like to thank Harry Ferguson and Sandy Faber for their insightful input and fruitful discussions.  We thank the anonymous referee for their insightful feedback and corrections.

  

\bibliographystyle{mn2e_new}
\bibliography{map_biblio_pca}

\newpage

\appendix
\section{Galaxies with FLAG=1} \label{section:appendix}

\begin{figure}
      \includegraphics[scale=0.3]
      {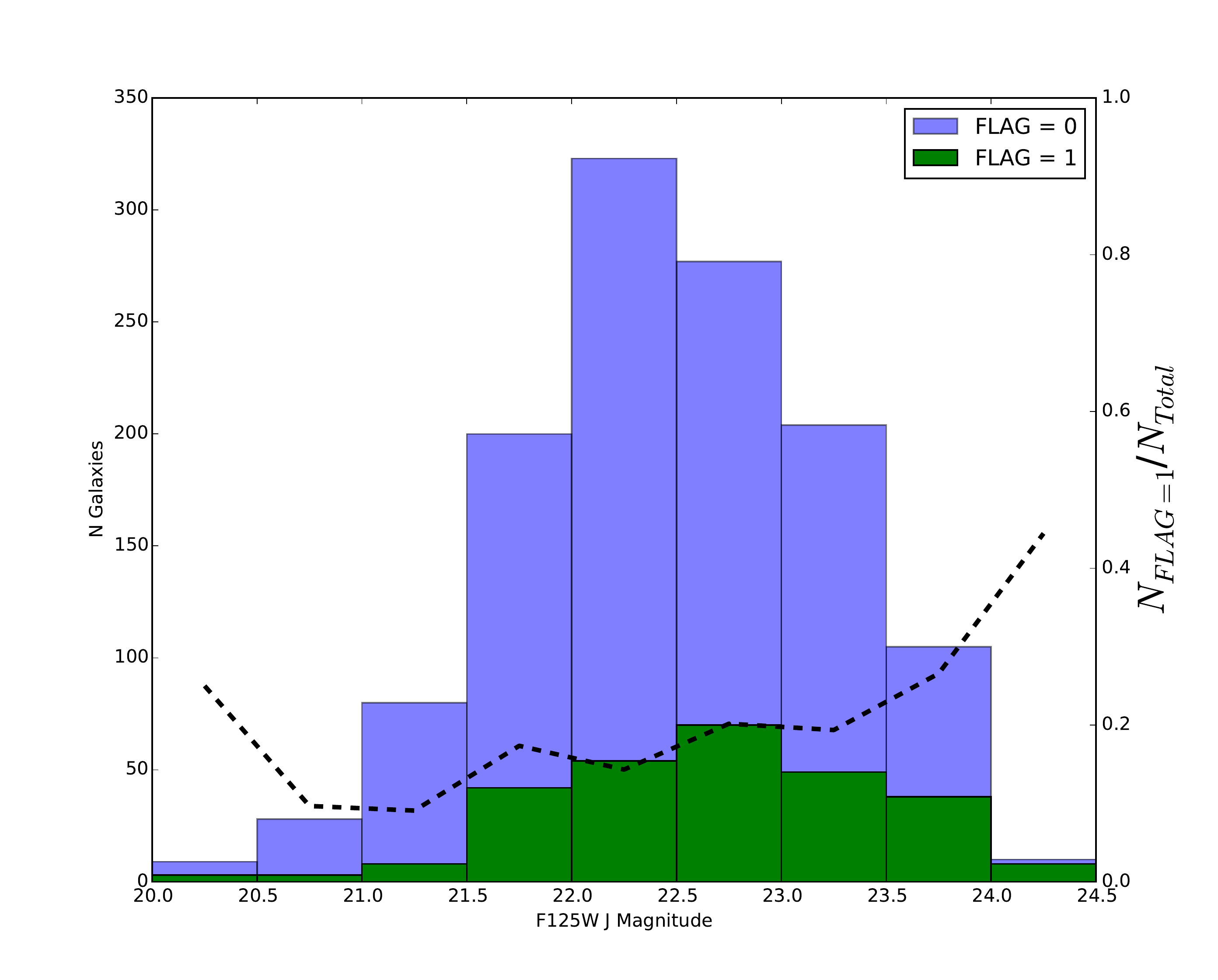}
      \centering
      \caption[Histogram of F125W J-band Magnitude for galaxies with FLAG=1 and FLAG=0]
      {Histogram of F125W J-band Magnitude for galaxies with FLAG=1 and FLAG=0 and a plot of the fraction of all galaxies with FLAG = 1 designation per magnitude bin (\textit{black dashed line}).}
      \label{fig:flag1_hist}
    \end{figure}

Galaxies with non-contiguous segmentation maps receive a FLAG=1 designation.  The disconnected segmentation maps could be the result of a few factors: the light of a nearby bright galaxy encroaching on a galaxy, low surface brightness or low signal-to-noise.  For this reason their non-parametric morphology measurements are likely to be unreliable.  Fig. \ref{fig:flag1_hist} is the normalized histogram of magnitudes for galaxies with either FLAG=0 or FLAG=1.  We also show the fraction of galaxies per magnitude bin.  The number of galaxies with FLAG=1 galaxies as a fraction of all galaxies increases up to magnitude 24.5, which is the brightness limit of the survey.  For these reasons we leave these galaxies out of our sample in this work, but we will investigate these galaxies in a future work.


\section{Concentration - S\'ersic Index Relationship} \label{section:c_n}

\begin{figure*}
  \centering
    \includegraphics[scale=0.4]
    {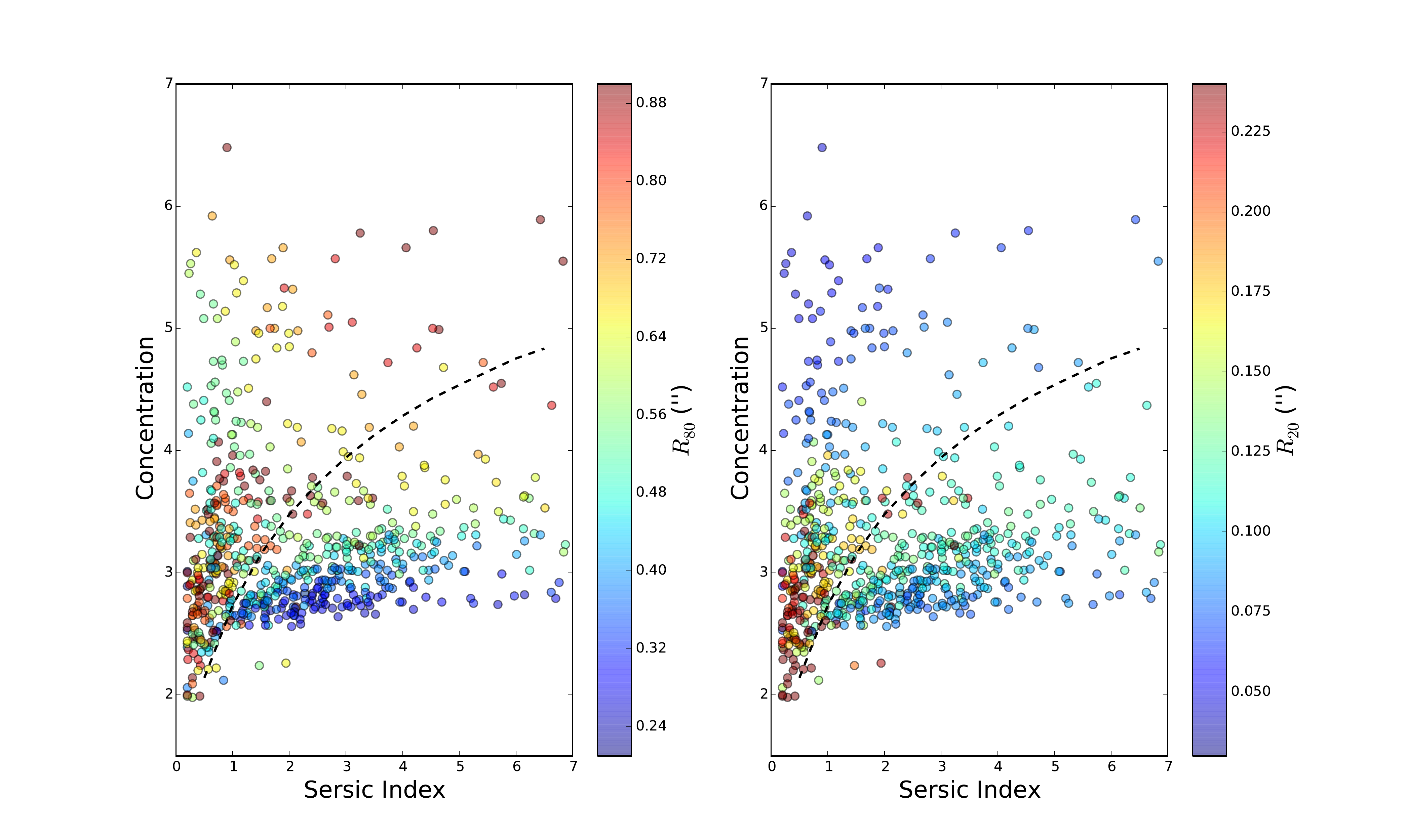}
      \caption[Concentration - S\'ersic index for sample]
      {WFC3 125W measured concentration versus F125W S\'ersic index \citep{vanderWel12} color coded by (\textit{left panel}) $R_{80}$ and (\textit{right panel}) $R_{20}$ for the entire sample.  Our $z\sim$1.5 galaxies generally follow a shallower relation than previously shown in \citealt{Andrae11} (\textit{black dashed line}).}
      \label{fig:c-n}
    \end{figure*}

\citet{Andrae11} demonstrated the correlation between concentration and S\'ersic-n.  However, this relationship does not appear to hold for our high redshift sample. Fig. \ref{fig:c-n} shows a less established relationship for concentration and S\'ersic-n in our galaxy sample.   We show that concentration is biased low for very small ($r_e <$ 2 kpc), high S\'ersic n galaxies ($n >$ 2.5) which represents $\sim$14 per cent of our sample.  We also find many z$\sim$1.5 galaxies with high concentration and low S\'ersic-n that deviate from the Andrae et al. (2011) relation and are not easily explained by measurement bias. 

The PSF for F125W has a full width half-maximum (FWHM) of $\sim$0.135''.  For many galaxies, $r_{20}$ is smaller than the PSF (and in some cases even $r_e$ is smaller than the PSF).  Fig. \ref{fig:c-n} shows that high S\'ersic galaxies make up some of the smallest objects in our sample.  These small galaxies can have $r_{80}$ $\sim$ 0.48'', which is only a few times larger than the PSF.  


We wish to test the effect of the size of the PSF can have on measuring the concentration index, particularly for small galaxies.  To accomplish this we take a pure S\'ersic surface brightness light profile I $\sim \exp[(r/r_e)^{1/n}]$ with $r_e$ = 10 kpc and calculate the Petrosian Radius (Eq. \ref{eq:petrosian_r}), $r_{80}$, $r_{20}$ and thus concentration.  We convolve the pure S\'ersic profile with a gaussian with the same FWHM as the PSF.  This convolution has little effect on the concentration for large galaxies.  However, we noticed in Fig. \ref{fig:c-n} that many galaxies have very small $r_e$ values which could lead to why concentration values are lower than anticipated.  To test this hypothesis we convolve the S\'ersic surface brightness profile of a small galaxy ($r_e$ = 1 kpc and 2 kpc) with a gaussian with the FWHM of the PSF.   This will allow us to observe the effect of convolving the surface brightness profile of a small galaxy with a PSF of comparable size.

\begin{figure}
  \centering
    \includegraphics[width=0.5\textwidth]
    {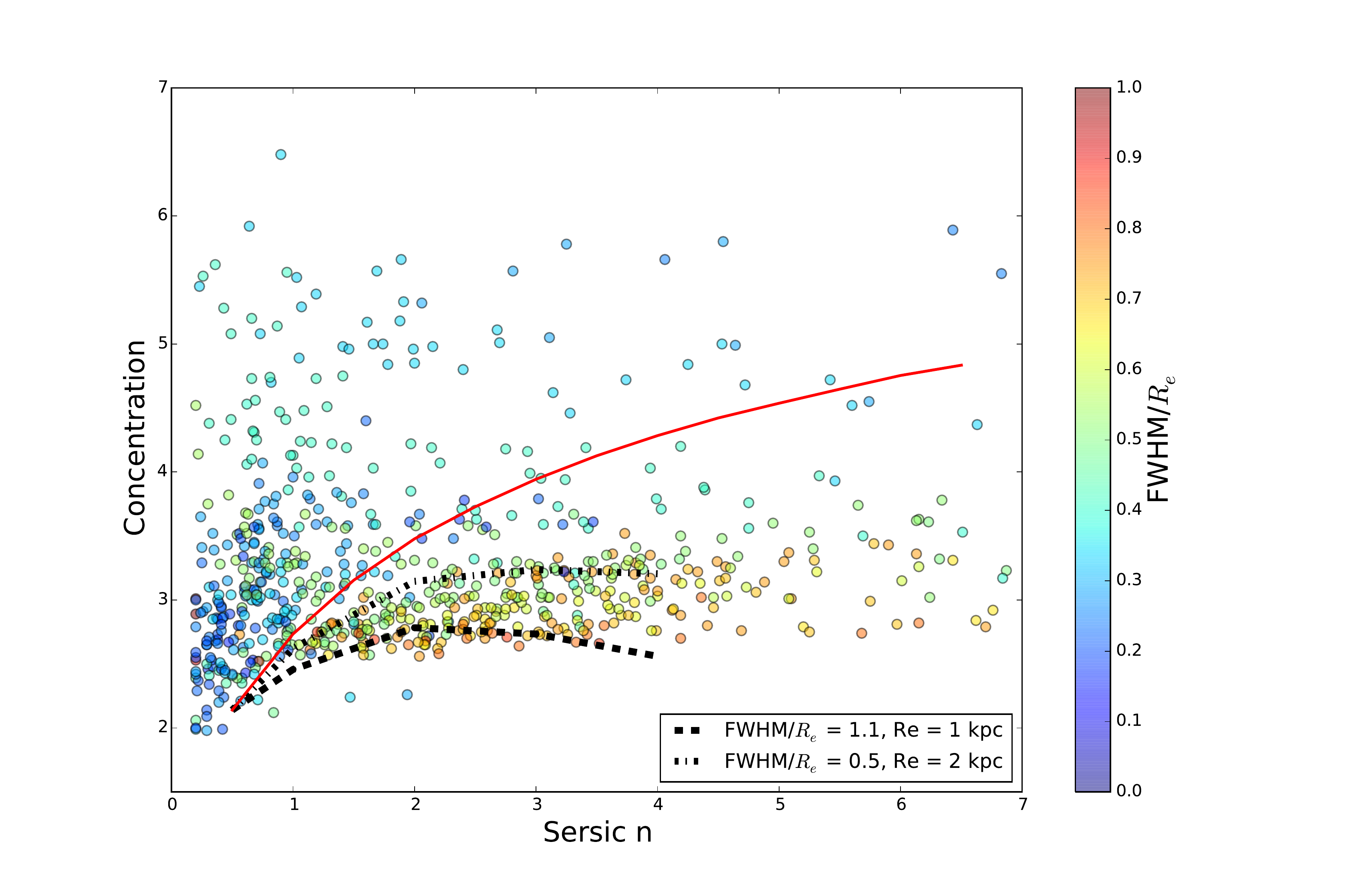}
      \caption[Concentration - S\'ersic Including PSF Effects]
      {Concentration versus S\'ersic index relation color-coded by the ratio of the PSF FWHM ($\sim$0.135'') to the effective radius of a galaxy.  We plot the numerically defined relationship for a pure Sersic profile (\citealp{Andrae11}., \textit{red line}) and the corrections to the pure Sersic profile when the PSF FWHM is 50 per cent the size of a 2 kpc galaxy (\textit{thick-thin dashed line}) and when the PSF is as large as a 1 kpc galaxy (\textit{thick dashed line}).  The relative size of the PSF to a galaxy has a large impact on the concentration values for galaxies with higher S\'ersic indices.}
      \label{fig:c-n-corrected}
    \end{figure}

Fig. \ref{fig:c-n-corrected} shows the concentration - S\'ersic relation present in our galaxy sample and is color coded by the ratio of the size of the PSF to the effective radius of a galaxy. The solid red line in Fig. \ref{fig:c-n-corrected} shows the relation between concentration and S\'ersic calculated for a pure S\'ersic surface brightness profile with $r_e$ = 10 kpc (first demonstrated in \citealt{Andrae11}).   The thin-thick and thick dashed lines in Fig. \ref{fig:c-n-corrected} show the concentration - S\'ersic relation for a surface brightness profile (of a $r_e$ = 1 kpc or 2 kpc galaxy) convolved with a gaussian with the FWHM of the PSF.    Galaxies with high FWHM/$R_e$ ratios (i.e. the galaxy has a comparable physical size to the PSF) fall noticeably below the concentration-S\'ersic relation for a pure S\'ersic surface brightness profile.  The flatter concentration-S\'ersic relation of the small galaxy surface brightness profiles convolved with the PSF closely follows the concentration and S\'ersic values we measure for our sample.  As the physical size of a galaxy decreases the concentration values are increasingly depressed.  We take this as evidence that small galaxies (those with physical sizes similar to the PSF, $r_e \sim$ 1-2 kpc) are most affected by the PSF.  Thus the reason the concentration values for our galaxies are smaller than the relation of \citet{Andrae11} is likely due to the small physical sizes of many galaxies in our sample.


Up to 14 per cent of our total sample maybe be quite small (roughly the size of the PSF, $r_e <$ 2 kpc) and have a high S\'ersic index ($n >$ 2.5) leading to an artificially depressed concentration value.  Many of these galaxies ($\sim$80 per cent) are in group 6.  After correcting the concentration values these galaxies would instead be classified into group 0.  This suggests that a portion of the group 6 galaxies would instead be group 0 if we had higher resolution images.  However, this implies only $\sim$26 per cent of all group 6 galaxies would be reclassified as group 0 so there is still a notable distinction between these two groups.

\label{LastPage}
\end{document}